\documentclass{aa}  
\usepackage{makecell}
\usepackage[utf8]{inputenc}
\usepackage{booktabs}
\usepackage{relsize}
\usepackage{physics}
\usepackage{xcolor}
\usepackage[scientific-notation=true]{siunitx}
\usepackage{threeparttable}
\usepackage{amsmath}
\usepackage{graphicx}
\usepackage{txfonts}
\usepackage{subcaption} 
\usepackage[draft]{hyperref}
\usepackage[utf8]{inputenc}
\usepackage[english]{babel}
\hypersetup{
    colorlinks=true,
    linkcolor=blue,
    filecolor=magenta,      
    urlcolor=teal,
    citecolor=black
}

\begin{document} 

   \title{TRAPPIST-1: Global Results of the Spitzer Exploration Science Program {\it Red Worlds} }
   \author{E. Ducrot,
          \inst{1}\fnmsep\thanks{\email{educrot@uliege.be}}
          M. Gillon\inst{1},
          L. Delrez\inst{1,5,14},
          E. Agol\inst{10},
          P. Rimmer\inst{2,3,4},
          M. Turbet\inst{5},
          M. N.\ G{\"u}nther\inst{6,7},
          B-O. Demory\inst{8},
          A. H. M. J. Triaud\inst{9},
          E. Bolmont\inst{5},
          A. Burgasser\inst{12},
          S. J. Carey\inst{13},
          J. G. Ingalls\inst{13},
          E. Jehin\inst{14},
          J. Leconte\inst{15},
          S. M. Lederer\inst{16},
          D. Queloz\inst{3,5},
          S. N. Raymond\inst{15},
          F. Selsis\inst{15},
          V. Van Grootel\inst{14},
          J. de Wit\inst{11}
          }

   \institute{Astrobiology Research Unit, Université de Liège, Alléee du 6 Aout 19C, B-4000 Liège, Belgium
        \and
             Department of Earth Sciences, University of Cambridge, Downing St, Cambridge CB2 3EQ, UK
        \and
             Cavendish Laboratory, JJ Thomson Avenue, Cambridge, CB3 0H3, UK 
        \and
             MRC Laboratory of Molecular Biology, Cambridge Biomedical Campus, Francis Crick Ave, Cambridge CB2 0QH, UK
        \and
             Observatoire astronomique de l’Université de Genève, 51 chemin de Pégase, 1290 Sauverny, Switzerland
        \and
            Department of Physics, and Kavli Institute for Astrophysics and Space Research, Massachusetts Institute of Technology, Cambridge, MA 02139, USA
        \and
            Juan Carlos Torres Fellow
        \and
             University of Bern, Center for Space and Habitability, Gesellschaftsstrasse 6, CH-3012, Bern, Switzerland
        \and
             School of Physics \& Astronomy, University of Birmingham, Edgbaston, Birmimgham B15 2TT, UK
        \and
            Astronomy Department, University of Washington, Seattle, WA 98195 USA
        \and
             Department of Earth, Atmospheric and Planetary Science, \\Massachusetts Institute of Technology, 77 Massachusetts Avenue, Cambridge, MA 02139, USA 
        \and 
            Center for Astrophysics and Space Science, University of California San Diego, La Jolla, CA, 92093, USA
        \and 
            IPAC, California Institute of Technology, 1200 E California Boulevard, Mail Code 314-6, Pasadena, California 91125, USA
        \and
            Space Sciences, Technologies and Astrophysics Research (STAR) \\	Institute, Université de Liège, Allée du 6 Août 19C, B-4000 Liège, Belgium
        \and
            Laboratoire d’astrophysique de Bordeaux, Univ. Bordeaux, CNRS, B18N, Allée Geoffroy Saint-Hilaire, F-33615 Pessac, France
        \and
             NASA Johnson Space Center, 2101 NASA Parkway, Houston, Texas, 77058, USA
             }
             
  %\date{Received }

  \abstract
    {With more than 1000 hours of observation from Feb 2016 to Oct 2019, the Spitzer Exploration Program Red Worlds (ID: 13067, 13175 and 14223) exclusively targeted TRAPPIST-1, a nearby (12pc) ultracool dwarf star, finding that it is orbited by seven transiting Earth-sized planets. At least three of these planets orbit within the classical habitable zone of the star, and all of them are well-suited for a detailed atmospheric characterization with the upcoming JWST.}% aims heading (mandatory)
    {The main goals of the Spitzer Red Worlds program were (1) to explore the system for new transiting planets, (2) to intensively monitor the planets' transits to yield the strongest possible constraints on their masses, sizes, compositions, and dynamics, and (3) to assess the infrared variability of the host star. In this paper, we present the global results of the project. }
    {We analyzed 88 new transits and combined them with 100 previously analyzed transits, for a total of 188 transits observed at 3.6 or 4.5 $\mu$m. For a comprehensive study, we analyzed all light curves both individually and globally. We also analyzed 29 occultations (secondary eclipses) of planet b and eight occultations of planet c observed at 4.5 $\mu$m to constrain the brightness temperatures of their daysides.}
    {We identify several orphan transit-like structures in our Spitzer photometry, but all of them are of low significance. We do not confirm any new transiting planets. We do not detect any significant variation of the transit depths of the planets throughout the different campaigns. Comparing our individual and global analyses of the transits, we estimate for TRAPPIST-1 transit depth measurements mean noise floors of $\sim$35 and 25 ppm in channels 1 and 2 of Spitzer/IRAC, respectively. We estimate that most of this noise floor is of instrumental origins and due to the large inter-pixel inhomogeneity of IRAC InSb arrays, and that the much better interpixel homogeneity of JWST instruments should result in  noise floors as low as 10ppm, which is low enough to enable the atmospheric characterization of the planets by transit transmission spectroscopy. Our analysis reveals a few outlier transits, but we cannot conclude whether or not they correspond to spot or faculae crossing events. We construct updated broadband transmission spectra for all seven planets which show consistent transit depths between the two Spitzer channels. Although we are limited by instrumental precision, the combined transmission spectrum of planet b to g tells us that their atmospheres seem unlikely to be $\mathrm{CH}_{4}$-dominated. We identify and model five distinct high energy flares in the whole dataset, and discuss our results in the context of habitability. Finally, we fail to detect occultation signals of planets b and c at 4.5 $\mu$m, and can only set 3$\sigma$ upper limits on their dayside brightness temperatures (611K for b 586K for c).}
    {}
    
    \titlerunning{TRAPPIST-1: Global Results of the Spitzer Exploration Science Program}
    \authorrunning{E. Ducrot}
    \maketitle
%-------------------------------------------------------------------
\section{Introduction}
    
    Thanks to their small size, mass, and luminosity combined with their relatively large infrared brightness, the nearest ultracool dwarf stars (spectral types M7 and later) represent promising targets for the detailed study of potentially habitable transiting exoplanets with upcoming giant telescopes such as JWST or ELTs \citep{Kaltenegger2009,Malik2019,Mansfield2019,Koll:2019}. This fact motivated the development of the new ground-based transit survey called the Search for habitable Planets EClipsing ULtra-cOOl Stars (SPECULOOS). SPECULOOS aims to explore the $\sim$1000 nearest ultracool dwarf stars for transiting rocky planets as small as the Earth \citep{Burdanov2017, Gillon2018, Delrez2018b}. While SPECULOOS only started its operations in 2019, a prototype version of the survey has been ongoing since 2011 on the TRAPPIST-South telescope \citep{Gillon2013} in Chile. As of 2015, this prototype survey has initially detected three transiting temperate Earth-sized planets around an isolated M8 dwarf star at $\simeq$ 12 parsecs from Earth, named the TRAPPIST-1 system \citep{Gillon2016}. The intensive ground-based photometric follow-up of this system suggests that it hosted several other transiting planets, while leaving their actual number and orbital periods ambiguous. In this context, the Spitzer Exploration Science Program \textit{Red Worlds} was initiated a 20-day long (i.e., 480 hrs) near-continuous monitoring of the system at 4.5 $\mu \mathrm{m}$, which revealed that it hosted no less than seven planets \citep{Gillon2017}. This was followed by intense, high-precision monitoring of the eclipses of the known planets at 3.6 $\mu \mathrm{m}$ and 4.5 $\mu \mathrm{m}$ from 2017 to 2018 (520 hrs).  The initial 20-day observation campaign
    revealed only one transit of the outermost planet ($h$), making follow-up
    impossible due to the unknown orbital period until four additional transits were detected with the K2 spacecraft,
    which enabled further monitoring with Spitzer \citep{Luger2017a}.

    One of the primary goals of this ambitious Spitzer program was to create a complete inventory of the transiting objects (planets, moons, Trojans) of the inner system of TRAPPIST-1, not only to constrain its dynamical properties, history, and stability, but also to identify more objects well-suited for detailed atmospheric characterization with next-generation telescopes. It also aimed to perform a thorough assessment of the infrared variability of the star. Finally, it aimed to determine the masses and constrain the orbital parameters of the planets through the transit timing variation method \citep[TTV; ][]{Agol2005, Holman2005}. The precise and accurate determination of the masses and radii of the planets - and the resulting constraints on their bulk compositions - is indeed critical for their thorough characterization, notably for the optimal exploitation of future atmospheric observations \citep{Morley2017}. 
    
    Since its discovery, a large number of research investigations have been dedicated to the study of the TRAPPIST-1 system. While a fully comprehensive list of all observational and theoretical results published to date is not practical for this work, we overview several important characteristics of this system in the following paragraph.\\
    
    First, the host star is an old M8V type star ($7.6 \pm 2.2$ Gyr, \citealt{Burgasser2017}) with a moderate flaring activity (about 1 or  2 flares per week, \citealt{Gillon2017}) and its (putative) stellar rotation period derived from K2 observations by \cite{Luger2017a} is $3.30 \pm 0.14$ days. 
    Recent study by \cite{Gonzales2019} presented a distance-calibrated SED for the star and found, from band-by-band comparisons, that TRAPPIST-1 exhibits a blend of field star and young star spectral features. 
    Its XUV luminosity is similar to the Sun's, which, when considering its past  evolution - notably its $\sim$2 Gyr-long premain sequence phase \citep{VanGrootel2018} and the small orbital distances of the planets (between 0.01 and 0.06 au) - potentially drove extreme atmospheric erosion and water  loss \citep{Wheatley2016,Bolmont2016,Bourrier2017,Fleming2019}. In short, if the habitable zone planets originally had primordial $\mathrm{H}/\mathrm{He}$ envelopes, XUV evaporation may have rendered the planets habitable  \citep{Luger2015b,Owen2016}.
    Three planets orbit within the habitable zone of the star \citep{Gillon2017}, and planet e is the most likely to harbour liquid water on its surface \citep{Wolf2017,Wolf:2018apj,Turbet:2018aa,Fauchez:2019gmd}.\\
    The seven planets form the longest resonant chain known to date \citep{Luger2017b, Papaloizou:2018}.
    Some works modeled the planet formation process from small dust grains to full-sized planets, while keeping track of their water content using pebble and planetesimal accretion mechanisms \citep{Ormel2017,Schoonenberg2019,Coleman:2019aa}. \\
    The planets are good potential targets for atmospheric characterization with JWST \citep{LustigYaeger2019,Lincowski2018, Lincowski2019, Wunderlich2019, KrissansenTotton2018, Barstow2016, Fauchez:2019apj}. 
    Preliminary atmospheric prescreening was performed with HST/WFC3 and the resulting low-resolution transmission spectra acquired in the 1.1-1.7 $\mu \mathrm{m}$ spectral range made it possible to exclude clear hydrogen-dominated atmospheres for six of the seven planets \citep{deWit2016, deWit2018, Wakeford2018}. \\
    Lastly, some works have suggested that the heterogeneous photosphere of the host star could overwhelm planetary atmospheric absorption features in transit transmission spectra because of the so-called transit light source (a.k.a. stellar contamination) effect \citep{Apai2018,Rackham2018}. Associated models predict important spot and faculae covering fractions (\citealt{Zhang2018} find faculae covering $\simeq$ 50\%, spots covering $\simeq$ 40\%, while
    \citealt{Wakeford2018} find $5800$K hot spots covering $<$ 3\% and
    $3000$ K hot spots covering $\simeq$ 35\%) for TRAPPIST-1. \\
    
    While the TRAPPIST system is gradually revealing its properties, some big questions still remain. For instance, we still can not  explain why the rotational modulation seen in K2 data is not detected in Spitzer light  curves \citep{Delrez2018, Luger2017b,Morris2018a}. Neither do we know if the host star's high-energy incident flux on the planets can jeopardize their habitability \citep{Roettenbacher2017,Vida_2017} or if it can alternatively drive chemical processes needed for life's origin, through, for example, CME-driven generation of prebiotically relevant molecules \citep{Air2016}, and by increasing NUV flux for the production of life's building blocks \citep{Ranjan2017,Rimmer2018}. We are also uncertain on the information content will we be able to retrieve from the planetary transmission spectra and how significant the impact of stellar contamination may be on their interpretation.
    
    In this context, our work aims to meet the initial expectations of the \textit{Red Worlds} Spitzer exploration program and present them within the framework of the most recent studies on stellar contamination, atmospheric retrieval, and habitability. \\
    
    This paper is structured as follows. In Section \ref{dataanalysis}, we present the observations, their reduction, and describe the data analysis. In Section \ref{discussion}, we discuss the results brought by the many analyses carried out, notably the evolution of the measured transit depths over time, the transmission spectra of the planets, their interpretation in terms of stellar contamination and atmospheric spectral features retrieval, the transit timing variations, the occultation signals and emission spectra, the impact of flares on the potential habitability of the planets, and finally the presence of orphan structures in the photometry. We summarize our results in Section \ref{conclusion}.

%---------------------------------------------------------------------------------------------------------------

\section{Observations and data analysis} \label{dataanalysis}
\subsection{Observations and data reduction}

\begin{figure*}[ht!]
    \centering 
    \includegraphics[width=\textwidth]{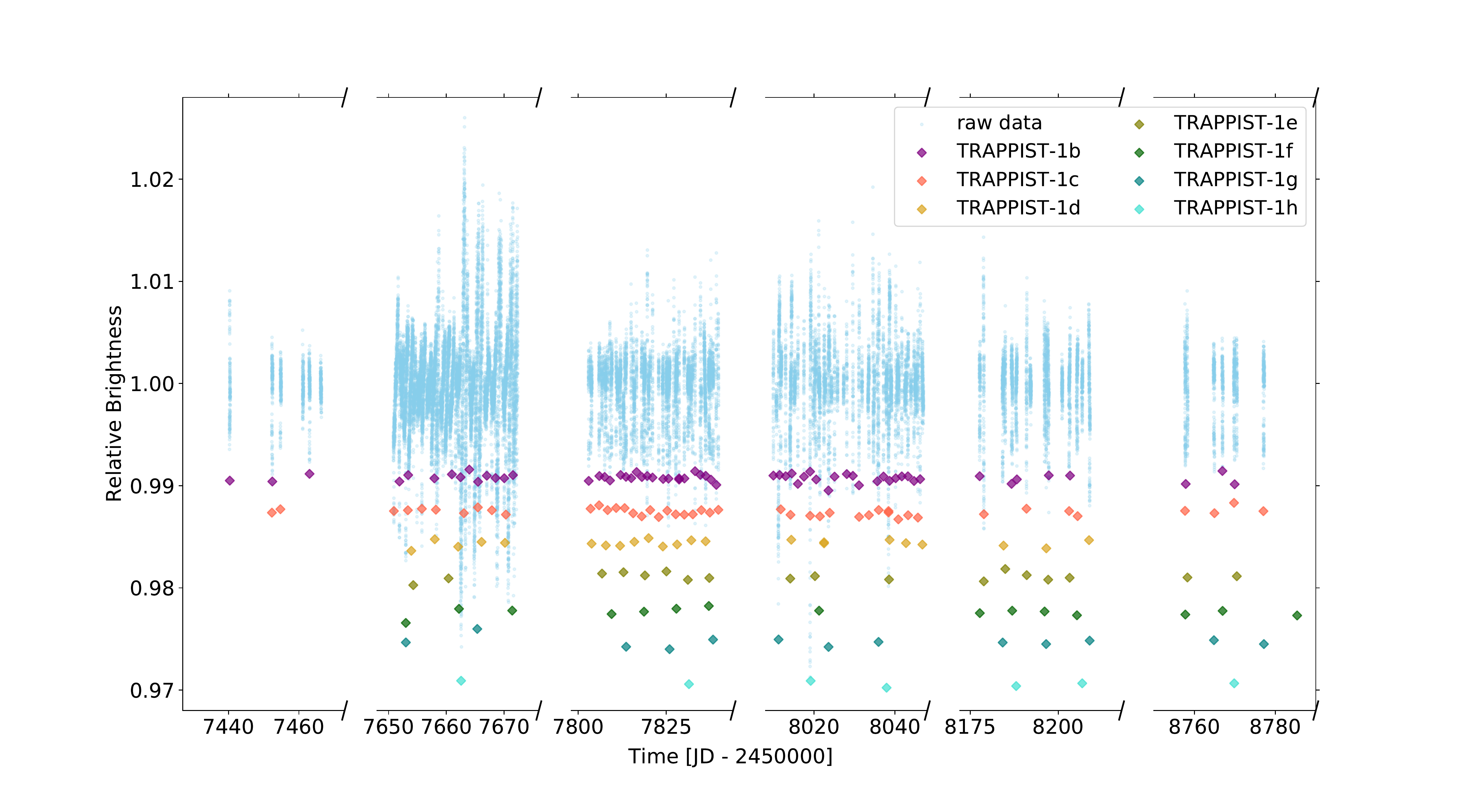}
    \caption{Spitzer photometric measurements (sky blue) resulting from observations of the star from February 2016 to Oct 2019 cleaned of data gaps between the four campaigns. Colored diamonds show the positions of the transits of the different planets with their corresponding depth + a constant offset by planet for clarity.}
    \label{all_LCs}
\end{figure*}

The dataset used in this work includes all time-series observations of TRAPPIST-1 carried out by Spitzer/IRAC since the discovery of its planetary system: 45hrs of observations gathered within the DDT program 12126 in Feb and March 2016 (\citealt{Gillon2017}, \citealt{Delrez2018}, hereafter D2018), and all data (1080hr) taken within the Spitzer Exploration Science program Red Worlds (ID 13067) between Feb 2017 and Oct 2019 (see Figure \ref{all_LCs}) including data from the DDT program 13175 (PI: L. Delrez) targeting occultations of the two inner planets and data from the DDT program 14223 (PI: E. Agol) taken in Oct 2019 to better constrain the masses of the planets and to tighten the ephemeris forecast for observations with JWST, see \cite{Agol2020}. All these data can be accessed through the online Spitzer Heritage Archive database\footnote{\url{http://sha.ipac.caltech.edu}}. This extensive dataset includes 65, 47, 23, 18, 15, 13, and 7 transits of planets b, c, d, e, f, g, and h, respectively. Among these 188 transits, 88 are "new", that is, they were observed in fall 2017 and fall 2019, and were not included in the analysis discussed in D2018 which presented data taken by Spitzer through March 2017. Our aim is to give an overview of the exploration of TRAPPIST-1 system with the Spitzer space telescope, we therefore did not include transits observed with other telescopes, but the results of the analysis of those additional observations can be found in existing papers: \cite{Luger2017b,Grimm2018} for K2 observations, \cite{deWit2016,deWit2018,Wakeford2018} for HST observations, \cite{Ducrot2018} for SPECULOOS and Liverpool telescope observations, \cite{Burdanov2019} for VLT, AAT and UKIRT observations. Nevertheless, we use those results later in the paper to constructed the transit transmission spectra of the planets, in Section.\ref{transit_transmission}

Back on the Spitzer dataset, we identified 29 blended transits (that is to say transits of multiple planets simultaneously) or partial transits (see Table \ref{nb_transits}) which were analyzed individually but not included in our global analysis presented in Section 3. Indeed, shapes of blended + partial transits are less constraining than well isolated full transits so we chose not to include them in our global analysis of all transits to have a better convergence. We also did not include them in our global planet by planet analyses for similar reasons when the transit was partial, and because we wished to deal with only one planet in these analyses.

Besides, we targeted 28 occultations of TRAPPIST-1b and 9 of TRAPPIST-1c with Spitzer/IRAC channel 2 with the aim to detect a signal or at least obtain an upper bound on the occultation depth and consequently derive the first empirical constraints on the planets' thermal emission. Indeed, as the orbital planes of the seven TRAPPIST-1 planets are aligned to $<$ $0.3\deg$ at $90\%$ confidence \citep{Luger2017b} their orbital tracks overlap over large fraction of their orbit \cite{Luger2017a}, hence we expect all planets to undergo secondary ecplise. Given the small size of the planets, our best chance to catch an occultation signal is to phase-fold several occultations, which is why focusing on the inner planets (with the smallest periods) is the wisest choice. Furthermore, an updated TTV analysis using all transits observed by Spitzer, HST, K2, and ground-based transits observed up to 2019 \citep{Agol2020} confirms that the expected eccentricities are very small (eccentricity < 0.01 for all planets). In this context, we note that we did not spot any sign of planet-planet eclipses in our analyses of the blended transits, and for none of them was a planet “caught up” by a more inner one during its crossing of the stellar disk.

As described by \cite{Gillon2017}, the star was observed nearly-continuously from 19 Sep to 10 Oct 2016 within the program 13067 (480hrs). The rest of the dataset (1061hrs) is composed of sequences of a few hrs corresponding to the observations of one or several transit(s) and/or occultation(s). For all observations in both bandpasses, each frame is composed of 64 subexposures each of 1.92 seconds on the target plus an additional 0.8s for read out, which gives a cadence of a point every 2.06 minutes. 

\begin{table*}[h!]
    \centering                          % used for centering table
    \begin{tabular}{cccc}        % centered columns (4 columns)
        \hline\hline                 % inserts double horizontal lines
        Planet  & \# Isolated transits & \# Blended or partial transits & Total \\    % table heading 
        \hline                        % inserts single horizontal line
           TRAPPIST-1\,b  & 54  & 11  & 65 \\      % inserting body of the table
           TRAPPIST-1\,c  & 39  & 8 & 47\\  
           TRAPPIST-1\,d  & 20  & 3 & 23\\  
           TRAPPIST-1\,e  & 17  & 1 & 18\\  
           TRAPPIST-1\,f  & 13  & 2 & 15\\  
           TRAPPIST-1\,g  &  9  & 4 & 13\\  
           TRAPPIST-1\,h  &  6  & 1 & 7\\  
        \hline                                   %inserts single line
    \end{tabular}
    \caption{\label{nb_transits} Number of transits monitored by Spitzer from early 2016 to late 2019 for each TRAPPIST-1 planet}
\end{table*}

All these observations were obtained with the Infrared Array Camera (IRAC) \citep{Carey2004} of the Spitzer Space telescope in subarray mode (32 $\times$ 32 pixels windowing of the detector) with an exposure time of 1.92 s. No dithering was used (continuous staring), and the observations were all done using the `peak-up' mode \citep{Ingalls2016} to maximize the accuracy in the position of the target on the detector's sweet spot (as detailed in \href{https://irsa.ipac.caltech.edu/data/SPITZER/docs/irac/iracinstrumenthandbook/}{IRAC Instrument Handbook}) to minimize the so-called ‘pixel phase effect’ of IRAC detectors (e.g., \citealt{Knutson2008}). All the data were calibrated with the Spitzer pipeline S19.2.0, and delivered as cubes of 64 subarray images of 32 $\times$ 32 pixels (pixel scale = 1.2 arcsec). 

All 2016 and 2019 data were obtained with the 4.5 $\mu$m IRAC detector. In 2017 and 2018, additional observations were obtained at 3.6$\mu$m with the goal of further constraining the chromatic variability of the transit depths of the seven planets. 
The same method was used for the photometric extraction as that described by \cite{Gillon2017} and D2018. We converted the fluxes from MJy/sr to photon counts, and then we used the IRAF/DAOPHOT\footnote{IRAF is distributed by the National Optical Astronomy Observatory, which is operated by the Association of Universities for Research in Astronomy, Inc., under cooperative agreement with the National Science Foundation.} software \citep{Stetson1987} to measure the flux of TRAPPIST-1 within a circular aperture of 2.3 pixels. For each subarray image, the aperture was centered on the star's point-spread function (PSF) by fitting a 2D Gaussian profile, yielding measurements of the PSF width along the x- and y-axis in the process. We discarded subarray images corresponding to > 10 $\sigma$ discrepant measurements for the PSF center, target flux, and background flux, as described by \cite{Gillon2014}. We then combined the measurements per cube of 64 images. The photometric error of each cube (which is the standard error on the mean) was taken as the error of its average flux measurement. 

%---------------------------------------------------------------------------------------------------------------

\subsection{Data analysis}
Our data analysis was divided in three distinct steps. First, we performed individual analyses of each transit light curve to select an optimal photometric model and assess the variability of the photometry, see Section \ref{individual_analysis}. We also carried out an analysis aiming to refine the stellar parameters of TRAPPIST-1, see Section \ref{stellar_param}. Then, we performed several sets of global analyses: (a) one with the entire set of transits to refine the physical parameters of the system; (b) seven others (one for each planet) for which we allow the transit depths to vary in order to assess their stability; and (c) a repeat of the seven global analyses (planet by planet), this time to improve the errors on the timings, see Section \ref{global_analysis}. Finally, we carried out a global analysis of the light curves from program 13175 (PI: L. Delrez) obtained around the expected occultation times for planet b and c to search for occultation signals, see Section \ref{Occultations} for details. 

%---------------------------------------------------------------------

\subsubsection{Individual analyses}\label{individual_analysis}

 We used our adaptive Markov-Chain Monte Carlo (MCMC) code \citep{Gillon2010,Gillon2012,Gillon2014} to analyze each transit light curve individually (that is to say each individual transit). It uses the eclipse model of \cite{Mandel2002} as a photometric time-series, multiplied by a baseline model to represent the other astrophysical and instrumental systematics that could produce photometric variations.

First, we select a model to represent each light curve through the minimization of the Bayesian Information Criteria (BIC; \citealt{Schwarz1978}) which is given by the formula:

\begin{equation}
    BIC = \chi ^{2} + k \times log(N)
\end{equation}

where $k$ is the number of free parameters of the model, $N$ is the number of data points, and $\chi ^{2}$ is the best-fit chi-square. We tested a large range of baseline models to account for different types of external sources of flux variations/modulations (instrumental and stellar effects). This includes polynomials of variable orders in: PSF size and position on the detector (to account for the Spitzer "pixel-phase" effect  and the breathing of its PSF; \citealt{Gillon2017}); time (to correct for time dependent trends); and the logarithm of time (to represent the "ramp" effect, \citealt{Knutson2008}). For some light curves the "pixel-phase" effect was additionally corrected by complementing the position polynomial model with the bi-linearly-interpolated subpixel sensitivity (BLISS) mapping method presented by \cite{Stevenson2012}. To do so, we sampled the detector area probed by the PSF center in several sub-pixel box such that at least five measurements fell within the same box. Further details of the implementation of BLISS mapping in our MCMC code can be found in \cite{Gillon2014}. The details of the baseline model adopted for each transit light curve are given in  Table \ref{baseline_indiv}. Once the baseline was chosen, we ran a preliminary analysis with one Markov chain of 50 000 steps to evaluate the need for re-scaling the photometric errors through the consideration of a potential under- or over-estimation of the white noise of each measurement and the presence of time-correlated (red) noise in the light curve. The white noise is represented by the factor $\beta _{w}$ issued from the comparison of the rms of the residuals and the mean photometric errors. The red noise is represented by the scaling factor $\beta _{r}$ derived from the rms of the binned and unbinned residuals for different binning intervals ranging from 5 to 120 min, following the procedure detailed in \cite{Winn2009}. The values of $\beta _{w}$ and $\beta _{r}$ derived for each light curve are listed in Table \ref{baseline_indiv}.

The jump parameters that were randomly perturbed at each step of the Markov chains were:

\begin{itemize}%[noitemsep]
\item the mass $M_{\star}$, the radius $R_{\star}$, the effective temperature $T_{eff}$, and the metallicity [Fe/H] of the star, assuming the following prior probability distribution functions  (PDFs) for these stellar parameters: $M_{\star} \in  \mathcal{N}(0.089,0.007^{2})M_{\odot}$, $R_{\star} \in \mathcal{N}(0.121,0.003^{2})R_{\odot}$, $T_{eff} \in \mathcal{N}(2511,37^{2})$K and [Fe/H] $\in \mathcal{N}(0.04,0.08^{2})\mathrm{dex}$;
\item the planet/star area ratio $dF = (\frac{R_{p}}{R_{\star}})^{2}$, where $R_{p}$ and $R_{\star}$ are the radius of the planet and the star, respectively;
%Michael: you should give a reference for these priors.
%------------------------------------------------------
% \item the occultation depth(s) $dF_{occ}$ (one per filter) for planets b and c 
%\item the time of minimum light T (inferior conjunction)
\item the transit impact parameter $b$ for the case of a circular orbit, defined as $b = a \mathrm{cos}(i_{p})/R_{\star}$ where $a$ and $i_{p}$ are, respectively, the semi-major axis and inclination of the orbit;
\item the mid-transit time (inferior conjunction) for which we assumed a noninformative uniform prior PDF;
\item the transit duration $T_{14}$, assuming a circular orbit (justified by really small eccentricities for all planets \citep{Luger2017a,Luger2017b,Agol2020}), defined as follow \citep{Winn2010}:
\begin{equation}
    T_{14} = \frac{P}{\pi}\arcsin\left[\frac{R_{\star}}{a}\frac{\sqrt{\big(1+\frac{R_{p}}{R_{\star}}\big)^{2}-b^{2}}}{\sin i}\right]
\end{equation}
\item the linear combinations of the quadratic limb darkening coefficients ($u_{1}$, $u_{2}$) in Spitzer's 3.6 and  4.5 $\mu$m channels, defined as $c_1 = 2 \times u_1 + u_2$ and $c_2 = u_1-2 \times u_2$.
Values and errors for $u_{1}$ and $u_{2}$ in a given band pass were interpolated from the tables of \cite{Claret2012, Claret2013} 
basing on the  stellar parameters $T_{eff}$=2511  K $\pm$ 37 K, log$(g[\mathrm{cm\,sec^{-2}}])=5.18 \pm 0.06$, and $[\mathrm{Fe/H}]=0.04 \pm 0.08$ dex, \citep{Delrez2018}. The corresponding normal distributions were used as prior PDFs (for channel 1: $ u1 = \mathcal{N}(0.1633 ,0.0364^{2})$ and $u2 = \mathcal{N}(0.2549  ,0.0570^{2})$, for channel 2: u1 = $\mathcal{N}(0.1442 ,0.0324^{2}))$ and $u2 =\mathcal{N}( 0.2173  ,0.0482^{2}))$ . In terms of combined limb darkening coefficients those value translates as: for channel 1: $ c1 = \mathcal{N}(0.5815, ,0.0676^{2})$ and $c2 = \mathcal{N}(-0.3465,0.0676^{2})$, for channel 2: c1 = $\mathcal{N}(0.5057, 0.0581^{2}))$ and $c2 =\mathcal{N}( -0.2904 ,0.05801^{2}))$.

\end{itemize}
All of our priors come from the updated system parameters presented in D2018. We recognize that those values were derived from analyses carried out on a subset of the same data set, noting that in this section our intention is not to determine the physical parameters of the system but rather to assess the stability (or variability) of the transits parameters. This subset is sufficient for this assessment.

We then re-scaled the photometric errors by multiplying the error bars by the correction factor $CF = \beta _{w} * \beta _{r} $. Once the correction factor was applied, we ran two Markov chains of 100 000 steps each to sample the PDFs of the parameters of the model and the system's physical parameters \citep{Ford2006}, and assessed the convergence of the MCMC analysis with the Gelman \& Rubin statistical test \citep{Gelman1992}. Our threshold for convergence was a Gelman-Rubin statistic lower than 1.1 for every jump parameter, measured across the two chains.

For all of the analyses, the resulting values and error bars for the jump and system parameters as well as the complete details on the assumed baseline and on the correction factor applied can be found in Table \ref{baseline_indiv}. In addition to setting the baseline to use for each light curve, proceeding to individual analyses is also a way to search for variability in the transits, notably spot/faculae crossing or flares events (see \ref{time_dep_variations}).

\subsubsection{Stellar parameters}\label{stellar_param}
\begin{enumerate}%[noitemsep]

    \item \textbf{Stellar parameters:}
In this paper, we derived stellar parameters using 142 of the 171 TRAPPIST-1's planets transits observed with Spitzer (the transits not included were either partial or multiple). We proceeded as follows: first we inferred the density of the star $\rho_{\star}$ and its error through a global MCMC analysis of all stacked transit (detailed in the following paragraph). Then we derived the mass of the star $M_{\star}$ following the empirical relationship between $M_{K_{s}}$ (magnitude in K band) and $M_{\star}$ (with $M_{\star}$ spanning from 0.075$M_{\odot}$ < $M_{\star}$ < 0.70$M_{\odot}$) derived from 62 nearby binaries by \cite{Mann2019}. The mass and its error were estimated by taking the metallicity of the star (from \cite{VanGrootel2018}) into account and through the use of the GitHub repository \href{https://github.com/awmann/M_-M_K-}{$M_--M_K-$} provided by \cite{Mann2019}, which accounts for systematic errors. With this estimate of the mass of the star and its error, we derived the radius of the star $R_{\star}$ from our posterior probability distribution function (PDF) on the density:
\begin{equation}
    R_{\star}=\Bigg[\frac{3M_{\star}}{4\pi \rho_{\star}}\Bigg]^{1/3}. 
    \label{eq:radius}
\end{equation}

Using the exquisite parallax value ($d = 80.4512 \pm 0.1211$ pc) from Gaia DR2 \citep{Lindegren2018} and the integrated flux derived from \cite{Filippazzo2015}, we computed the luminosity of the star, with no correction for extinction.:
%Michael: give error on the parallax, #solve give reference for Gaia DR2 #solve
\begin{equation}
    L_{\star} = 4 \pi d^{2} \int_{0 \mu \mathrm{m}}^{1000 \mu \mathrm{m}} F_{\lambda}(t) d\lambda,
    \label{eq:luminosity}
\end{equation}

Finally, we derived the effective temperature of TRAPPIST-1 from its luminosity and its radius following the Stefan-Boltzmann Law:

\begin{equation}
    T_{eff, \star} = \Big(\frac{L_{\star}}{4 \pi R_{\star}^{2} \sigma_{SB}}  \Big)^{1/4},
    \label{eq:teff}
\end{equation} where $\sigma_{SB}$ is the Stefan-Boltzmann constant.

We note that the computations of $L_{\star}$, $R_{\star}$ and $T_{eff}$ are straightforward, such that we don’t need to consider possible systematic errors. Their errors were then computed through error propagation on equations \ref{eq:radius}, \ref{eq:luminosity} and \ref{eq:teff}.The stellar parameters derived from this approach are presented in Table \ref{table_stellar_param}.\\

\begin{table}[h!]
\centering
\begin{tabular}{lccccccc}        % centered columns (4 columns)
\hline
\hline
\textbf{Quantity} & \textbf{Value }\\
\hline
\hline
Density   $\rho_{\star}$ ($\rho_{\odot}$) & $52.31 \pm 2.2$  \\
Mass $M_{\star}$ ($M_{\odot}$) & $0.0898 \pm 0.0024 $  \\
Radius  $R_{\star}$ ($R_{\odot}$) &  $0.1197 \pm 0.0017 $  \\
Luminosity   $L_{\star}$ ($L_{\odot}$) & $0.000553 \pm 0.000019 $  \\
Effective temperature $(\mathrm{K})$ & $2557 \pm 47 $ \\
%Metallicity [Fe/H] $(\mathrm{dex})$\tablefootmark{1} & $0.04 \pm 0.08 $ \\
\hline  
%inserts single line
\end{tabular}
%\tablefoot{ \tablefoottext{1}{The metallicity is taken from models by \cite{VanGrootel2018}.}}
\caption{Updated stellar parameters of TRAPPIST-1 from the approach detailed in Section \ref{stellar_param}. \ref{updated_param}. }             % title of Table
\tablefoot{ \tablefoottext{1}{Those parameters are not the final one, the final stellar parameters from this work are given in Table.}}

\label{table_stellar_param}
\end{table}

    \item \textbf{Density inferred from individual planets:}
    
    In the preceding paragraph we mentioned that we derived the stellar density through a global MCMC analysis of all stacked transit. This method uses the transits shapes and Kepler third's law to constraint the stellar density \citep{Seager2003}. 
    However, in this particular case, the TRAPPIST-1 system is composed of 7 planets, that is to say 7 different sets of transit parameters. Hence, it is legitimate to investigate whether there are noticeable differences between the stellar density value inferred from each individual planet's analysis and the one inferred from all transits together. The level of agreement between the two values would allow us to justify the use of the globally derived stellar density.  Figure \ref{fig:densities_per_planet} shows the stellar density value as obtained from individual planet analysis with its error bars and a color code for the number of transits used in each analysis. Table \ref{table:densities_per_planet} presents the corresponding values. In those analysis, for the star, $M_{\star}$, $R_{\star}$, $T_{eff,\star}$, [Fe/H], and the linear combinations c1 and c2 of the  quadratic limb-darkening coefficients (u1,u2) for each bandpass were jump parameter, with informative priors on $M_{\star}$,$T_{eff,\star}$, [Fe/H], u1 and u2. And for each the planet, the transit depth 4.5$\mu m$, the impact parameter and the transit depth difference between Spitzer/IRAC 3.6$\mu m$ and Spitzer/IRAC 4.5$\mu m$ channels were jump parameters.

\begin{figure}
\begin{minipage}[b]{\columnwidth}
    \centering
    \renewcommand{\arraystretch}{1.0}
    \setlength{\tabcolsep}{3pt} % Default value: 6pt
    \begin{tabular}{ccc}       % centered columns (4 columns)
        \hline               % inserts double horizontal lines
        Planet  & \# transits & $\rho_{\star}$ ($\rho{\odot}$) \\    
        \hline  
        \hline % inserts single horizontal line
           TRAPPIST-1\,b  & 54 & $44.6^{-3.9}_{+4.4}$   \\ 
           TRAPPIST-1\,c  & 39 & $48.2^{-4.4}_{+3.7}$  \\  
           TRAPPIST-1\,d  & 20 & $46.5^{-12.0}_{+5.2}$  \\  
           TRAPPIST-1\,e  & 17 & $48.5^{-8.6}_{+5.8}$  \\
           TRAPPIST-1\,f  & 13 & $57.3^{-7.2}_{+3.7}$  \\  
           TRAPPIST-1\,g  & 9 &  $58.5^{-9.1}_{+6.0}$  \\  
           TRAPPIST-1\,h  & 6 &  $53.8^{-18.0}_{+10.0}$  \\  
        \hline                                   %inserts single line
    \end{tabular}
    \captionof{table}{\label{table:densities_per_planet} Stellar density from individual planets's MCMC analyses with its $1-\sigma$ uncertainty. For each planets the number of transits used in the analysis is indicated in the second columns.}
\end{minipage}
\begin{minipage}[b]{\columnwidth}
    \centering
    %\rule{6.4cm}{3.6cm}
    \includegraphics[width=0.8\columnwidth]{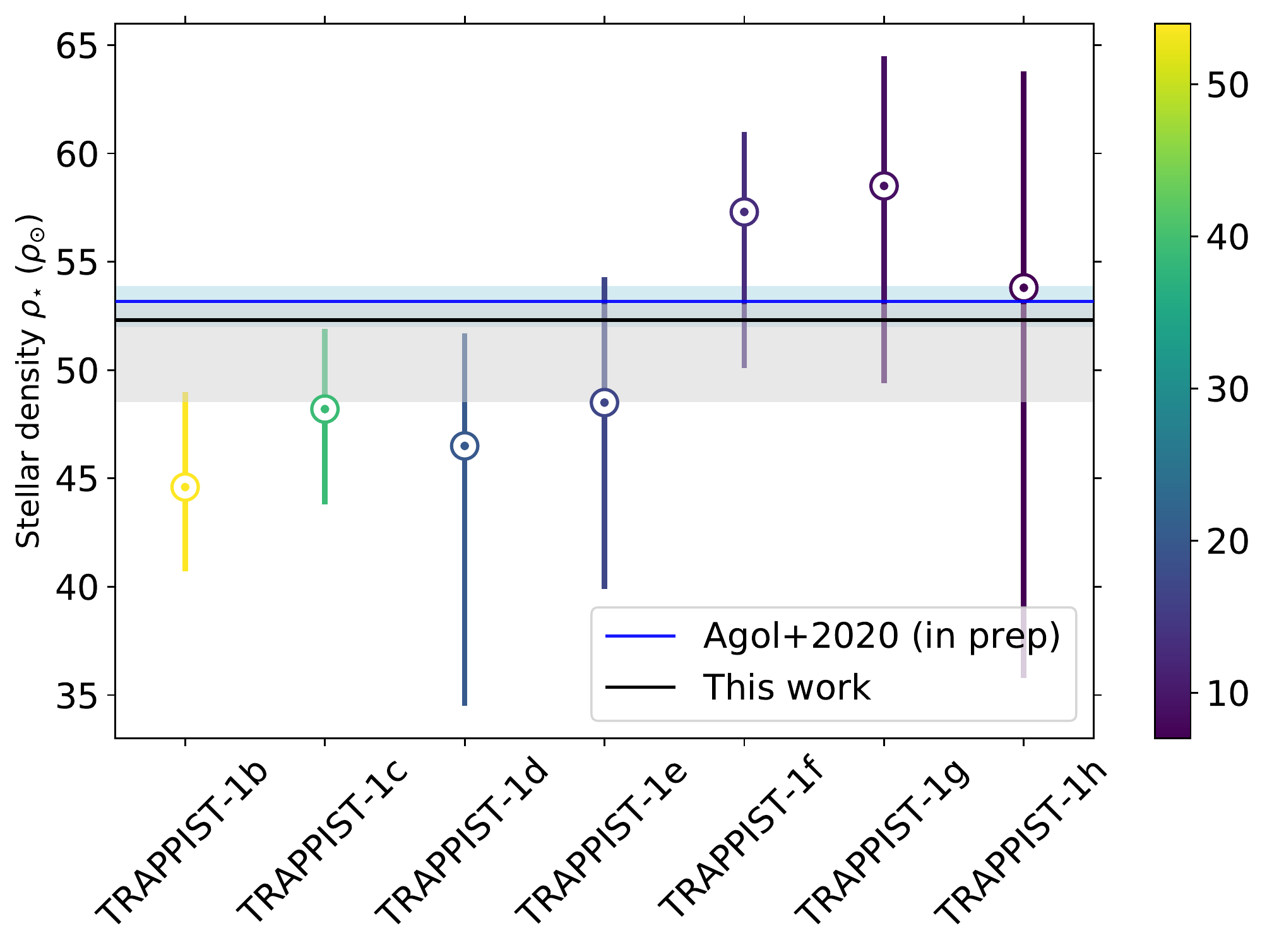}
    \captionof{figure}{\label{fig:densities_per_planet} Coloured dots gives the stellar density derived from MCMC analyses of transits from a single planet, solid black line gives the density derived from a global analysis of all transits observed with its $1-\sigma$ uncertainty in gray shades. Colorbar shows the number of transits used in each analysis. Solid blue line give the stellar density value computed by \cite{Agol2020} using a photodynamical model created with the mass-ratios and orbital parameters derived from a transit-timing analysis, with its $1-\sigma$ uncertainty in blue shades. }
\end{minipage}
\end{figure}

%\vspace{\belowdisplayskip}

From Figure \ref{fig:densities_per_planet}, it appears that the inner planets prefer a lower stellar density than the outer ones. This could be translated as a correlation between period $P$ and the inferred stellar density $\rho_{\star}$. We carried out a linear regression between $\rho_{\star}$ and $P$,  and computed the Akaike information criterion (AIC) and Bayesian information criterion (BIC) to identify the best fit, see Figure \ref{fit_densities}.

\begin{figure}[h!]
    \centering
    \includegraphics[width=0.8\columnwidth]{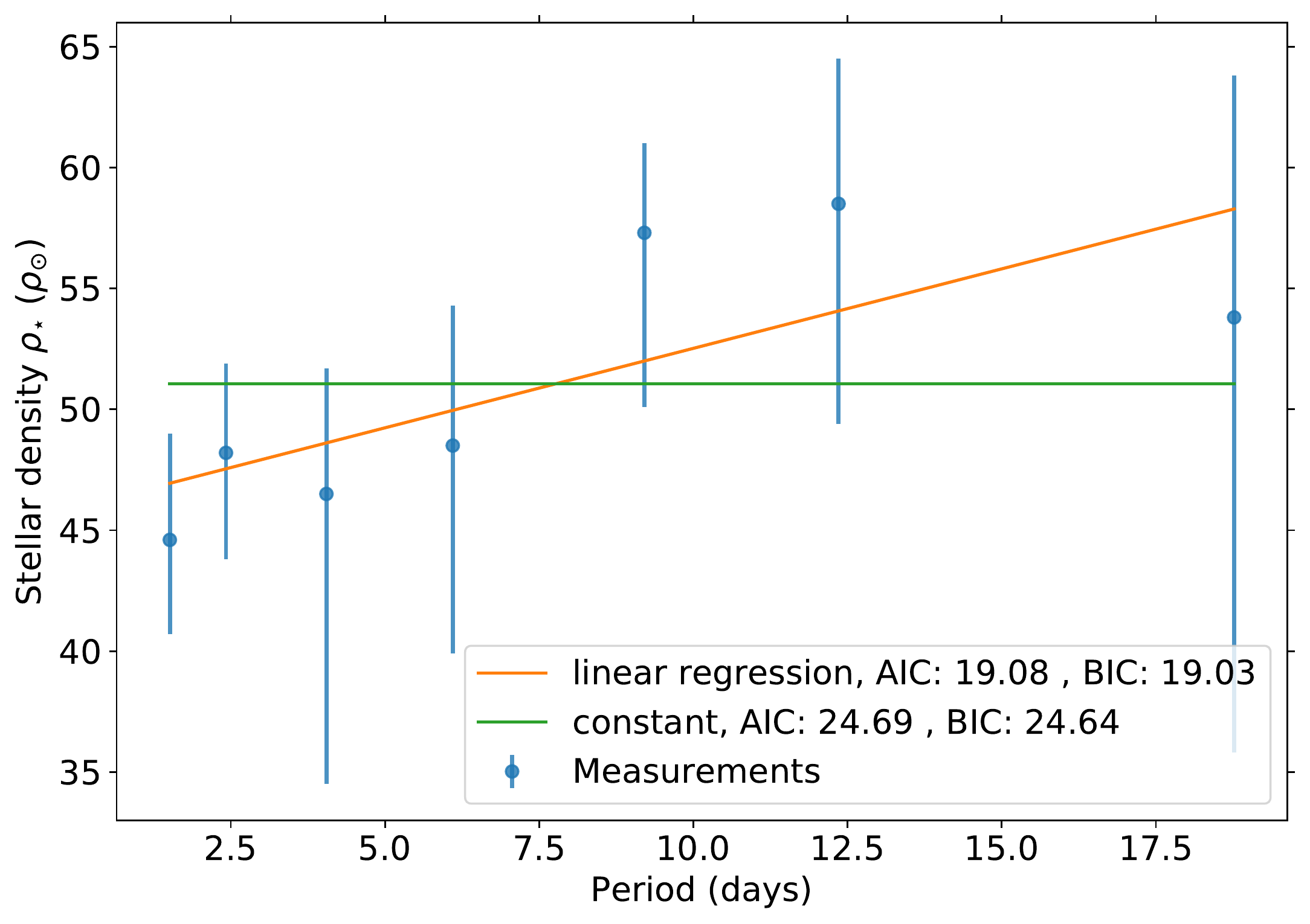}
    \caption{Stellar density inferred from individual planet's analysis versus period of the corresponding planet and its linear and constant fit with their corresponding AIC and BIC values.  }
    \label{fit_densities}
\end{figure}

According to Figure \ref{fit_densities}, it turns out that a linear relation between the period and the stellar density is slightly preferred over a constant density (with a functional form of $\rho_{\star} = 0.65 * P + 45.95$). This suggests that a correlation may exist between density and orbital period (and therefore b and $T_{14}$). However, the amplitude of this bias is smaller than the $1-\sigma$ error bars on the stellar density from the individual planet analyses, and therefore we conclude it is insignificant. On the origin of this weak trend in stellar density versus period, it could be the result of a trend of orbital eccentricity with orbital period. However, the eccentricities computed by \cite{Agol2020} do not seem to confirm this scenario. Another explanation could be that it is the consequence of an observational bias, as we have less transits for the outer planets (details on the number of transits per planets in Table \ref{nb_transits}). As a whole, this comparison shows the advantage of having a multiplanetary system where the planets sample different parts of the stellar disk. We conclude that using the stellar density derived from a global analysis of all transits, as we did in the previous paragraph, is appropriate. 

Furthermore, in Figure \ref{fig:densities_per_planet} we have added the stellar density computed by \cite{Agol2020} using a photodynamical model created with the mass-ratios and orbital parameters derived from a transit-timing
analysis. We observe that the value derived from our global analysis is in excellent agreement with the one by \cite{Agol2020}, which strengthens our confidence in using this value as the final one.

\end{enumerate}

\subsubsection{Global analyses} \label{global_analysis}

Once we selected a baseline model for each light curve (see Section \ref{individual_analysis}), we continued to the next steps of our analysis. First, we carried out a global analysis of all the light curves together to refine the transit parameters. This is an update of the parameters presented in Table 1 of D2018 with the advantage that a global analysis with transits in both channels enables us to lift a part of the degeneracy between the transit parameters and the assumed limb-darkening coefficients. Second, we conducted 2*7 global analyses of transits for each planet to focus on computing first the transit timing variations and then the transit depth variations.

\begin{enumerate}%[noitemsep]
    \item \textbf{All light curves} \label{test}
    This analysis consisted of a preliminary run of one 50 000 step Markov chain to estimate the correction factors $CF$ that are applied to the photometric error bars (see Section \ref{individual_analysis}) and a second run with two Markov chains of 500 000 steps for which we used the Gelman-Rubin test to assess the convergence. The relatively large number of steps for the two chains is necessary for a data set of this size. This method for analysis is identical to the one conducted by D2018, but includes an increased number of transits observed at 4.5 $\mu \mathrm{m}$ for all planets and newly observed transits at 3.6 $\mu \mathrm{m}$ for planets c, d ,e, f, g, and h.  The jump parameters
    were $R_{\star}$, $M_{*}$, $T_{eff}$, [Fe/H], the linear combinations $c_{1}$ and $c_{2}$ of the quadratic limb-darkening coefficients ($u_{1}$, $u_{2}$) for each bandpass. For each planet, parameters include:\\
    - the transit depth at $4.5\mu \mathrm{m}$, $dF_{4.5\mu \mathrm{m}}$\\
    - the impact parameter, $b$\\
    - the transit depth difference between Spitzer/IRAC 3.6 $\mu \mathrm{m}$  and Spitzer/IRAC 4.5 $\mu \mathrm{m}$ channels, $ddF = dF_{3.6\mu \mathrm{m}} - dF_{4.5\mu \mathrm{m}}$\\
    - the transit timing variation (TTV) of each transit with respect to the mean transit ephemeris derived from the individual analyses\\
    
    For the mass of the star, we used a normal prior PDF based on the value derived in Section \ref{stellar_param} ($M_{\star}$ = 0.0898 $\pm$ 0.0024 $M_{\odot}$). Then, for the metallicity and the limb-darkening coefficients in both channels we assumed the same normal prior PDFs as in Section \ref{individual_analysis}. For the rest of the jump parameters, we assumed uniform noninformative prior distributions.
    We did not set the transit duration as a jump parameter because it is defined for each planet by its orbital period, transit depth, and impact parameter, combined with the stellar mass and radius \citep{Seager2003}. Furthermore, dynamical models predict rather small amplitudes of variation for the transit durations \citep{Luger2017a}. The convergence of the chains was checked with the Gelman-Rubin statistic \citep{Gelman1992}. The value of the statistic was less than 1.1 for every jump parameter, measured across the two chains, which indicates that the chains are converged. From the jump parameters the code deduced the physical parameters of the system at each step of the MCMC. In particular, the value of the effective temperature, $T_{eff}$, was derived at each MCMC step from the $R_{\star}$ and $L_{\star}$ values given in Section \ref{stellar_param}. Then, for each planet, values for the radius of the planet $R_{p}$, its semi-major axis $a$, its inclination $i$, its irradiation $S_{p}$, and its equilibrium temperature $T_{\mathrm{eq}}$ were deduced from the values for the stellar and transit parameters. Table \ref{updated_param} presents the outputs from this analysis.

    \begin{table*}[ht!]
    \begin{threeparttable} 
    \centering 
    \tiny
    \setlength{\tabcolsep}{1.pt} % Default value: 6pt
    \renewcommand{\arraystretch}{1.3} % Default value: 1
    \begin{tabular}{cccccccc}        % centered columns (8 columns)
    \hline
    \textbf{Parameters} & & & & \textbf{Value }& & &  \\
    \hline
    \textbf{Star} & & & & \textbf{TRAPPIST-1}  & & &  \\
    Mass~\tnote{a}~ $M_{\star}$ ($M_{\odot}$) & & & & $0.0898 \pm 0.0023 $  & & &  \\
    Radius   $R_{\star}$ ($R_{\odot}$) & & & & $0.1234 \pm 0.0033 $ & & &  \\
    Density   $\rho_{\star}$ ($\rho_{\odot}$) & & & & $47.98 \pm 3.90 $ & & &  \\
    Luminosity~\tnote{a}~    $L_{\star}$ ($L_{\odot}$) & & & & $0.000552\pm 0.000018 $ & &  & \\
    Effective temperature $(\mathrm{K})$ &  & & & $2520 \pm 39 $ & & &  \\
    Metallicity~\tnote{a}~  [Fe/H] $(\mathrm{dex})$ &  & & & $0.0535 \pm 0.088 $ & & &  \\
    \thead{LD coefficient, $u_{1,3.6\mu \mathrm{m}}$~\tnote{a}~ } & & && $0.168 \pm 0.016$ & & &    \\
    \thead{LD coefficient, $u_{2,3.6\mu \mathrm{m}}$~\tnote{a}~ } & & & & $0.245 \pm 0.019$ & & &  \\
    \thead{LD coefficient, $u_{1,4.5\mu \mathrm{m}}$~\tnote{a}~ }  & &  & & $0.141 \pm 0.016 $ &  &  & \\
    \thead{LD coefficient, $u_{2,4.5\mu \mathrm{m}}$~\tnote{a}~ } & & & & $0.198 \pm 0.018 $   & &  & \\
    \thead{Combined LD coefficient, $c_1,3.6\mu m$}  & & & & $0.581 \pm 0.039$ &  &  & \\
    \thead{Combined LD coefficient, $c_2,3.6\mu m$} & & & &  $-0.322 \pm 0.045$ & & &\\
    \thead{Combined LD coefficient, $c_1,4.5\mu m$} & & & &  $0.482 \pm 0.031$ & &  & \\
    \thead{Combined LD coefficient, $c_2,4.5\mu m$} & & & &  $-0.256 \pm 0 .044$ & & & \\
    \hline                 % inserts double horizontal lines
    \hline
    \textbf{Planets}  & \textbf{b} & \textbf{c} & \textbf{d} & \textbf{e} & \textbf{f} & \textbf{g} & \textbf{h}   \\    % table heading 
    %\hline                        % inserts single horizontal line
    \# of transits & 54 & 39 & 20 & 17 & 13 & 9 & 6\\
    Period (days) & \thead{1.51088432 \\ $\pm$ 0.00000015} & 
    \thead{2.42179346 \\$\pm$ 0.00000023} & 
    \thead{4.04978035 \\$\pm$ 0.00000266} & 
    \thead{6.09956479 \\$\pm$ 0.00000178} & 
    \thead{9.20659399 \\$\pm$ 0.00000212} & 
    \thead{12.3535557 \\$\pm$ 0.00000341} & 
    \thead{18.7672745 \\$\pm$ 0.00001876} \\
    
    \thead{Mid-transit time \\ $T_{0}$ - 2450000 ($\mathrm{BJD}_{\mathrm{TDB}}$) } & \thead{7322.514193 \\ $\pm$ 0.0000030} & 
    \thead{7282.8113871 \\$\pm$ 0.0000038} & 
    \thead{7670.1463014 \\$\pm$ 0.0000184} & 
    \thead{7660.3676621 \\$\pm$ 0.0000143} & 
    \thead{7671.3737299 \\$\pm$ 0.0000157} & 
    \thead{7665.3628439 \\$\pm$ 0.0000206} & 
    \thead{7662.5741486 \\$\pm$ 0.0000913} \\
    %MICHAEL: use 2-digit precision for all parameters (periods of d-e-f-g)
    
    \thead{Transit depth ($R_{p}^{2}/R_{\star}^{2}$) \\ at $4.5\mu \mathrm{m} (\%)$ } & 
    $0.7236 \pm 0.0072 $ & 
    $0.7027 \pm 0.0068$ & 
    $0.3689 \pm 0.0067$ & 
    $0.4936 \pm 0.0081$ & 
    $0.6313 \pm 0.0091 $ & 
    $0.745 \pm 0.011$ & 
    $0.351 \pm 0.012$ \\
    
    \thead{Transit depth ($R_{p}^{2}/R_{\star}^{2}$) \\ at $3.6 \mu \mathrm{m} (\%)$ } & 
    $0.7209 \pm 0.0067$ & 
    $0.721 \pm 0.014$ & 
    $0.351 \pm 0.016$ & 
    $0.491 \pm 0.011$ & 
    $0.655 \pm 0.019$  & 
    $0.724 \pm 0.024$ & 
    $0.313 \pm 0.027$ \\
    
    \thead{Transit impact \\ parameter $b$ ($R_{*}$)} & 
    $0.254^{+0.110}_{-0.085} $  & 
    $0.254^{+0.110}_{-0.087} $ & 
    $0.235^{+0.120}_{-0.094} $ & 
    $0.299^{+0.085}_{-0.072} $ & 
    $0.391 \pm 0.056$  & 
    $0.430 \pm 0.049$ & 
    $0.448 \pm 0.054$\\
    
    \thead{Transit duration \\ $T_{14}$ (min) } & 
    36.309 $\pm$ 0.093 & 
    42.42 $\pm$ 0.12 &
    49.37 $\pm$ 0.32 & 
    56.31 $\pm$ 0.25 & 
    63.28 $\pm$ 0.31 & 
    69.10 $\pm$ 0.36 &  
    76.28 $\pm$ 0.81 \\
    
    \thead{$R_{p}/R_{\star}$ at $4.5\mu \mathrm{m}$ } & 
    \thead{$0.085062$ \\ $\pm 0.000042 $} & 
    \thead{$0.083827$ \\ $\pm 0.000040$} & 
    \thead{$0.06073$ \\ $\pm 0.00056$} & 
    \thead{$0.07025$ \\ $\pm 0.00058$} & 
    \thead{$0.07945$ \\ $\pm 0.00057 $} & 
    \thead{$0.08632$ \\ $\pm 0.0062$} & 
    \thead{$0.05927$ \\ $\pm 0.0099$} \\
    
    \thead{$R_{p}/R_{\star}$ at $3.6 \mu \mathrm{m}$ } & 
    \thead{$0.084903$ \\ $\pm 0.00040$} & 
    \thead{$0.08495$ \\ $\pm 0.00086$} & 
    \thead{$0.0593$ \\ $\pm 0.0013$ } & 
    \thead{$0.07009$ \\ $\pm 0.00075$} & 
    \thead{$0.0809$ \\ $\pm 0.0013$}  & 
    \thead{$0.0851$ \\ $\pm 0.0014$} & 
    \thead{$0.0559$ \\ $\pm 0.0025$} \\

    Inclination $i$ ($^{\circ}$) & 
    $89.28 \pm 0.32$ & 
    $89.47 \pm 0.24$ & 
    $89.65 \pm 0.15$ & 
    $89.663 \pm 0.092$  & 
    $89.666 \pm 0.059$ & 
    $89.698 \pm 0.044$ & 
    $89.763 \pm 0.037$ \\
    
    \thead{Semi major axis \\ a ($10^{-3}$AU)} & 
    $11.534_{-0.092}^{+0.099}$ & 
    $15.79_{-0.13}^{+0.14}$ & 
    $22.26_{-0.18}^{+0.19}$ & 
    $29.24_{-0.23}^{+0.25}$ & 
    $38.7740 _{-0.31}^{+0.33} $& 
    $46.81528_{-0.37}^{+0.40}$ &
    $61.8656_{-0.49}^{+0.53}$ \\
    
    Scale parameter $a/R_{\star}$ & 
    $20.13_{-0.55}^{+0.46}$ &  
    $27.57_{-0.76}^{+0.62}$ &  
    $38.85_{-1.1}^{+0.88}$ & 
    $51.0_{-1.4}^{+1.2}$ & 
    $67.1_{-1.9}^{+1.15}$ & 
    $81.7_{-2.3}^{+1.8}$ & 
    $107.9_{-3.0}^{+2.4} $\\
    
    Irradiation $S_{p}$ ($S_{\odot}$) & 
    $4.15  \pm 0.16$ &  
    $2.211 \pm 0.085$  & 
    $1.114 \pm 0.043$ & 
    $0.645 \pm 0.025 $ & 
    $0.373 \pm 0.014$ & 
    $0.252  \pm 0.0097$ & 
    $0.144 \pm 0.0055$ \\
    
    \thead{Equilibrium \\ temperature $T_{\mathrm{eq}}$ (K)\tnote{b} }  & 
    $397.6 \pm 3.8$ & 
    $339.7 \pm 3.3$&  
    $286.2 \pm 2.8$  & 
    $249.7 \pm 2.4 $& 
    $217.7 \pm 2.1$   & 
    $197.3 \pm 1.9$ & 
    $171.7 \pm 1.7$ \\
    
    \thead{Radius $R_{p,3.6\mu m}$ ($R_{\oplus}$)} & 
    $1.1407 \pm 0.035$ & 
    $1.141 \pm 0.037$ &  
    $0.799 \pm 0.026$ & 
    $0.944 \pm 0.025$ & 
    $1.087 \pm 0.027$  & 
    $1.147 \pm 0.041$  & 
    $0.752 \pm 0.037$ \\
    
    \thead{Radius $R_{p,4.5\mu m}$ ($R_{\oplus}$)} & 
    $1.144  \pm 0.027$ & 
    $1.128  \pm 0.027$ &  
    $0.817  \pm 0.022$ & 
    $0.945  \pm 0.026$ & 
    $1.068  \pm 0.028$  & 
    $1.161  \pm 0.030$  & 
    $0.797  \pm 0.025$ \\
    \hline      
    \hline
    \end{tabular}
    \begin{tablenotes}
        \item[a] Informative prior PDFs were assumed for these stellar parameters
        \item[b]  where $T_{eq}$ is computed from $T_{eq} =\Bigg[ \frac{(1-A)*S_{p}}{4*\sigma}\Bigg] ^{1/4}$, assuming a null Bond albedo

    \end{tablenotes}   
    \end{threeparttable}
        \caption{Updated system parameters: median values and 1-$\sigma$ limits of the posterior PDFs derived from our global MCMC analysis of all nonblended and partial transits of Trappist-1 planets observed by Spitzer.} 
          % title of Table
    \label{updated_param}
    \end{table*}
    
    We note that prior to this paper different stellar parameters for TRAPPIST-1 were published. In 2019 \cite{Gonzales2019} presented a distance-calibrated SED of TRAPPIST-1 using a new NIR FIRE spectrum and parallax from the Gaia DR2 data release from which they derived updated fundamental parameters for the star. Back in 2018, \cite{VanGrootel2018} derived stellar parameters from two distinct approaches to compute the mass of the star, first via stellar evolution modeling, and second through an empirical derivation from dynamical masses of equivalently classified ultracool dwarfs in astrometric binaries. The stellar parameters we derived are in agreement with those from previous studies, as shown in Table \ref{table:compa_stellar_param}. \\
    
    \begin{table*}[h!]
    \centering
        \setlength{\tabcolsep}{5pt} 
        \begin{tabular}{|lccc|} 
            \hline 
            \hline
            Quantity & Gonzales +2019\tablefootmark{1}  & Van Grootel +2018 & This paper \\
            \hline
            \hline
            %Density   $\rho_{\star}$ ($\rho_{\odot}$) & &&$51.05 \pm 1.20 $  \\
            Mass $M_{\star}$ ($M_{\odot}$) & 0.0859 $\pm$ 0.0076 & 0.0889 $\pm$ 0.0060 & $0.0898 \pm 0.0023 $  \\
            Radius  $R_{\star}$ ($R_{\odot}$) & 0.1164 $\pm$ 0.0030 & 0.1182 $\pm$ 0.0029 & $0.1234 \pm 0.0033 $  \\
            Luminosity   $L_{\star}$ ($L_{\odot}$) & 0.000608 $\pm$ 0.000022 & 0.000522 $\pm$ 0.000019 & $0.000552 \pm 0.000018 $  \\
            Effective temperature $(\mathrm{K})$ & 2628 $\pm$ 42 &2516 $\pm$ 41 & $2520 \pm 39 $ \\
            Parallax (mas) & 80.45 $\pm$ 0.12 & 82.4 $\pm$ 0.8 & 80.45 $\pm$ 0.12\\
            %Metallicity [Fe/H] $(\mathrm{dex})$\tablefootmark{1} & &&$0.04 \pm 0.08 $ \\
            \hline                                   %inserts single line
            \hline
        \end{tabular}
        \caption{Comparison of stellar parameters value from various studies. }             % title of Table
        \tablefoot{ \tablefoottext{1}{Derived for age range $0.5$ to $10$ Gyr, the field age constraint from \cite{Filippazzo2015}, see \cite{Gonzales2019}.}}

        \label{table:compa_stellar_param}
    \end{table*}

    \item \textbf{Planet by planet}
    
    For each planet, the analysis itself was divided in three: a global analysis to extract the TTV of each transit, a global analysis with a free variation of the transit depth allowed to monitor its evolution with time, and finally an analysis of the occultation observations for planets b and c.
    
    \begin{enumerate}%[noitemsep]
        \item \textbf{Transit timing variations:} 
        We used nearly the same priors and jump parameters as in the individual analyses although we fixed the time of transit for epoch zero ($T_{0}$) and period $P$ for each planet, and set TTV as a jump parameter for each transit. The priors value for $T_{0}$ and $P$ are extracted from D2018. In this analysis, the same depth was assumed for all transits observed in the same bandpass. For each transit, we assumed the same baseline as the one obtained from its individual analysis.d Then, after one 50 000 step Markov chain, we re-scaled our photometric errors with the resulting correction factor and ran two Markov chains of 100 000 steps. Transits timings and their corresponding TTVs are reported in Table.\ref{transit_timing_variations} and displayed on Figure \ref{fig:ttv}. From those, for each planet we performed a linear regression of the timings as a function of their epochs to derive an updated mean transit ephemeris, i.e., an updated value of the mid-transit time $T_{0}$ and the orbital period $P$ for each planet, see Table \ref{updated_param} (as done in D2018).  Finally the median of the global MCMC posterior PDF of the transit depth in both channels  are given in Table \ref{depth_global_ttv}, those results are discussed in Section \ref{discussion} and used to construct the planetary transmission spectra in Section \ref{transit_transmission}.
        
        \begin{figure}[ht!]
            \centering
            \includegraphics[width=0.9\columnwidth]{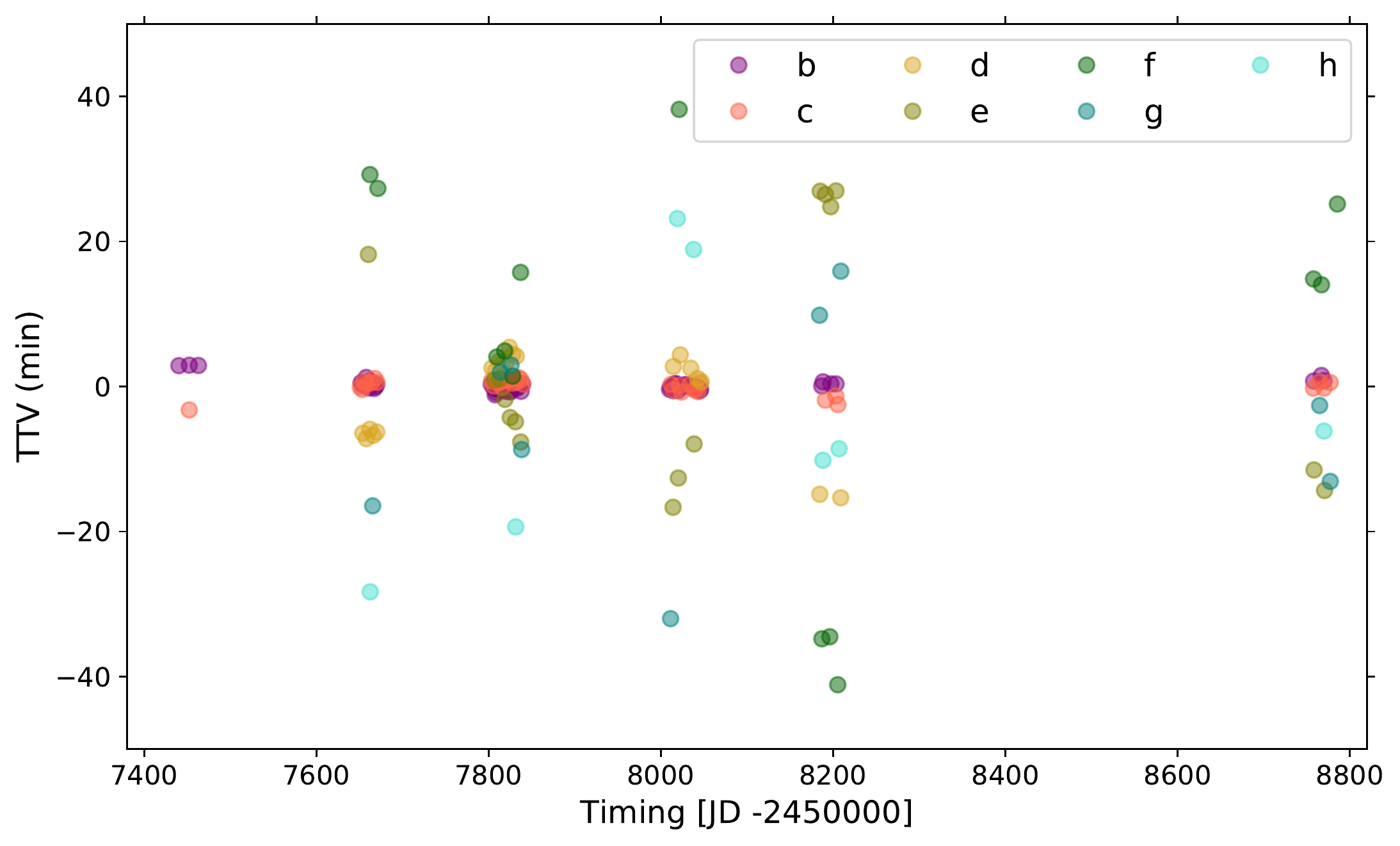}
            \caption{TTVs measured for the seven planets as obtained from our global planet by planet analyses (see Section \ref{global_analysis}) relative to the ephemeris ($T_{0}$ and $P$) given in Table \ref{updated_param}.}
            \label{fig:ttv}%
        \end{figure}
        
        \begin{table}[h!]
        \setlength{\tabcolsep}{6.5pt} % Default value: 6pt
        \renewcommand{\arraystretch}{1.2} % Default value: 1
        \centering
                             % used for centering table
        \setlength{\tabcolsep}{5pt} 
        \begin{tabular}{|c|c c c|c c c|} 
        % centered columns (4 columns)   
        \hline \multicolumn{1}{|c|}{\textbf{ Planet }} & \multicolumn{3}{c|}{\textbf{\thead{Transit depth \\ $dF_{3.6 \mu \mathrm{m}}$ $\pm$ $1\sigma$ - $3\sigma$ (\%)}}} & \multicolumn{3}{c|}{\textbf{\thead{Transit depth \\ $dF_{4.5\mu \mathrm{m}}$ $\pm$ $1\sigma$ - $3\sigma$ (\%)}}} \\ \hline 

           b  & 0.7179 & 0.0058 & 0.021 & 0.7195 & 0.0069 & 0.021\\      %
           c  & 0.7211 & 0.0130 & 0.039 & 0.6996 & 0.0058 & 0.018\\  
           d  & 0.3407 & 0.0150 & 0.042 & 0.3653 & 0.0070 & 0.021 \\  
           e  & 0.4889 & 0.0010 & 0.027 & 0.4950 & 0.0075 & 0.023 \\  
           f  & 0.6463 & 0.0175 & 0.047 & 0.6240 & 0.0093 & 0.029 \\  
           g  & 0.7049 & 0.0330 & 0.094 & 0.7449 & 0.0110 & 0.024 \\  
           h  & 0.3120 & 0.0210 & 0.069 & 0.3478 & 0.0130 & 0.039 \\ 
        \hline                                   %inserts single line
        
        \end{tabular}
        \caption{Median of the global MCMC posterior PDF of the transit depth derived from global analyses of all transits, planet by planet, with no transit depth variations allowed. Those values are used to construct transmission spectra in Section \ref{transit_transmission}}             % title of Table
        \label{depth_global_ttv}
        \end{table}

        \item \textbf{Transit depth variations:} 
        Here, we also used similar priors as in the individual analyses except that we fixed the values of transit timings and periods $P$ but for jump parameters we set the TTV of each transit and $\delta$dF, the depth variations from one transit to another. Again we ran first a 50 000 steps Markov chain to get the $CF$, and then two 100 000 steps chains. The evolution of the transit depths as a function of the epochs is presented for each planet in Figure \ref{fig:global_ddf}. For further comparison, these figures also display the medians of the global MCMC PDFs as obtained by the previous global analysis with TTV and no $\delta$dF variations , values from Table \ref{depth_global_ttv}. We compared the results obtained from the individual and global analyses of the transits and found them to be fully consistent. On Figure \ref{fig:global_ddf} we chose to plot the depth values obtained from the global analysis with $\delta$dF variations allowed instead of the individual analyses because the global analysis should be less impacted by systematic errors due to the red noise (i.e., the response of the pixels to time-varying illumination) in Spitzer photometry. \\
        
    % Constraining the transit shape through a global analysis slightly improves the errors on the depths or timings for some transits, while others have larger errors due to the clearer separation between signal and red noise.

        \begin{figure*}[ht]
            \centering
            %\subfloat{{\includegraphics[width = 0.45\textwidth]{Fig/ddf_36.pdf}}}%\qquad
            %\subfloat{{\includegraphics[width = 0.45\textwidth]{Fig/ddf_45.pdf}}}
            \includegraphics[width = 0.47\textwidth]{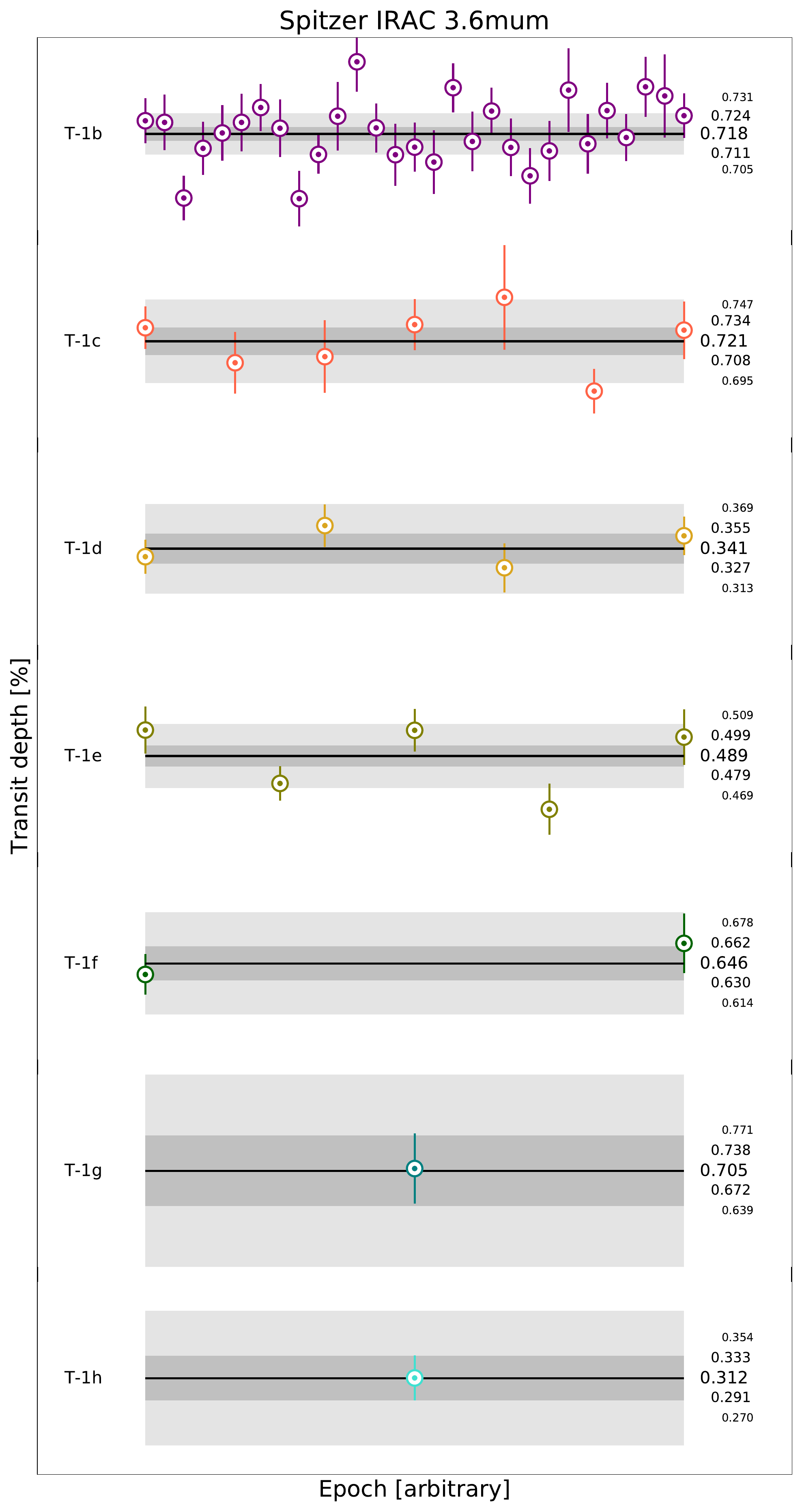}
            \includegraphics[width = 0.47\textwidth]{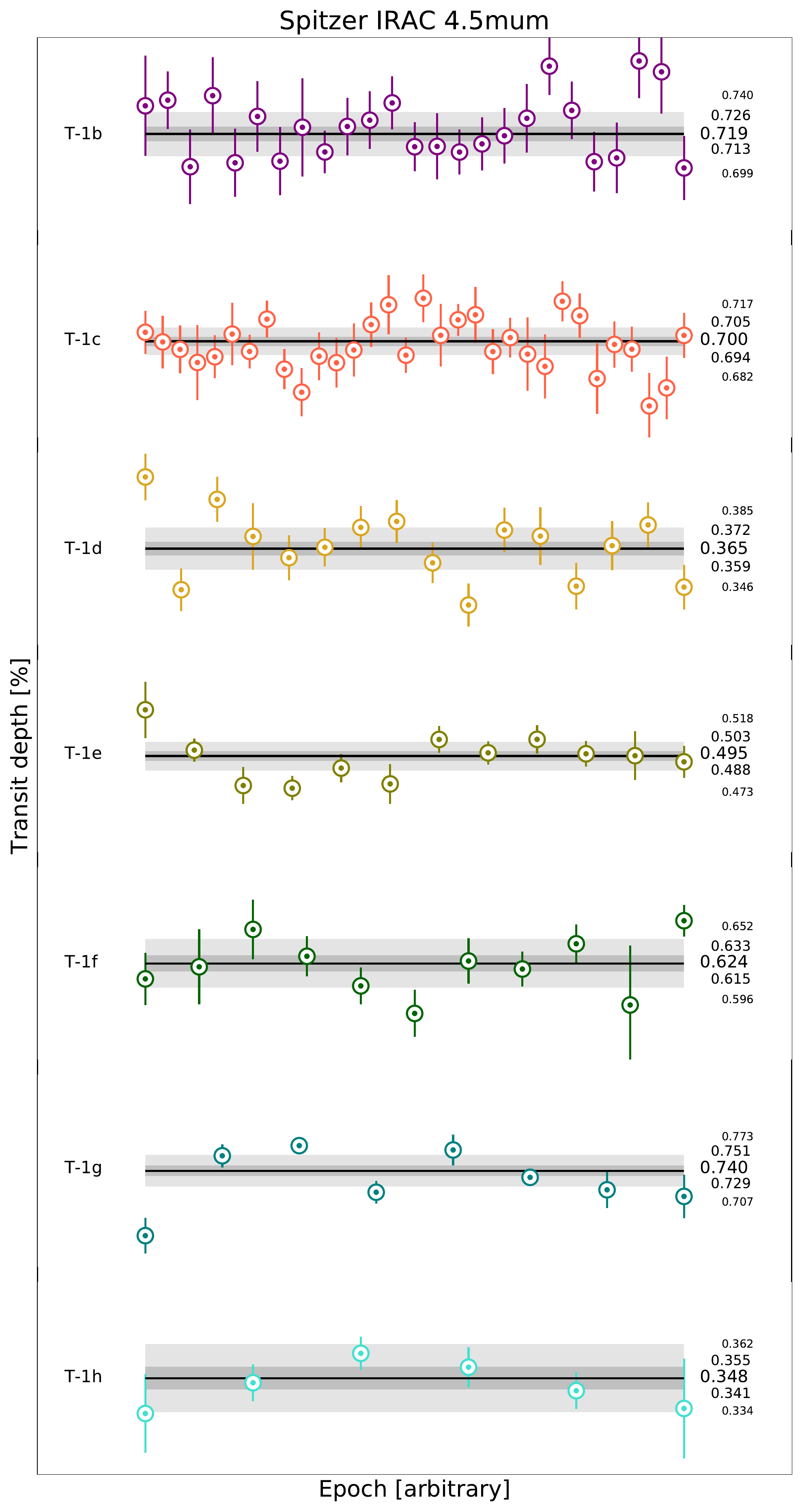}
            \caption{\textit{Left}: Evolution of the measured transit depths from the planet- per- planet global analyses of transit light curves at 3.6 $\mu \mathrm{m}$. The horizontal black lines show the medians of the global MCMC posterior PDFs from the planet- per- planet analyses with TTV and no transit depth variations allowed (with their 1$\sigma$ and 2$\sigma$ confidence intervals (see values in Table \ref{depth_global_ttv}) in shades of gray) Events are ranked in order of capture, left to right (but not linearly in time). \textit{Right}: Similarly, but for transits observed at 4.5 $\mu \mathrm{m}$. }
            \label{fig:global_ddf}
        \end{figure*}
        
        \item \textbf{Occultations:} We also used the eclipse model of \cite{Mandel2002} to represent occultations of TRAPPIST-1b and c as photometric time-series, multiplied by a baseline model to represent external sources of photometric variations (either from astrophysical or instrumental mechanisms). The details of this analysis are given in Section \ref{Occultations}, where we discuss the occultation observations and their interpretation. 
        In this work, we analyzed 29 occultations of planet b and 8 occultations of planet c. All windows were observed in channel 2 ($4.5\mu \mathrm{m}$) as part of the DDT program 13175 (PI: L. Delrez). Our aim was to constrain the day-side brightness temperature of the two inner planets from the occultation depths. 
        For both planets, we performed a global analysis of the occultation light curves, assuming as priors the Gaussian PDFs corresponding to the values for the stellar parameters and for the planets' transit  depths, impact parameters, mid-transit-times, and orbital periods derived from our global analysis of all Spitzer transit light curves (Table \ref{updated_param}). Circular orbits were assumed for both planets, and the occultation depth  was the only jump  parameter of the analyses for which a uniform prior PDF $[0, +\infty)$  was assumed.We justify circular orbits assumption from the fact that in a system with planets that have migrated in Laplace resonances orbital, like TRAPPIST-1, eccentricity are expected to be very small. Indeed simulations carried out by \cite{Luger2017a} show that within a few Myr eccentricities of each planet damped to less than 0.01. In addition, recent results by \cite{Agol2020} from TTV and photodynamical model confirm that all planets eccentricities are most likely inferior to 0.01. Furthermore, when calculating the timing of secondary eclipse using the eccentricities given by \cite{Agol2020} and their $3-\sigma$ uncertainties we compute a shift in time of 0.28 hours and 0.27 hours for planet b and c respectively. Considering that the out-of-secondary eclipse time is $\simeq$ 2 hours for each light curve of this DDT program, we can confidently state that we did not miss the time of secondary eclipse. As with the transit analysis reported above, we identified the most applicable baseline for each light curve and ran a first chain of 50 000 steps to get the CF coefficients, applied these coefficients to the photometric error bars, and then ran two MCMC chains of 100 000 steps. We ascertained the convergence of our analyses with the Gelman \& Rubin test (less than 1.1 for all jump parameters, as recommended by \cite{Brooks1998,Gelman2004}). Unfortunately, no significant occultation signal was detected, Table \ref{results_occ} gives the occultation depth as output by the MCMC analysis with its 1 sigma and 3 sigma uncertainty. Those results are discussed in Section \ref{Occultations}.
       
        \begin{table}[h!]
            \setlength{\tabcolsep}{7.pt} % Default value: 6pt
            \renewcommand{\arraystretch}{1.5} % Default value: 1
            \centering                         
            \begin{tabular}{c c c} \hline    
                Planet  & $\delta_{occ} \pm 1\sigma$  (ppm) & $\delta_{occ} \pm 3\sigma$  (ppm) \\ \hline
                TRAPPIST-1\,b  & $90_{\num{-53}}^{+\num{59}}$ & $90_{\num{-90}}^{+\num{180}}$   \\
               TRAPPIST-1\,c  & $74_{\num{-52}}^{+\num{80}}$ & $74_{\num{-74}}^{+\num{290}}$  \\
                \hline
            \end{tabular}
            \caption{ Median of the occultation depths global MCMC posterior PDF + their 1$\sigma$ and 3$\sigma$ uncertainities as derived from the global analysis of 28 occultations of TRAPPIST-1b and 9 occultations of TRAPPIST-1c observed at 4.5 $\mu \mathrm{m}$.}
            \label{results_occ}
        \end{table}
    \end{enumerate}
\end{enumerate}

%__________________________________________________________________

\section{Results and discussion}\label{discussion}\label{results}

%__________________________________________________________________
\subsection{Transits}\label{system_parameters}

\subsubsection{Mean transit ephemeris}

Our global analysis of the transits of the seven planets led to an updated mean ephemeris, given in Table \ref{updated_param}. The mean ephemeris for each planet was derived from a linear regression of the timings derived in Section \ref{global_analysis} - planet by planet -  as a function of their epochs.  The new ephemerides are consistent with the ones derived in D2018. They do not take into account the TTVs, but should be sufficiently precise  to forecast transit observations with an accuracy of better than $\sim$30 min (for the outer planets, and much better for the inner ones) within the next couple of years. 

\subsubsection{Noise floor}

From the transit depths globally derived in each band (Table \ref{depth_global_ttv}), and from the mean error on the depths of each transit given in Tables \ref{global_3-6} and \ref{global_4-5}, we can estimate the amplitude of the noise floor of Spitzer transit monitoring of TRAPPIST-1. To do so, we compute the mean depth error  $\overline{\mathcal  \sigma_{dF,i,c}}$ for the $i$th planet in each band, $c$, from Tables \ref{global_3-6} and \ref{global_4-5}. Assuming a purely white noise, the expected error for $n$ transits of planet $i$ in band $c$ should be $\sigma_{global,exp} = \frac{\overline{\mathcal  \sigma_{dF,i,c}}}{\sqrt{n}}$.  We then subtract in quadrature this expected value from the globally derived transit depth to estimate the Spitzer noise floor:

\begin{equation}
    \sigma_{noise floor} = \sqrt{\sigma_{global,obs}^{2} - \sigma_{global,exp}^{2}}.
    \label{eq:floor_noise}
\end{equation}

From equation \ref{eq:floor_noise} we calculate the noise floor for each planet and derive the mean noise floor in each channel. The resulting values are 36 ppm in channel 1 and 22 ppm in channel 2. These values are consistent -- and even lower actually-- than those derived for a sample of $\sim$20 bright sun-like stars by \cite{Gillon_RV}. Considering how small these values are, we can conclude that stacking dozens of transits of TRAPPIST-1 observed in the infrared does improve the precision nearly in a $\sqrt{n}$ manner. This agrees well with the high photometric stabilty of TRAPPIST-1 observed during its Spitzer 20d continuous monitoring \citep{Gillon2017, Delrez2018}. Furthermore, the  larger value of the  noise floor at 3.6 $\mu$m --as observed by \cite{Gillon_RV} for brighter Sun-like stars-- suggests that this floor is mostly of instrumental origin, as the pixel-phase effect is significantly larger in IRAC channel 1 than in channel 2 and requires more complex baseline models (see Table A.1).

These results are particularly encouraging for the upcoming atmospheric characterization of the planets by transit transmission spectroscopy with JWST \citep{Gillon2020}. Indeed, the detectors of the JWST instruments (HgCdTe for all except SiAs for MIRI) should all have a much better intrapixel homogeneity than the IRAC InSb arrays, which should result in much less severe position-dependent effects in the JWST spectrophotometric light curves. This is supported by the results obtained by \cite{Kreidberg2014}, who observed 15 transits of GJ1214b with HST/WFC3 (also an HgCdTe array, like NIRISS, NIRSPEC and NIRCAM) and obtained global transit depth errors consistent with a noise floor of $\sim10$ ppm. Based on these considerations, noise floors in the 10-20ppm range can thus be expected for JWST observations of TRAPPIST-1, low enough to enable the detection and characterization of compact atmospheres around the planets \citep[e.g.,][]{LustigYaeger2019}. 

%In comparison with D2018, we decreased the uncertainties on the transit depths by a factor of 1.10, 1.11, 1.04, 1.16, 1.04, 1.00, and 1.17 for planet -1b, -1c, 1-d, -1e, -1f, -1g, and -1h respectively at $4.5 \mu \mathrm{m}$ and 1.48 for -1b at $3.6\mu \mathrm{m}$. We also derived more precise values for the other transit parameters, notably the impact parameters, the duration and their ephemeris ($T_{0}$ and P see Table \ref{updated_param}).

%Furthermore, we derived stellar parameters in a similar manner to D2018; specifically, no informative prior was assumed for the stellar radius, and the stellar density was only constrained by the shape of the transits of the planets, while our prior on the stellar mass came from the value we derived in Section \ref{stellar_param}. Such an approach has the great advantage of being less model-dependent.
%We note that in a purely white noise scenario, we would expect the errors on P, $T_{0}$, b and $dF_{4.5}$ and $dF_{3.6}$ to be improved by the following factor:

\subsubsection{Time-dependent variations of the transit depths}\label{time_dep_variations}

\begin{figure}[ht!]
    \centering
    \includegraphics[width=0.8\columnwidth]{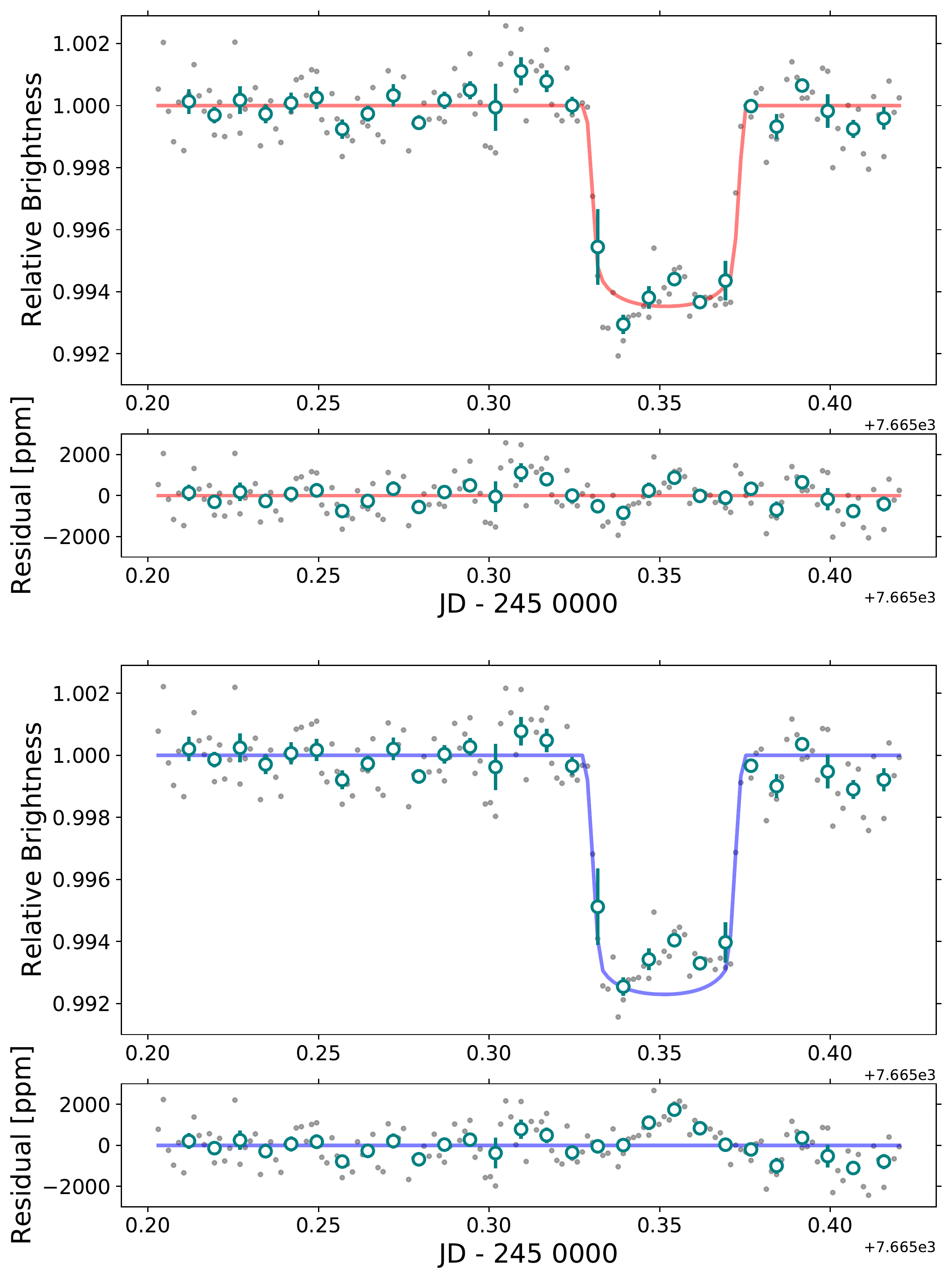}
    \caption{{\it Top panel:} Detrended light curve of the first isolated transit of TRAPPIST-1g observed by Spitzer with, superimposed in red, the best-fit model resulting from the global per-planet analysis with variations of the transit depth allowed.
    {\it Bottom panel:} Similar to Top panel but with the best-fit model resulting from the individual analysis assuming a constant depth for all transits superimposed in darkblue.}
    \label{1g_Spitzer_outlier}
\end{figure}

One possible way to gain insight into the host star of a planetary system is to use transits as a scan of the stellar photosphere \citep{Espinoza2018}. By comparing the transit depths at different epochs we can identify unusual events that could inform us about the (in-)homogeneity of the star. Spot and faculae crossings are typically the kind of signatures detected with this method. For this purpose, we looked for unusually low or high depth values in the results from the global planet-by-planet analyses (Figure \ref{fig:global_ddf}). We identified one clear outlier at $3 \sigma$ lower than the other measurements for planet g (first point of the plot at 4.5 $\mu \mathrm{m}$ on Figure \ref{fig:global_ddf}, epoch 0). The corresponding light curve and the fit for this epoch obtained from the global analysis are displayed in Figure \ref{1g_Spitzer_outlier}.

Yet, when we look at the same light curve as modeled in the global planet-by-planet analysis assuming a constant transit depth  (Figure \ref{1g_Spitzer_outlier} Bottom panel), the  first part of  the transit  seems to  be consistent  with the global  model,  while the  rest is affected by a significant flux increase as expected for a spot-crossing event \citep{Espinoza2018}.  

% \begin{figure}[ht!]
%     \centering
%     \includegraphics[width = \columnwidth]{Fig/1g_Spitzer_crossing_event.png}
%     \caption{Detrended light curve of the first isolated transit of TRAPPIST-1g observed by Spitzer with, over-imposed in darkblue, the best-fit model resulting from the individual analysis assuming a constant depth for all transits.}
%     \label{1g_Spitzer_crossing_event}
% \end{figure}

The discrepancy between individual and global fits is explained by the fact that in the global analyses per planet the model tries to optimize the fit with free transit depth variations allowed so the MCMC favor an unusually small depth for planet g to fit the unusual structure (see Figure \ref{1g_Spitzer_outlier}). From the individual fit, Figure \ref{1g_Spitzer_outlier} bottom panel, we see that the structure in transit is very large (almost as long as the transit duration). If it corresponds to a spot crossing event the latter must be very large and at quite a low latitude as planet g has an impact parameter of $\simeq$ 0.42 from Table \ref{updated_param}. To investigate the origin of this structure, first we checked all the planetary transits happening near the outlier, meaning several days before and several days after the event, and found no evidence of a similar structure in any of those transits including transits of TRAPPIST-1f, the planet with the closest impact parameter to planet g (see Table \ref{updated_param}). Nevertheless, it is worth mentioning that the closest transit of planet h to the event (happening 3 days before the event at 2457662.55449 JD precisely) is one of the outliers shown in Figure \ref{fig:ouliers} (see below for details of this Figure), yet this light curve is particularly noisy in- and out- of transit and therefore not reliable. Secondly, as this event was captured during the continuous observation of the system by Spitzer in 2016, we looked in the photometry for evidence of important variations in the amplitude of the stellar variability around this event as a sign of a sudden appearance of a massive spot that could explain the structure in planet g's epoch 0 transit. To do so we applied a time rolling window (of fixed size equal to 20min) on the residuals of the detrended light curve corresponding to several days before and after the event, and from this rolling window we calculated the standard deviation and amplitude of the residual in order to catch any significant increase. Unfortunately, there does not seem to be any correlation between the appearance of the structure in the transit light curve of g and the variability of TRAPPIST-1, and as the Spitzer space telescope underwent some tracking problems during this campaign, our interpretation is limited. In a nutshell, this event is most probably isolated, which weakens the spot-crossing hypothesis considering that a massive photospheric heterogeneity would be needed to explain the observations. Nevertheless, as this structure cannot be corrected with any detrending of the systematics, one could still hypothesize that planet g transited with a different stellar hemisphere facing Earth than the other planets, or at least compared with f and h (as they have similar transit chords), and that the expected changes in stellar variability for such a large spot is not significant enough in the near infrared to significantly influence the stellar variability.  However, this hypothesis is ruled out by the monotonic increase in the planets' transit duration and impact parameters with orbital period, which implies an extremely coplanar planet system \citep{Luger2017b}.

\begin{figure*}[ht!]
    \centering
    \includegraphics[width = 0.8\textwidth]{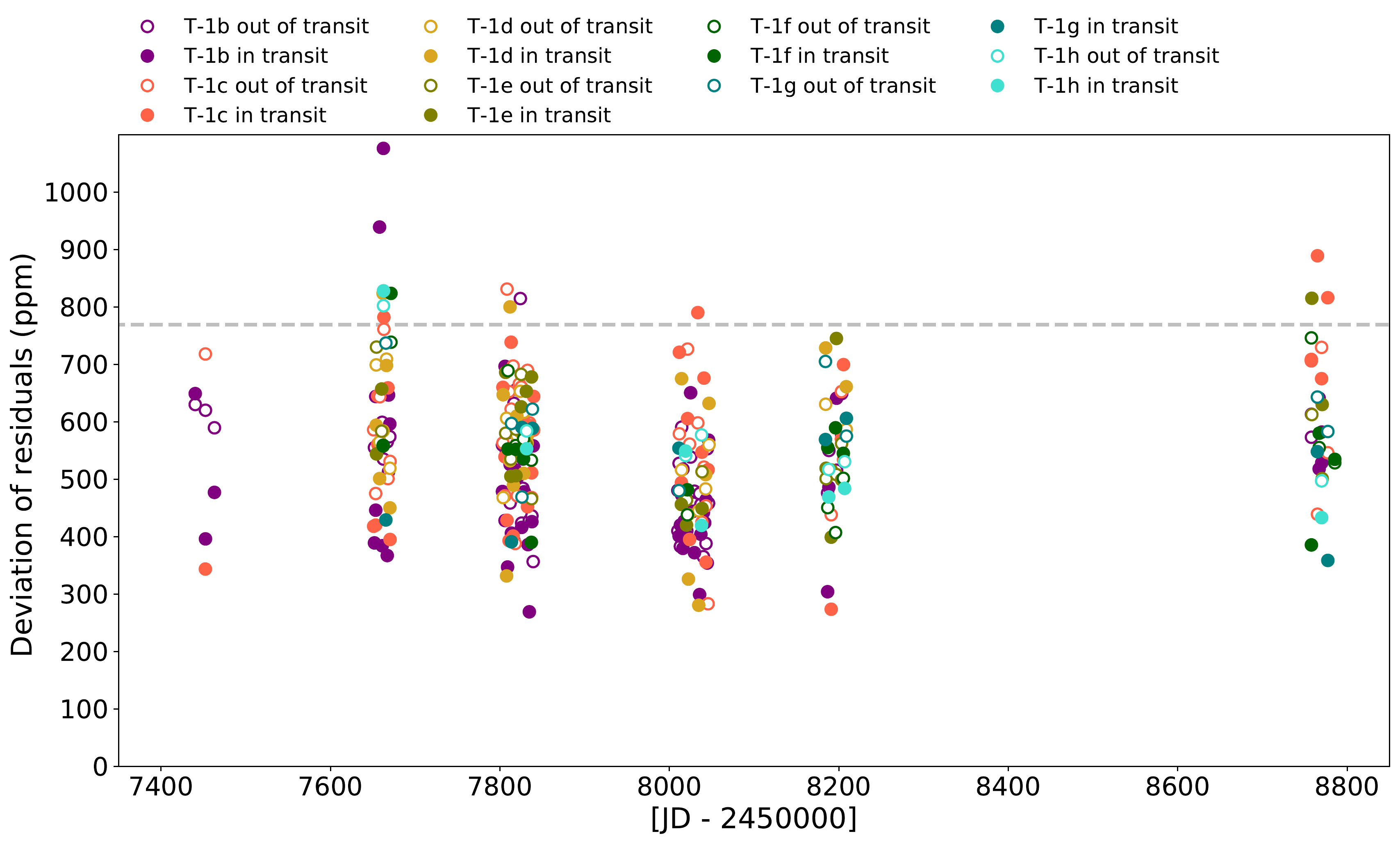}
    \caption{Standard deviation of the residuals (ppm) in- and out-of-transit for each planet (filled dots for in-transit data, empty dots for out-of-transit) such that for each transit's epoch there is one empty dot and one filled dot. Each color is associated with a planet: purple for planet b, orange for c, yellow for d, olive for e, dark green for f, teal for g and turquoise for h. The dashed gray line show the limit value above we label the transits as outliers.}
    \label{dev_res}
\end{figure*}

\begin{figure*}[ht!]
    \centering
    \includegraphics[width=0.9\textwidth]{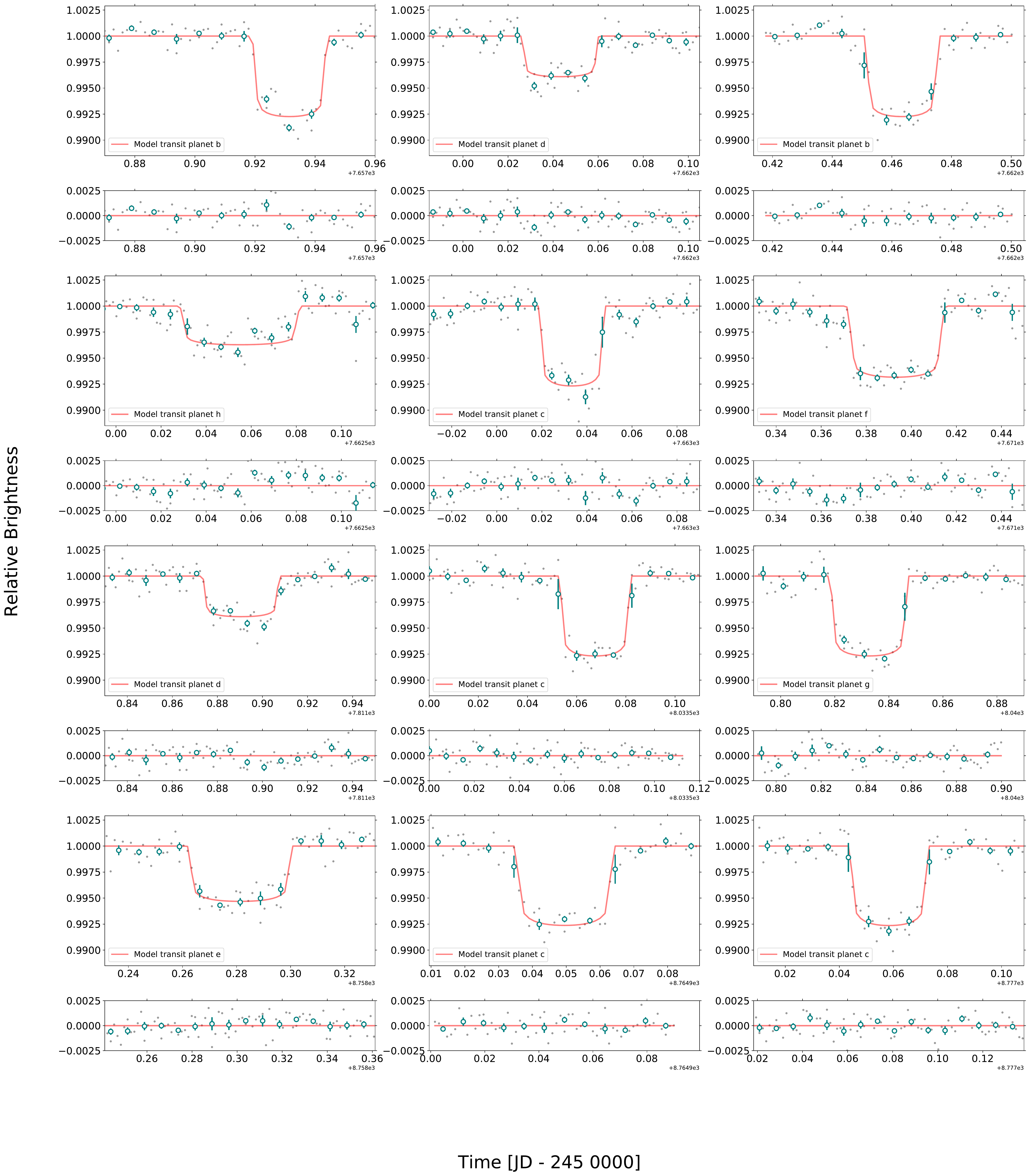}
    \caption{Transit light curves and their residuals for the 9 outliers identified on Figure \ref{dev_res}. Each outlier was attributed a number in chronological order from 1 to 9, and the corresponding transiting planet at the time is indicated within the brackets. }
    \label{fig:ouliers}
\end{figure*}

Based on this experience with planet g, we visually inspected all individual light curves associated with the other outlier values in the global analysis results of Figure \ref{fig:global_ddf}, but we did not find additional peculiar transits. To identify transit depth anomalies, we compute the median values and deviation of the photometric residuals in and out of transit (Table.\ref{table_dev_res}) as derived from the planet-by-planet global analyses (like in D2018). Figure \ref{dev_res} and Table \ref{table_dev_res} present the standard deviations obtained for the in- and out of transit residuals. Such statistics allow us to investigate the localized spot/faculae population through the "in-transit" variations and the global stellar activity more generally through the "out-of-transit" variations. We deal specifically with stellar flares in Section \ref{results_flares}.

Considering the scatter of the measurements throughout the observations, we choose to define outliers as measurements whose standard deviations of the residuals is above 769ppm (median of deviation in- and out-of-transit + 3 $\sigma$, dashed gray line on Figure \ref{dev_res}). Then, a careful look at all the light curves allows us to understand the source of uncertainty of those measurements.  We were particularly interested in cases where the standard deviation of the in-transit residuals is larger than the standard deviation of the out-of-transit residuals as this could correspond to spot or faculae crossing events. Yet, we kept in mind that a the  standard deviation of the residuals in-transit has a lower precision because it is calculated with fewer points than the standard deviation of out-of-transit residuals as the planets spend more time out-of-transit than in-transit, limiting the amount of data that can be collected in-transit.
In Figure \ref{dev_res}, we identify nine outliers, notably two transits of planet b which show a standard deviation of the in-transit residual of more than 1000ppm and 900 ppm respectively. The corresponding light curves are presented in Figure \ref{fig:ouliers}.

\begin{figure*}[h!]
    \centering
    \includegraphics[width = 0.7\textwidth]{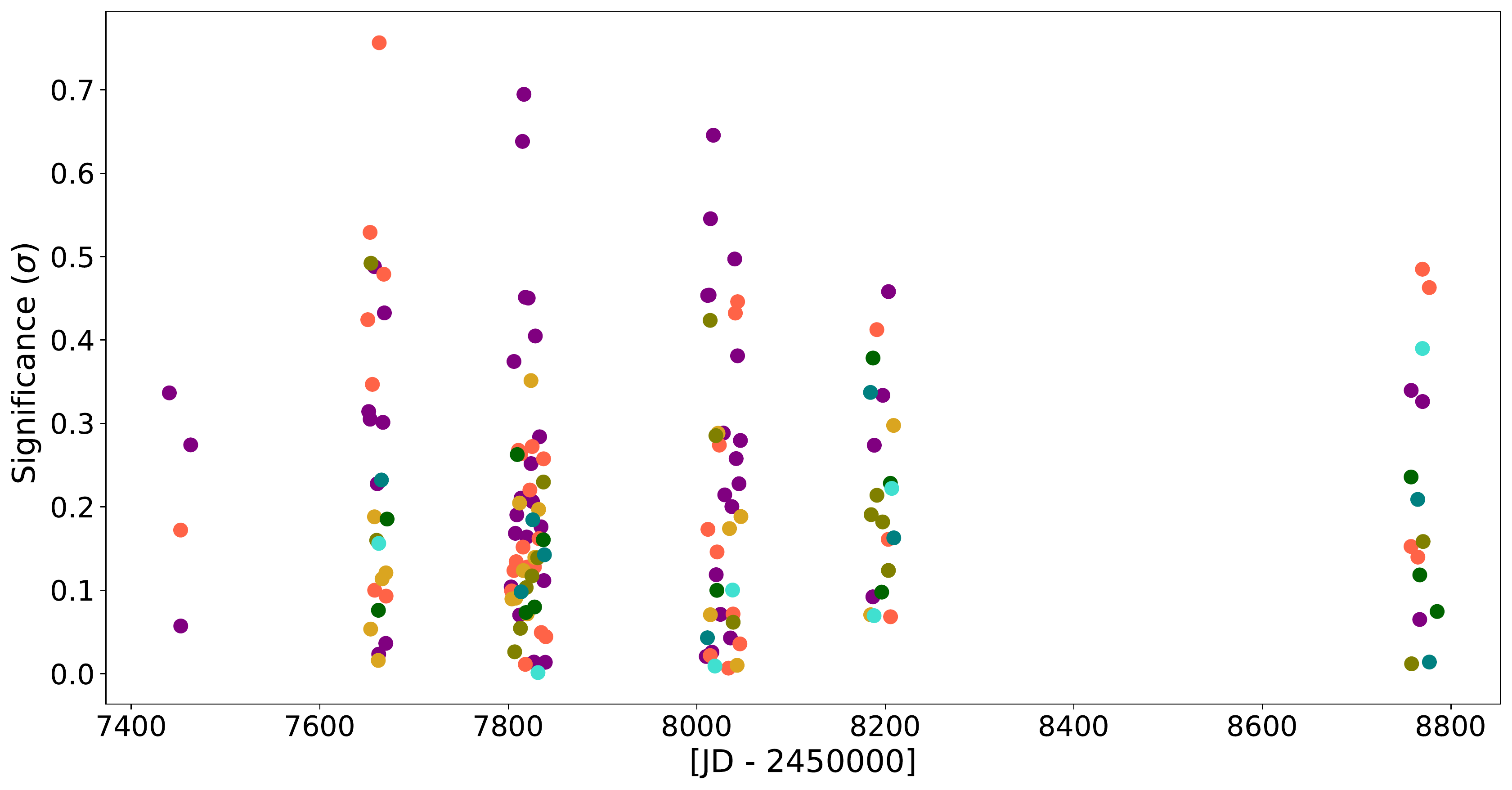}
    \caption{Significance  of the difference between the median of the residuals (ppm) in and out of transit for each planet computed as: significance = $\frac{\abs{\mathrm{median}_{\mathrm{in}}-\mathrm{median}_{\mathrm{out}}}}{\sqrt{\sigma_{\mathrm{in}}^{2}+\sigma_{\mathrm{out}}^{2}}}$. Each dot stands for one unique transit. Each color is associated to a planet: purple for planet b, orange for c, yellow for d, olive for e, dark green for f, teal for g and turquoise for h.}
    \label{sig}
\end{figure*}

In Figure \ref{fig:ouliers}, we observe that for some light curves the large value of the standard deviation of the in-transit residuals is explained by a structure that modifies the shapes of the transit. Such structures could indeed be due to the crossing of spots or faculae located within the transit chord of the planet at the time of transit. Light curves \#1, \#2, \#4, \#5, \#6 and \#9 could be interpreted as cases of bright spot crossing, while light curves \#3, \#7, and \#8 could be interpreted as cases of dark spot crossing. 
The potential presence of spots could be worrisome for a precise derivation of the radius of the planets which is an essential step toward their detailed characterization \citep{Roettenbacher2017}. To weight the relevance of those anomalies and leverage the statistical bias mentioned above, we calculated the significance of the difference between the median residual in- and out-of-transit which we define by the following formula:

\begin{equation}
    \mathrm{significance} = \frac{\abs{\mathrm{median}_{\mathrm{in}} - \mathrm{median}_{\mathrm{out}}}}{\sqrt{\sigma_{\mathrm{in}}^{2}+\sigma_{\mathrm{out}}^{2}}}
\end{equation}

\noindent
where $\mathrm{median}_{\mathrm{in}}$ and $\mathrm{median}_{\mathrm{out}}$ are the medians of the residuals in- and out-of- transit, and $\sigma_{\mathrm{in}}$ and $\sigma_{\mathrm{out}}$ are the absolute deviations of the residuals in- and out-of transit, respectively.  The results are presented in Table.\ref{table_dev_res} and on Figure \ref{sig} for clarity. 

We do not notice any significant difference between the in- and out-of-transit medians as they are comparable to within $1 \sigma$ for all transits, see Figure \ref{sig}. Those results do not favor the spot/faculae crossing hypothesis to explain the variability in transit depths that we discussed earlier, but rather systematic effects or some high-frequency stellar variability equally affecting in- and  out-of-transit data to explain those anomalies. In fact, most of the outliers identified previously belong to the second of the five campaigns, during which the Spitzer telescope had some known drifting issues due to the use of inaccurate pointing coordinates \citep{Gillon2017}. 
%Combined with the fact that the sizes and locations derived for the possible spots are not consistent with each other, 
We conclude that although it is hard to firmly discard this scenario, our results do not support the presence of stellar photospheric heterogeneities (spots and  faculae affecting the transit shape at 3.6 $\mu \mathrm{m}$ and 4.5 $\mu \mathrm{m}$).
However, one could argue that the lower contrast expected in the mid-IR may explain why we do not firmly detect any spot/faculae crossing event. Yet, recent results by \cite{Ducrot2018} failed to observe any spot crossing event in either the visible or the near-IR which favors a rather homogeneous stellar photosphere, at least for the portion transited by the seven planets. If numerous, spots would be expected to be relatively cool and small or out of the transits chords to agree with the very few events observed, see D2018, \cite{Ducrot2018}, and \cite{Morris2018c}.
Nevertheless, it is still worth mentioning that some techniques are being developed to recover the true radii of planets transiting spotted stars with axisymmetric spot distributions from measurements of the ingress/egress duration, on the condition that the limb-darkening parameters are precisely known, see \cite{Morris2018b}. The authors of the latter paper applied this technique to TRAPPIST-1 and concluded that active regions on the star seem small, low contrast, and/or uniformly distributed \citep{Morris2018a}. In any case, future JWST observations are expected to be more precise, less impacted by the limb darkening effect and therefore decisive for the confirmation of those conclusions, especially with an instrument like NIRSpec that will be able to cover a large spectral range (from 0.6 to 5.3 $\mu \mathrm{m}$, \citealt{Ferruit2014}).

\begin{figure*}[ht!]
    \centering
    \includegraphics[width =0.48\textwidth]{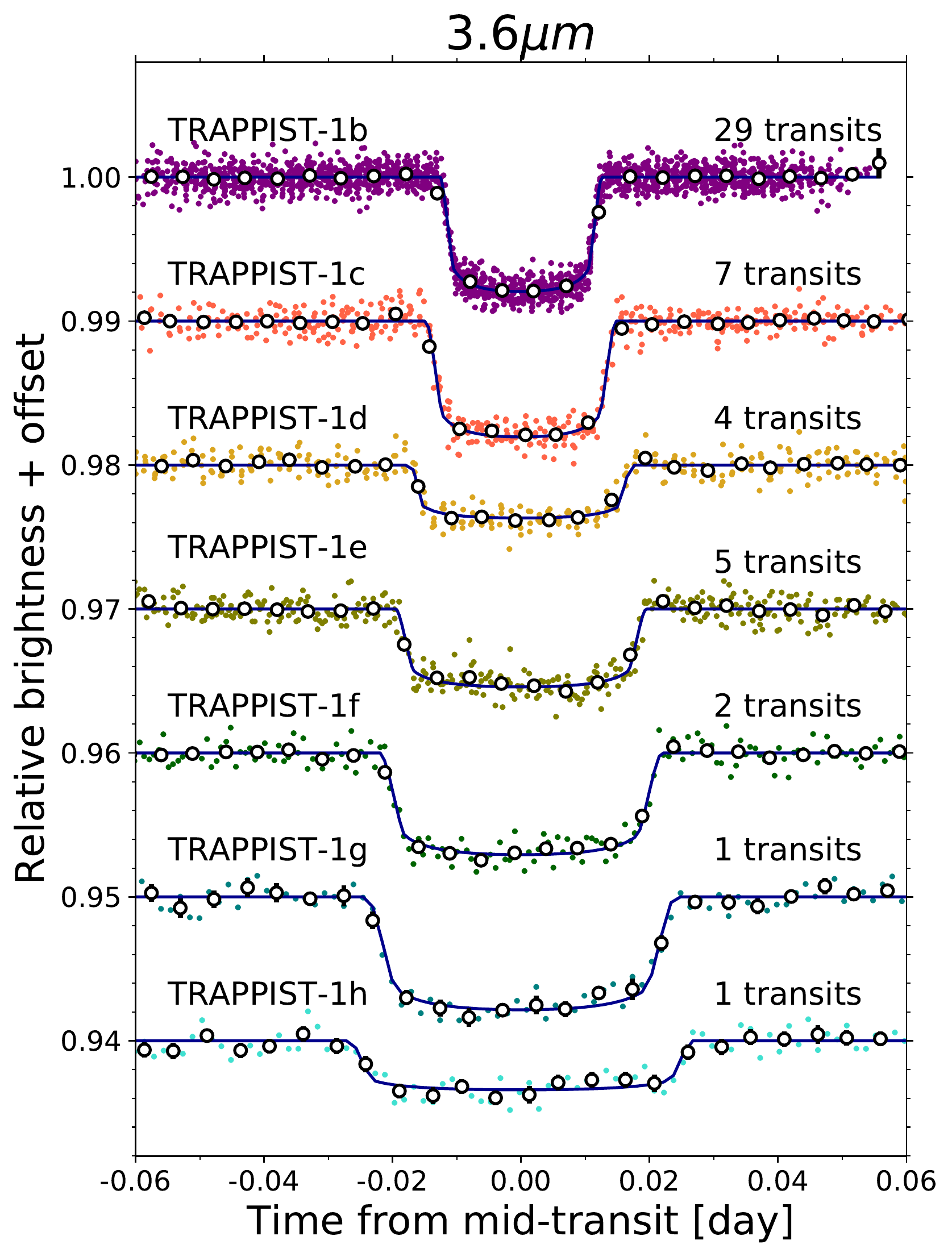}
    \includegraphics[width = 0.48\textwidth]{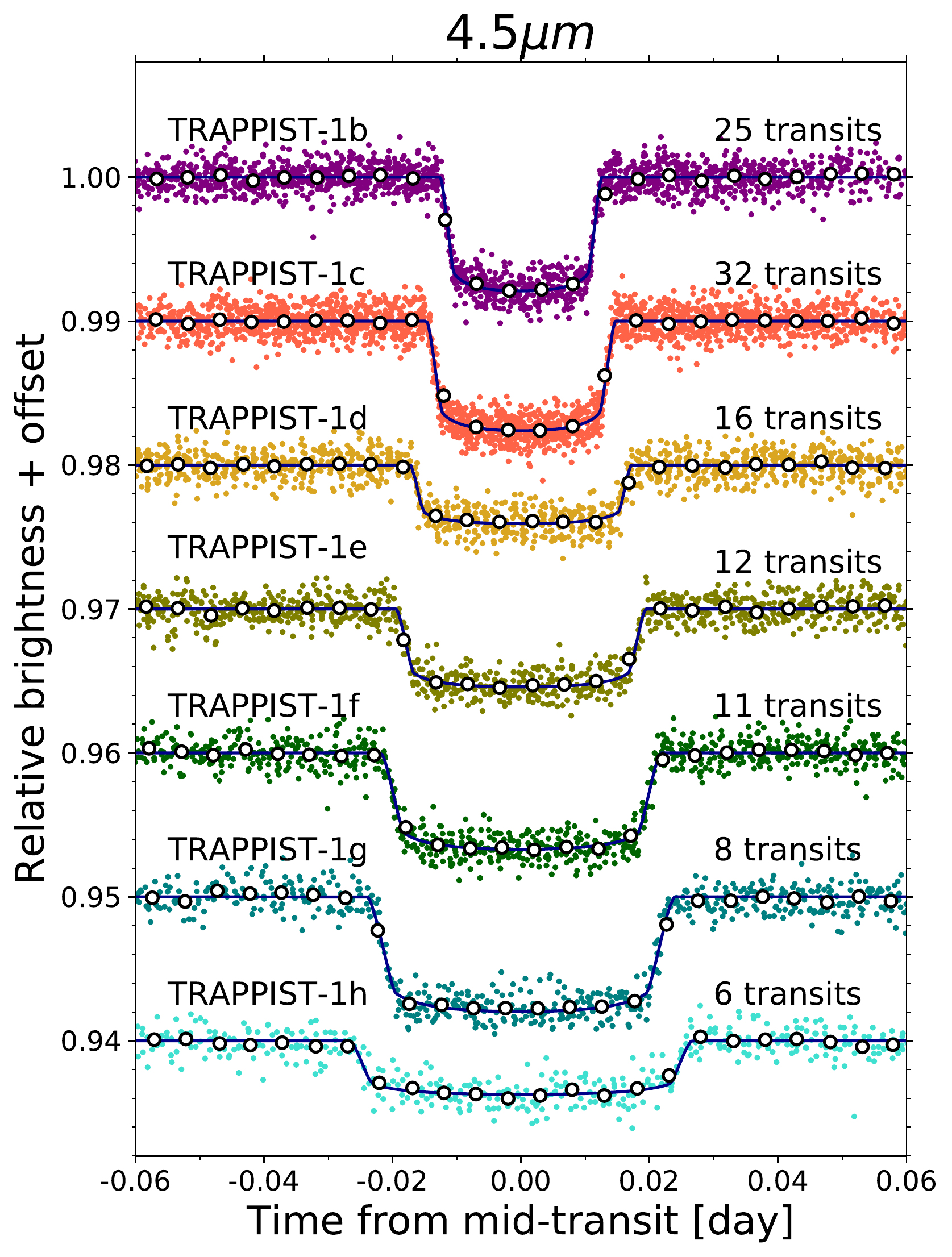}
    \caption{\textit{Left:} Period-folded photometric measurements obtained with Spitzer/IRAC channel  1 (at 3.6 $\mu \mathrm{m}$) near the transits of the seven planets, corrected for the measured TTVs. Colored dots show the unbinned measurements; open circles depict the measurements binned over 5 minutes for visual clarity. The best-fit transit models are shown as dark blue lines. The numbers of transits that were observed to produce these combined curves are written on the plot. \textit{Right}: Similarly at 4.5 $\mu \mathrm{m}$.}
    \label{fig:phase_folded_transits}
\end{figure*}

Figure \ref{fig:phase_folded_transits} shows the period-folded photometric measurements for all transits in both bands, corrected for the measured TTVs as well as the corresponding best-fit baseline models. We observe no recurrent structure for all planets. The limb darkening effect is less important at those wavelengths than in the visible or near-IR (see \citealt{Ducrot2018}) and the difference between the two channels is hardly noticeable by eye in Figure \ref{fig:phase_folded_transits}.
%MICHAEL: in fact, we can't! The figure is too small and the transits too shifted in y. #solve

\subsubsection{Transmission Spectra of the Planets}\label{transit_transmission}

\begin{enumerate}%[noitemsep]
    \item \textbf{Stellar contamination}\\
    As mentioned above, the TRAPPIST-1 planets are promising candidates for atmospheric characterization for several different reasons: the proximity of the system, the short periods of the planets, and their large sizes relative to their host. Yet, the host star itself might turn out to be an obstacle in characterizing the atmosphere of the planets if its photospheric inhomogeneity and evolution (formation, fading, and migration of spots/faculae) complicate the retrieval of atmospheric transmission signals. Consequently, a growing number of studies are dedicated to understanding the role played by the star's activity and by the heterogeneity of its photosphere when studying exoplanets with the transit method \citep[for studies of TRAPPIST-1, see ][]{Apai2018, Rackham2018, Zhang2018, Wakeford2018}. To confirm the detection of atmospheric spectral features of terrestrial planets it is essential to optimize the disentanglement of signals of planetary and stellar origin. As stated before, intensive photometric follow up is an efficient way to probe the time-variable component of stellar activity of the host star. In the previous section we focused on the impact of the presence of stellar photospheric heterogeneities within the transits chords only. Here we discuss the potential impact of spectral contamination from out-of-transits spots \& faculae on the chromatic variability of the transit depth through the transit light source effect (more details below) and how this can complicate the characterization of the planets. In particular, we construct the broadband transmission spectrum of each planet to estimate the amplitude of potential false spectral features introduced by star spots and faculae. 
    In Figure \ref{transmission_spectra}, we show the updated version of the transmission spectra of the seven planets presented by \cite{Burdanov2019}. This update consists of an additional point at $3.6\mu \mathrm{m} $ for planets c-h, updated values at $4.5\mu \mathrm{m}$, and updated weighted mean values for all planets (continuous lines in the plot). Figure \ref{transmission_spectra} combines results from \cite{deWit2016,deWit2018,Ducrot2018,Wakeford2018,Burdanov2019} and shows transmission spectra with the largest number of experimental measurements to date for the TRAPPIST-1 planets. We have decided not to include HST measurements \citep{deWit2016, deWit2018,Wakeford2018} to compute the weighted mean depth for each planet (black continuous line). This choice is justified by the fact that, although the transit transmission spectra measured in HST/WFC3 spectra are certainly reliable in relative terms, the derived absolute value of the transit depths themselves can be questioned because HST/WFC3 spectrophotometric observations are affected by orbit-dependent systematic effects which can result in diluted or amplified monochromatic transit depths, as implied by several previous studies \citep{deWit2016,Wakeford2018,Ducrot2018}. 
    
    \begin{figure*}[ht!]
        \centering
        \includegraphics[width = 0.68\textwidth]{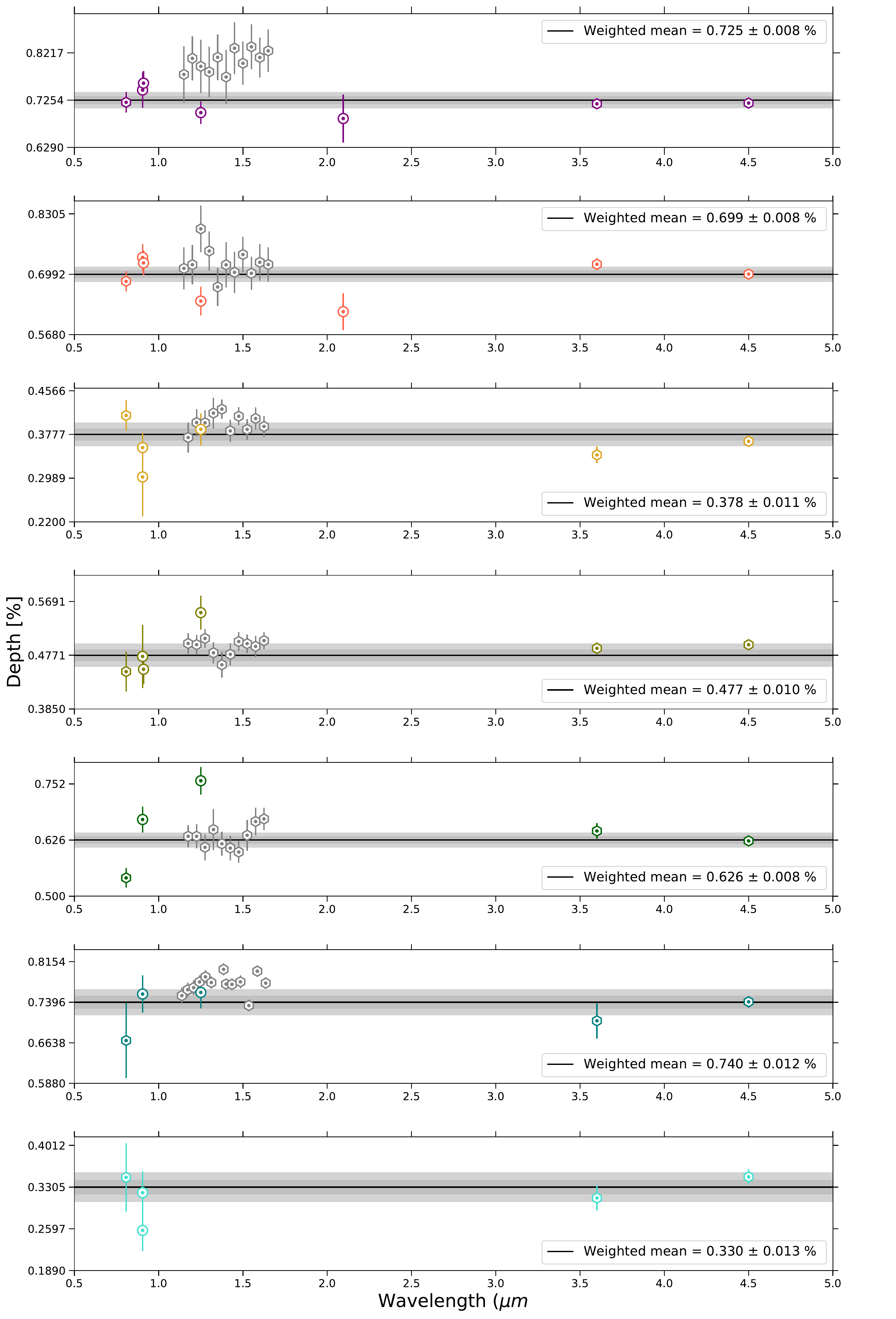}
        \caption{Updated version of the transit transmission spectra of the seven TRAPPIST-1 planets. In each subplot the continuous line is the weighted mean depth $\overline{\Delta f}$ of all non-HST measurements, with its 1$\sigma$ and 2$\sigma$ confidence intervals, in shades of gray calculated from the following formula $\sigma_{\overline{\Delta f}} = \frac{1}{\sum w_{i}} \sqrt{\sum (w_{i} \sigma_{i})^{2}}$, $\sigma_{i}$ and $w_{i}$ being respectively the error of measure i and its weight (i.e number of transits observed at each wavelength). HST measurements are presented as gray points. Coloured dots stand for the measured transit depth at the effective wavelength of the instrument, ground based measurement are symbolized by circles and space based measurement by hexagons. Each point is associated with a particular observation, in ascending order of wavelength: one point for K2 (value from \cite{Ducrot2018}), one for SSO (value from \cite{Ducrot2018}), one for LT (value from \cite{Ducrot2018}), one in the J-band for UKIRT/WFCAM and/or AAT (value from \cite{Burdanov2019}), one for the NB-2090 filter band, one point for VLT/HAWK-I only for planet b and c (value from \cite{Burdanov2019}), in IR 11 points for b and c, 10 for d, e, f and 13 for g taken with HST/WFC3 (values from \cite{deWit2016,deWit2018,Wakeford2018}) and two points for Spitzer/IRAC channel 1 and 2 (values from this work, Table \ref{depth_global_ttv}). }
        \label{transmission_spectra}
    \end{figure*}
    
    Concerning the Spitzer observations only, in Figure \ref{transmission_spectra} we observe that for all planets there are no significant differences between the $3.6\mu \mathrm{m}$ to $4.5\mu \mathrm{m}$ measurements (particularly in comparison with visible and near-IR variations), both agreeing with each other better than 2-$\sigma$ for all planets (value given in Table \ref{depth_global_ttv}).
    When we next consider all of the observational points, the depths measured at different wavelengths are all consistent with each other at better than 1-$\sigma$ for planet b, and better than 2-$\sigma$ for planets d and g. However, for planets c, e, f, and h the transmission spectra show a scatter larger than expected based on the measurement errors alone. 
    For planet h, only one point exceeds the two sigma confidence, the one derived from the Liverpool Telescope (LT) dataset. But, it is worth mentioning that the effective wavelength for LT observations ($0.9046$ $\mu \mathrm{m}$) is very close to that of SPECULOOS ($0.9102$ $\mu \mathrm{m}$) and yet the SPECULOOS value is not discrepant with the others. Furthermore those data were obtained at the same period on SPECULOOS and LT. Therefore, the difference in depth measurements between those two facilities is probably more of a systematic rather than a physical origin. 
    Yet, for planet c and e, the points that are the most inconsistent with the weighted mean value are the measurements obtained from observations carried out in the near-Infrared, with either UKIRT, VLT, AAT, or HST. Such spectroscopic transit depth variations could be the result of photospheric heterogeneity on the host star, such as unocculted active regions, which could alter the observed transit depths through the transit light source effect \citep{Rackham2018}. This effect refers to the case where a difference between the disk-averaged spectrum and the spectrum of the transit chord - which is the actual light source from the measurement - imprints spectral features on the observed transmission spectrum \citep{Rackham2019}. In that regard, \cite{Zhang2018} modeled the transit light source effect by adapting spots parameters to fit the empirical transit depth of the TRAPPIST-1 planets derived from K2, SPECULOOS, LT, HST, and previous Spitzer observations (\citealt{Ducrot2018}, \citealt{deWit2016,deWit2018,Wakeford2018}, D2018). The new depths that we derived at $3.6 \mu \mathrm{m}$ for planets c, d, e, f, g and the updated value at $4.5\mu \mathrm{m}$ for all planets allow us to compare two new observational values with the theoretical predictions of \cite{Zhang2018}. Figure \ref{fig:zhang_vs_obs} shows the combined transmission spectrum for b+c+d+e+f+g+h and best-fitting contamination model formerly computed by \cite{Zhang2018} using K2+SSO+HST+Spitzer (IRAC channel2, D2018 ).
   We computed the $\chi^{2}$ statistic value of the best-fit contamination model published in \cite{Zhang2018} to our observational points (that contains updated and new points compared to the values used in \cite{Zhang2018}). We obtain a $\chi^{2}$ of 17.5 with a p-value \citep{Pearson1900} of $0.29$, indicating that the model is not ruled out by the data. The next step would be to see if those new data points could be better fitted with a new run of the model presented by \cite{Zhang2018} or with any other model, but this is out of the scope of this paper.

    \begin{figure*}[ht!]
        \centering
        \includegraphics[width = 0.85\textwidth]{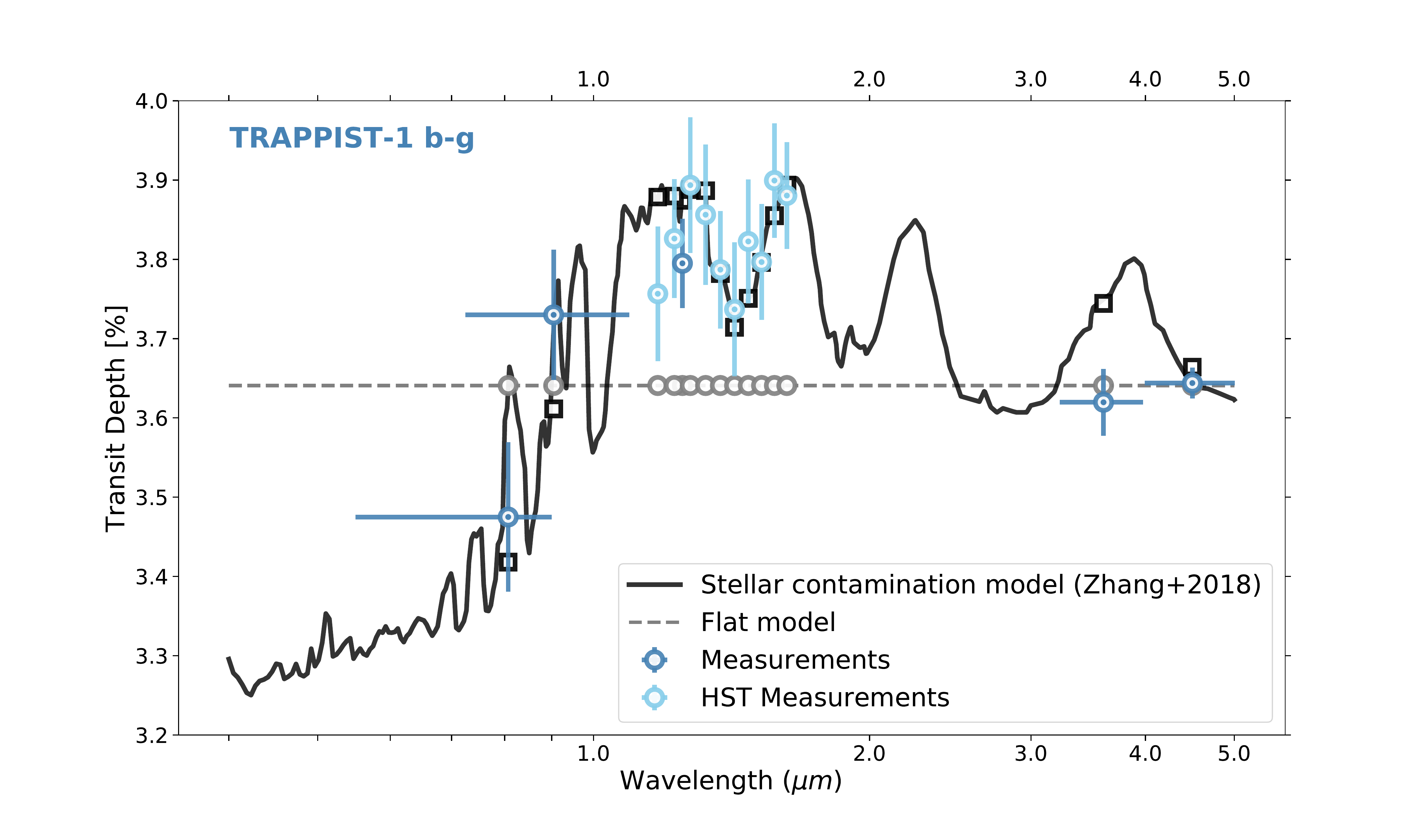}
        \caption{TRAPPIST-1 combined transmission spectra for planets b+c+d+e+f+g (blue points) constructed from individual spectra (see Figure \ref{transmission_spectra}), over-plotted with the stellar contamination model (black solid line) derived by \cite{Zhang2018} from the fit of previously published K2 + SPECULOOS-South (\citealt{Ducrot2018}) + HST (\citealt{deWit2016,deWit2018,Wakeford2018}) + Spitzer (D2018) data. The black squares are the integrated depth values as predicted by the stellar contamination model of \cite{Zhang2018} in each spectral band for which observations were performed, and gray dashed lines is the flat model at the weighted mean value of the transit depth. Wavelength is in log scale. }
        \label{fig:zhang_vs_obs}
    \end{figure*}
    
    Based on their analysis, \cite{Zhang2018} concluded that TRAPPIST-1 should be covered at $\simeq 50\%$ with spots and $\simeq 40\%$ with faculae overall. Yet, their model predicted some location-dependent spot covering fraction over the star. For instance it favored a scenario where the region transited by the planets was less spotted than the whole disk, with a spot covering fraction of only $\simeq$ 10\% of the transit chord. The authors interpreted this mismatch as the presence of an active region left unprobed by the transit chords, such as an active high-latitude or circumpolar spot. Similar structures had already been observed on fully convective M-dwarfs \citep{Barnes2015,Barnes2017}. In light of the previous section which showed that neither our work nor D2018 nor \cite{Ducrot2018} detected clear spot/faculae crossing events, while the transit chords represent 56\% of the hemisphere (D2018), and considering the results of \cite{Zhang2018}, it could indeed be possible that spots appear at preferential latitudes, like the poles. As mentioned by \cite{Zhang2018}, one way to confirm this could be to use Doppler tomography with the upcoming E-ELT/HIRES instrument, as has already been done for a few brighter stars \citep{Barnes2005}, or to use spectral template fitting to constrain spot sizes and population through molecular band observations \citep{Vogt1979}. Meanwhile, analyses driven by \cite{Morris2018c} on the Spitzer dataset (observations carried out from 2016 up to 2017) using the self-contamination method from \cite{Morris2018b} suggest that the mean photosphere of TRAPPIST-1 was rather similar to the photosphere occulted by the planets, even if they could not rule out a scenario of small-scale magnetic activity analogous in size to the smallest sunspots present within (or outside) of the transit chords. Such a scenario could also agree with our empirical results but cannot be confirmed with the current photometric precision of the existing instruments.
    Last, it is important to mention that a method to separate the planetary transmission spectrum from stellar molecular features using the out-of-transit stellar spectra, planetary transit geometries, and planetary atmospheric models has been developed by \cite{Wakeford2018}. After discarding several scenarios, \cite{Wakeford2018} concluded that a three component flux model composed of the photosphere, hotter spots ($\simeq$ 35\%) and some faculae (<3\%), with an additional small fraction of flux (1\%) from magnetic activity would be the most likely scenario for TRAPPIST-1, and that the planetary transmission spectra were likely not contaminated by any stellar spectral features \citep{Wakeford2018}. \\
    
    Considering next the transit spectra of planets c and e, it should be emphasized that, as mentioned in section \ref{time_dep_variations}, \cite{deWit2016,deWit2018} warned us that HST/WFC3 spectrophotometric observations are affected by orbit-dependent systematic effects, which can alter the monochromatic absolute transit depth values. Furthermore, concerning the measurements obtained from UKIRT, VLT and AAT, \cite{Burdanov2019} highlighted a few data gaps during some observations that could have influenced the derived transit depth values, and also emphasized how strong the influence of water vapor can be at those wavelengths for ground-based observations. Finally, those observations are also the least numerous ones, with less than 3 combined transits for all planets used to derive the values plotted in Figure \ref{transmission_spectra} in the corresponding bands. For these reasons, it is important to remain cautious about the relevance of our interpretation of planet c and e transmission spectra.
    For planet f, however, it is different as the dispersion of the measurements is even larger, with clear outliers in J band and in the visible. On one side, we could again argue for low statistics in the near-IR band to justify the large gap between UKIRT measurement and the weighted mean value, as only 3 transits were used to compute this value. But in the visible, K2's surprisingly low depth value seems more robust as it was derived from the combined analysis of 6 transits. The transmission spectrum of planet f is really intriguing and more observations in the near-IR and visible are required to draw proper conclusions on the origin of its larger scatter.
    
    All things considered, we can confidently state that the observed transmission spectra of the TRAPPIST-1 planets are not flat, but we cannot conclude on weather those variations could be attributed to stellar contamination or low statistic data. Nonetheless, these Spitzer observations provide unique statistics and show no discrepancy greater than $400$ ppm peak-to-peak and consistency at better than 2-$\sigma$ for all planets. Recent papers are converging toward a three spectral component TRAPPIST-1 photosphere as the most likely scenario \citep{Rackham2018,Zhang2018,Wakeford2018}, and authors are proposing different approaches to constrain this stellar photosphere, such as: time resolved spectral decomposition \citep{GullySantiago2017}; joint retrievals of stellar and planetary properties \citep{Pinhas2018}; visual transmission spectroscopy \citep{Rackham2017a}; transit-crossing events \citep{Espinoza2018}; or the use of out-of-transit stellar spectra to reconstruct the stellar flux \citep{Wakeford2018}.
    The combination of multiple approaches toward the study of the star's photosphere represents a  promising path toward the disentangling of the planetary atmospheric features from the stellar signals and therefore the optimization of future transit transmission spectroscopy with the eagerly-awaited JWST. 
    \\
    \item \textbf{Comparison with atmospheric models}\\
    One of the most ambitious results that the exoplanet community wishes to achieve with the upcoming JWST is the first detection of an atmosphere around a terrestrial exoplanet \citep{Madhusudhan2019}. For the reasons discussed earlier, the TRAPPIST-1 system is particularly favorable for the achievement of this goal via transit transmission spectroscopy \citep{Barstow2016,Morley2017,Batalha2018,KrissansenTotton2018,LustigYaeger2019,Fauchez:2019apj}, and offers the opportunity to probe atmospheres not only around terrestrial planets but also around temperate terrestrial planets within the habitable-zone of their host star. In the previous section, we discussed the impact of stellar contamination on the planet transmission spectra, and we concluded that several solutions are being developed to optimize the retrieval of planetary atmosphere features. In this section, we do not consider stellar contamination and only discuss potential detections of atmospheric features of the TRAPPIST-1 planets. Here, we limit our discussion to include only the cases of TRAPPIST-1b, c, e and g because (a) b and c have the smallest periods - that is to say the most transits and therefore the greatest precision on measurements, (b) planet e is arguably the most promising candidate for habitability, for the reasons given in \cite{Wolf2017,Wolf:2018apj}, \cite{Turbet:2018aa} and \cite{Fauchez:2019gmd}, and (c) planet g was the most observed with HST/WFC3 \citep{Wakeford2018}. 
    
    Combining ground and space-based observations, we can construct the broadband transmission spectra obtained in various wavelengths for each planet and compare them to recent atmospheric models of the TRAPPIST-1 atmospheres computed by \cite{Lincowski2018}, see Figure \ref{spectra_atmos_model}. To construct this figure, we have added a vertical offset to \cite{Lincowski2018}’s models to optimally overlap the observations. These offsets correspond physically to the difference between the assumed radius for TRAPPIST-1b and the solid body radius assuming a model atmosphere and its associated absorbing radius above the surface \citep{Lincowski2018}. We have applied this offset such that the models crossed the measured transit depth at the value of the sum of the weighted mean depth of each planet (shown in black solid line on Figure \ref{transmission_spectra}). For the reasons mentioned above we also applied an offset to adjust the mean level of each HST/WFC3 spectra to the weighted mean depth for each planet. By doing this we can benefit from the trustful information given by HST/WFC3 measurements on relative depths and use it to better constrain atmospheric properties.
    
    %4.5 $\mu \mathrm{m}$ because this is the point for which we have the most transits and the best photometric precision for all planets.

    \begin{figure*}[ht]
        \centering
        \includegraphics[width = 1\textwidth]{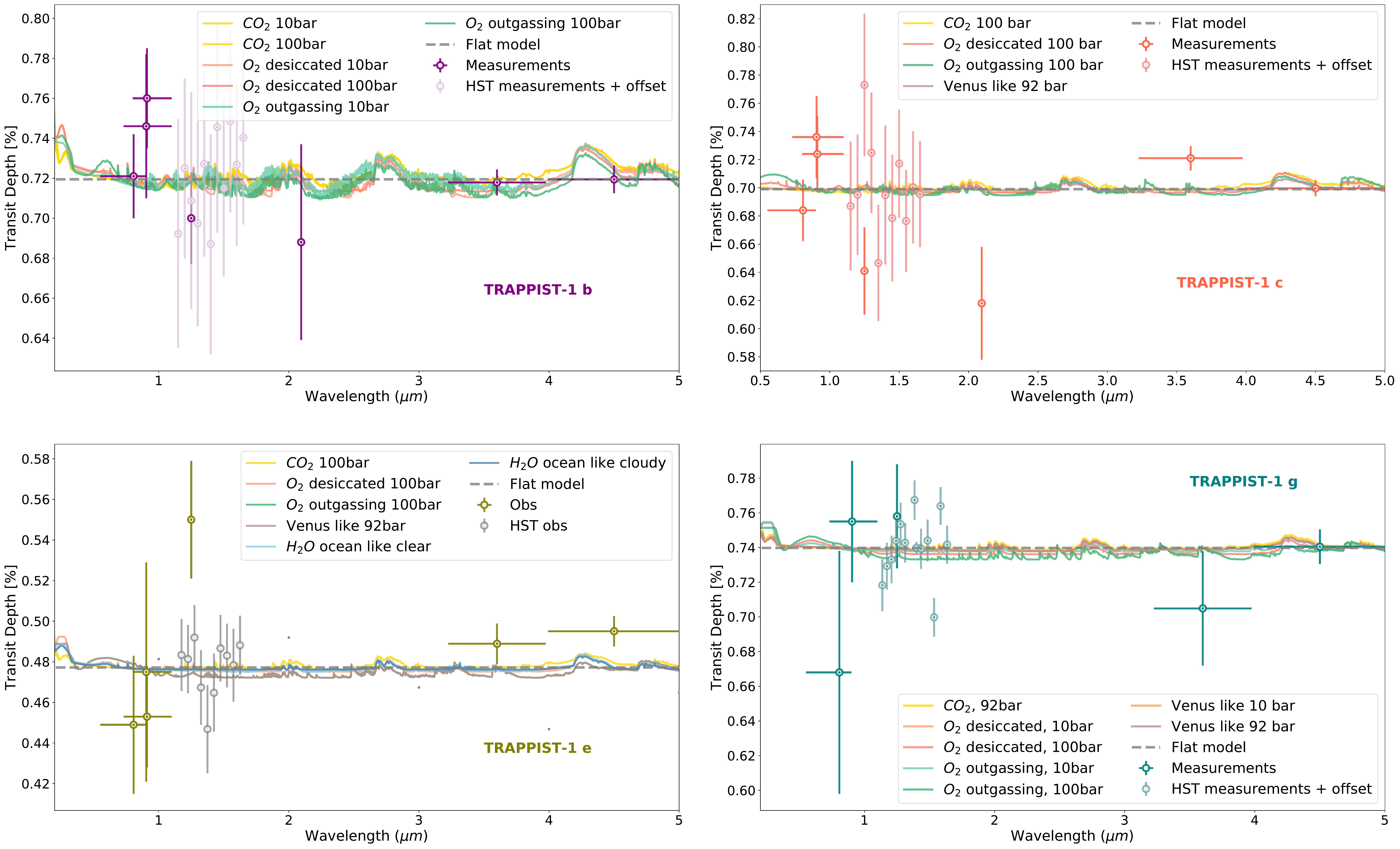}
        \caption{\textit{Top-left:} Transit transmission spectrum of TRAPPIST-1b from observations (similar Figure \ref{transmission_spectra}, except for HST-measurement on which an offset has been applied) compared with simulated transit transmission spectra derived by \cite{Lincowski2018} for different terrestrial atmospheres. Each color is associated with a different scenario: - gold stands for 10 or 100 bars $\mathrm{CO}_{2}$-rich atmospheres - salmon stands for 10 or 100 bars $O_{2}$-rich desiccated atmospheres - green stands for 10 or 100 bars $O_{2}$-rich outgazing atmospheres - brown stands for 10 or 92 bars Venus-like atmospheres - and blue stands for an aqua planet with either clear or cloudy sky. \textit{Top-right:} Similarly but for TRAPPIST-1c. \textit{Bottom-left:} Similarly but for TRAPPIST-1e. \textit{Bottom-right:} Similarly but for TRAPPIST-1g.}
        \label{spectra_atmos_model}
    \end{figure*}
    
    Those spectra illustrate our current knowledge of the transit transmission spectra gained from follow-up observations. The wavelength range that has been probed since the discovery of the system goes from $\simeq$ 0.6 $\mu \mathrm{m}$ to $\simeq$ 5 $\mu \mathrm{m}$. In this spectral range the strongest molecular features that we could expect in the absence of clouds and haze - and in a plausible planetary environment - are $\mathrm{CO}_{2}$, $\mathrm{CH}_{4}$, $\mathrm{H}_{2}\mathrm{O}$ and $\mathrm{CO}$ \citep{Tennyson2012,Gordon2017,Morley2017}. Considering the distribution of the effective wavelength of the different instruments we can only look for some localized, strong features: - the $\mathrm{CO}_{2}$ 4.3~$\mu \mathrm{m}$ spectral feature in the 4-5 $\mu \mathrm{m}$ channel of Spitzer/IRAC (width of the bandpass is 1.015~$\mu \mathrm{m}$ for that channel) - the $\mathrm{CO}_{2}$ 2.1~$\mu \mathrm{m}$ spectral feature in the VLT/HAWK-I's NB-2090 filter bandpass (width of the bandpass is 0.020~$\mu \mathrm{m}$ for NB-2090 filter) - and the $\mathrm{CH}_{4}$ 3.3~$\mu$m spectral feature in the 3.15-3.9 $\mu \mathrm{m}$ channel of Spitzer/IRAC (width of the bandpass is 0.750~$\mu \mathrm{m}$ for that channel).
    
    We deliberately did not consider models of hydrogen-dominated atmospheres as there is now plenty of evidence that all TRAPPIST-1 planets are unlikely to host this kind of atmospheres. First, transmission spectroscopy with HST/WFC3 has shown that most of the planets in the system are unlikely to have cloud-free H2-rich atmospheres \citep{deWit2016,deWit2018,Wakeford2018}. Although transmission spectroscopy cannot rule out H2-dominated atmospheres containing high-altitude aerosols \citep{Moran2018}, such configuration is in fact unlikely. This stems from the fact that any small variation of hydrogen content between planets, as expected from (1) variations in the hydrogen-rich gas accretion rates during the planet formation phase \citep{Hori2020} and from (2) variations in H2 escape rates \citep{Owen2016,Bolmont2017,Bourrier2017}, are expected to produce large variations in density between planets \citep{Turbet:2020} that are not observed \citep{Grimm2018,Agol2020}.
    
    The transmission spectrum of TRAPPIST-1b - HST/WFC3 measurements excluded - can be relatively well fit by the models. It contains the observational points with the best precision with an error bar as low as 58 ppm (Table \ref{depth_global_ttv}) in Spitzer/IRAC channel 2 measurement (thanks to the combination of 28 transits).  Yet, even with 28 transits combined the precision reached is still of same order as the expected amplitude for detectable atmospheric features on TRAPPIST-1b ($\simeq 100\%$ \citep{Morley2017,LustigYaeger2019}. 
    TRAPPIST-1c's spectrum shows a greater scatter than the one of b with an apparently poor fit to the models, yet the uncertainties on the measurements are large (at least relatively to the expected atmospheric features) and those variations are not significant at more than 3-$\sigma$.
    Then, for planets e and g, the expected spectral features are even shallower than for b and c, and the observations are less precise because of the smaller number of transits analyzed; thus it is impossible to speculate on the presence of any molecular species.
    
    One possibility to gain some precision in the measured transit depths is to study the combined transit transmission spectrum of several planets. Figure~\ref{fig:zhang_vs_obs_model} shows a transmission spectrum constructed from the combination of planets b, c, d, e, f and g's transmission spectra. 
    \begin{figure*}[ht!]
        \centering
        \includegraphics[width = 0.85\textwidth]{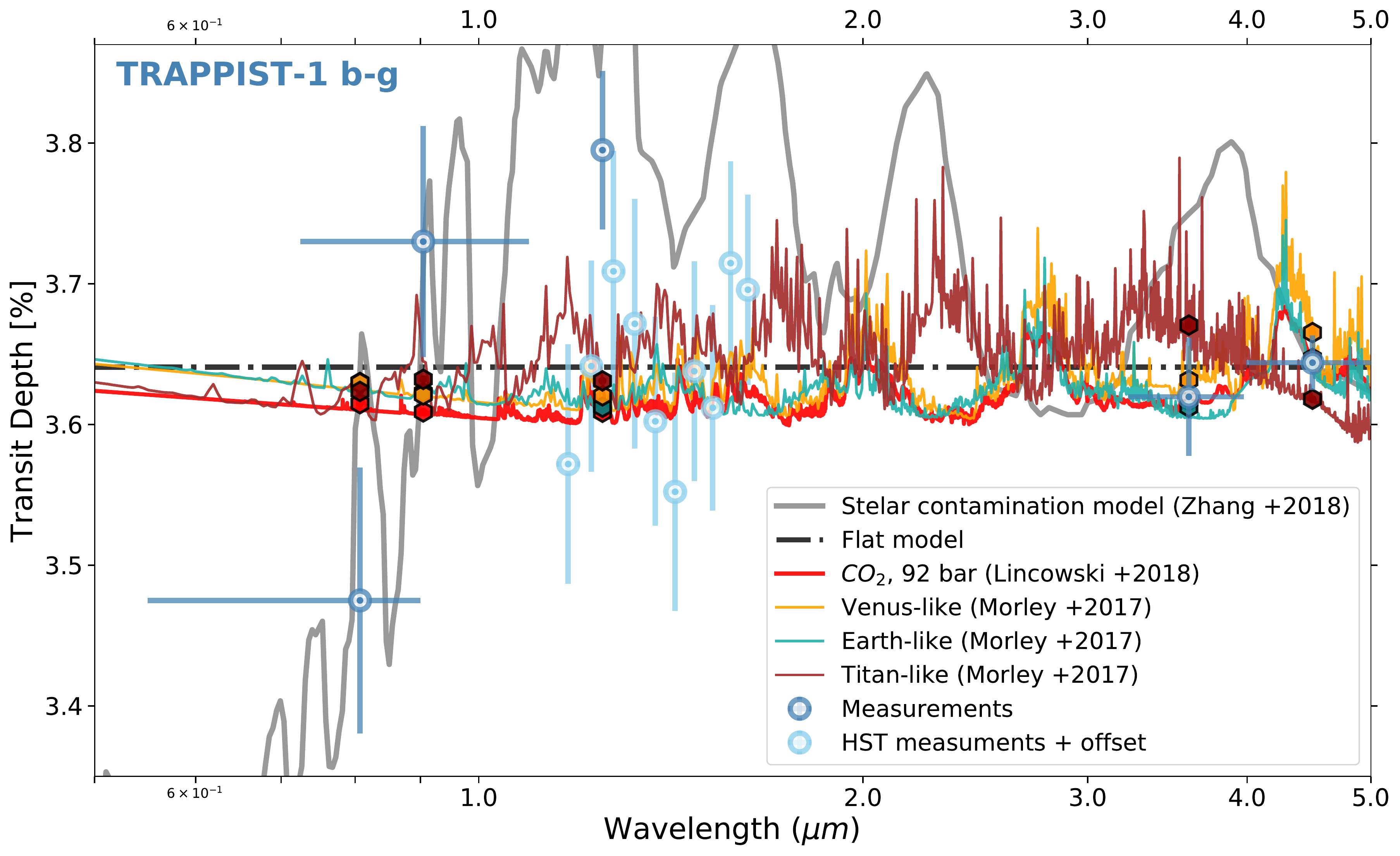}
        \caption{Similar to Figure \ref{fig:zhang_vs_obs} with simulated combined transmission spectra. Observations and their error bars are in blue. Simulated combined transit transmission spectra expected for $\mathrm{CO}_{2}$ dominated atmospheres is in red, for Earth-like atmospheres in blue-green, for Venus-like atmospheres in orange and for Titan-like atmospheres in brown. Corresponding coloured hexagons give the value expected from the models at the wavelengths of the observations. Wavelength is in log scale. }
        \label{fig:zhang_vs_obs_model}
    \end{figure*}    
    
    This figure is similar to Figure \ref{fig:zhang_vs_obs} , except from the fact that we have now put an offset on HST-measurements to adjust the mean level to the weighted mean depth calculated from the rest of the observations, and the have over-plotted several simulated combined transmission spectra from \cite{Lincowski2018} and \cite{Morley2017}. For the reasons we mentioned earlier, we have added a vertical offset to the atmospheric models to optimally overlap the observations. The offset value is such that the models crossed the value of the sum of the weigthed mean depth of each planet (shown in black dotted line on Figure \ref{fig:zhang_vs_obs}). From the observations, only points derived from the Spitzer dataset analyses have a precision of comparable magnitude than the variations expected in presence of an atmosphere. Table \ref{chi_square_atmos} gives the reduced chi square $\chi^{2}_{\nu} = \chi^{2}/\nu$ for each model if its aim was to fit the observations, with $\nu$ the number of degrees of freedom and $\chi^{2}$ is defined by:
    
    \begin{equation}
        \chi^{2} = \mathlarger{\sum_{i}} \frac{ \big (pred(i) - obs(i) \big )^{2}}{\sigma(i)^{2}}
        \label{chi_square_eq}
    \end{equation}
    
    where $obs(i)$ is the measured depth at wavelength i, $\sigma(i)$ its error, and $pred(i)$ is the depth predicted by the model for wavelength i. 
    
    \begin{table}[h!]
        \setlength{\tabcolsep}{7.pt} % Default value: 6pt
        \renewcommand{\arraystretch}{1.2} % Default value: 1
        \centering                          % used for centering table
        \begin{threeparttable}  
            \centering
            \begin{tabular}{|c|c|}
                \hline            % inserts double horizontal lines
                \textbf{Model fitted to the data}  & \textbf{Reduced $\chi^{2}$}  \\    % table heading 
                \hline  
                \hline  
                No atmosphere & 1.1920 \\
                Venus-like atmosphere \tnote{c} & 1.4376  \\ 
                $CO_{2}$, 92bar \tnote{b} & 1.5086 \\ 
                Earth-like atmosphere \tnote{c} & 1.5554 \\ 
                Titan-like atmosphere \tnote{c} & 1.8542 \\ 
                \hline                            
                \hline
            \end{tabular}
            \begin{tablenotes}
                \item[a] \cite{Zhang2018}
                \item[b] \cite{Lincowski2018}
                \item[c] \cite{Morley2017}
            \end{tablenotes}
        \end{threeparttable}
        \caption{\label{chi_square_atmos} Reduced $\chi^{2}$ values for different atmospheric models. The number of degrees of freedom used in calculating the reduced-chi-squared values listed here equal to the size of the observation sample - 1.} 
    \end{table}
    
    On Figure \ref{fig:zhang_vs_obs_model}, Earth-like \citep{Morley2017}, Venus-like \citep{Morley2017} and $\mathrm{CO}_{2}$-dominated \citep{Lincowski2018} atmospheric scenarios seem to agree reasonably well with Spitzer experimental values, showing notably a larger depth in IRAC channel 2 than in channel 1 like the data suggest. In contrast, in  a  Titan-like (i.e., CH$_4$-dominated) scenario the depth measured in  channel 1 is expected to be larger than the one measured in channel 2. This explains by the fact that a Titan-like atmospheres exhibit a strong, broad CH$_4$ absorption feature centered at 3.3$\mu$m that produces a deeper transit depth in the Spitzer channel 1 than in the channel 2. We note that the discrepancy between Titan-like atmospheres and the measured Spitzer channels 1-2 transit depths would be even greater if we assume that stellar contamination occurs at these wavelengths.
    Looking at the value reported in Table \ref{chi_square_atmos} we indeed confirm 
    that a Titan like atmosphere appears to be the less likely considering our current observational points and their errors bars. As a very preliminary estimation, we could predict that, assuming high-mean-molecular weight atmospheres the TRAPPIST-1 planets, it is rather unlikely that most of the TRAPPIST-1 planets possess a $\mathrm{CH}_{4}$-dominated atmosphere. 
    
    Yet, what we can also note from Table \ref{chi_square_atmos} is that the most likely scenario, given the current observations, is a model with no atmosphere where the transit depth is equal to the sum of the weighted mean depth at all wavelengths. However, we cannot drawn any clear conclusion because as we mention before we are extremely limited by the precision on our each measurements, even for Spitzer/IRAC channel 1 and 2.
    
    In conclusion, from Figure \ref{spectra_atmos_model} we can only lament that our current level of precision is unfortunately not high enough to draw proper conclusions about the existence of compact, high mean-molecular weight atmospheres around the TRAPPIST-1 planets. Even the combination of 22 transits of planet b at 4.5~$\mu \mathrm{m}$ and 28 transits at 3.6~$\mu \mathrm{m}$ with the Spitzer space telescope cannot reduce our error bars to sufficient precision. Nevertheless, the combined transmission spectrum of planets b to g presented in Figure~\ref{fig:zhang_vs_obs_model} tells us that the atmospheres of the TRAPPIST-1 planets are unlikely to be methane-dominated. Yet, this interpretation is made from only two observational points (Spitzer IRAC channels 1 and 2) and requires further investigation. 
    A more rigorous study of the planets' atmospheres will likely have to wait for JWST. 
    In particular, the Prism mode of the NIRSPEC instrument shows a high potential to detect compact atmospheres around the planets \citep{Batalha2018,Lincowski2018,LustigYaeger2019,Lincowski2019,Fauchez:2019apj}. Several independent simulations predict that it could take less than 10 transits for the seven planets to detect the dominant absorber \citep{Morley2017,KrissansenTotton2018,LustigYaeger2019,Wunderlich2019,Batalha2018}. This number may increase if clouds and/or photochemical hazes are present \citep{Fauchez:2019apj}. Moreover, more and more studies focus on understanding how JWST could provide us with insight into the planets' potential habitability, either through the presence of biogenic oxygen in their atmospheres \citep{Lincowski2018,Morley2017,Meadows2018}, or via the detection of anoxic biosignatures such as $\mathrm{CH}_{4}$ + $\mathrm{CO}_{2}$ minus $\mathrm{CO}$ \citep{KrissansenTotton2018}, while keeping in mind the importance of false positives/negatives \citep{Harman2015,Reinhard2017}.

\end{enumerate}

\subsubsection{Transit timing variations}

While the near-infrared (nIR) spectropolarimeter / velocimeter for the Canada-France-Hawaii Telescope (CFHT) SPIRou \citep{Donati2018} is expected to soon provide us  with the first radial velocity detection of the TRAPPIST-1 planets \citep{Klein2019}, transit timing variations (TTVs) are still the most useful tool for precisely estimating the masses of the planets. Combining the measured sizes (from transits) and masses (from TTVs) of the planets, we can deduce their densities and draw some inferences about their bulk compositions \citep{Grimm2018,Dorn:2018,Turbet:2019aa}. Furthermore, TTVs allow us to comprehend the complex dynamics that exists between those seven fast orbiting planets. Our measured timings as obtained from the global analysis (planet-by-planet) are presented in Table \ref{transit_timing_variations}, and will be used in a subsequent paper to improve the determination  of the masses of the planets through the TTV method. For any future use of the transit timings of the TRAPPIST-1 planets from Spitzer observations, we recommend to use the ones derived from the global analyses planet by planet rather than from individual analysis as they are less impacted by systematic errors due to the red noise.

\subsection{Occultations}\label{Occultations}

\begin{figure*}[ht]
    \centering
    %\subfloat{{\includegraphics[width = 0.8\columnwidth]{Fig/occultation_1b.pdf}}}
    %\qquad
    %\subfloat{{\includegraphics[width = 0.8\columnwidth]{Fig/occultation_1c.pdf}}}
    \includegraphics[width = 0.5\textwidth]{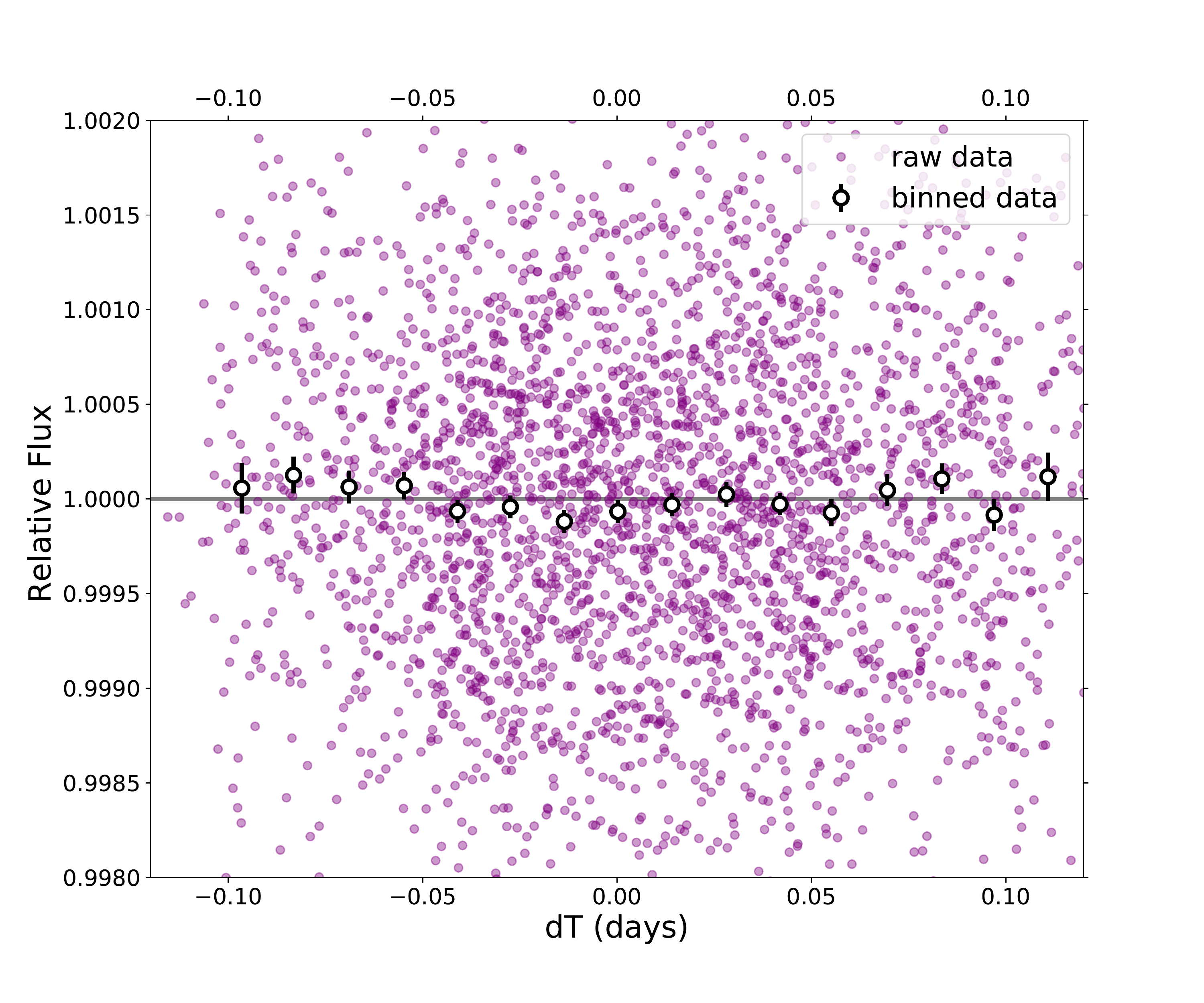}\includegraphics[width = 0.5\textwidth]{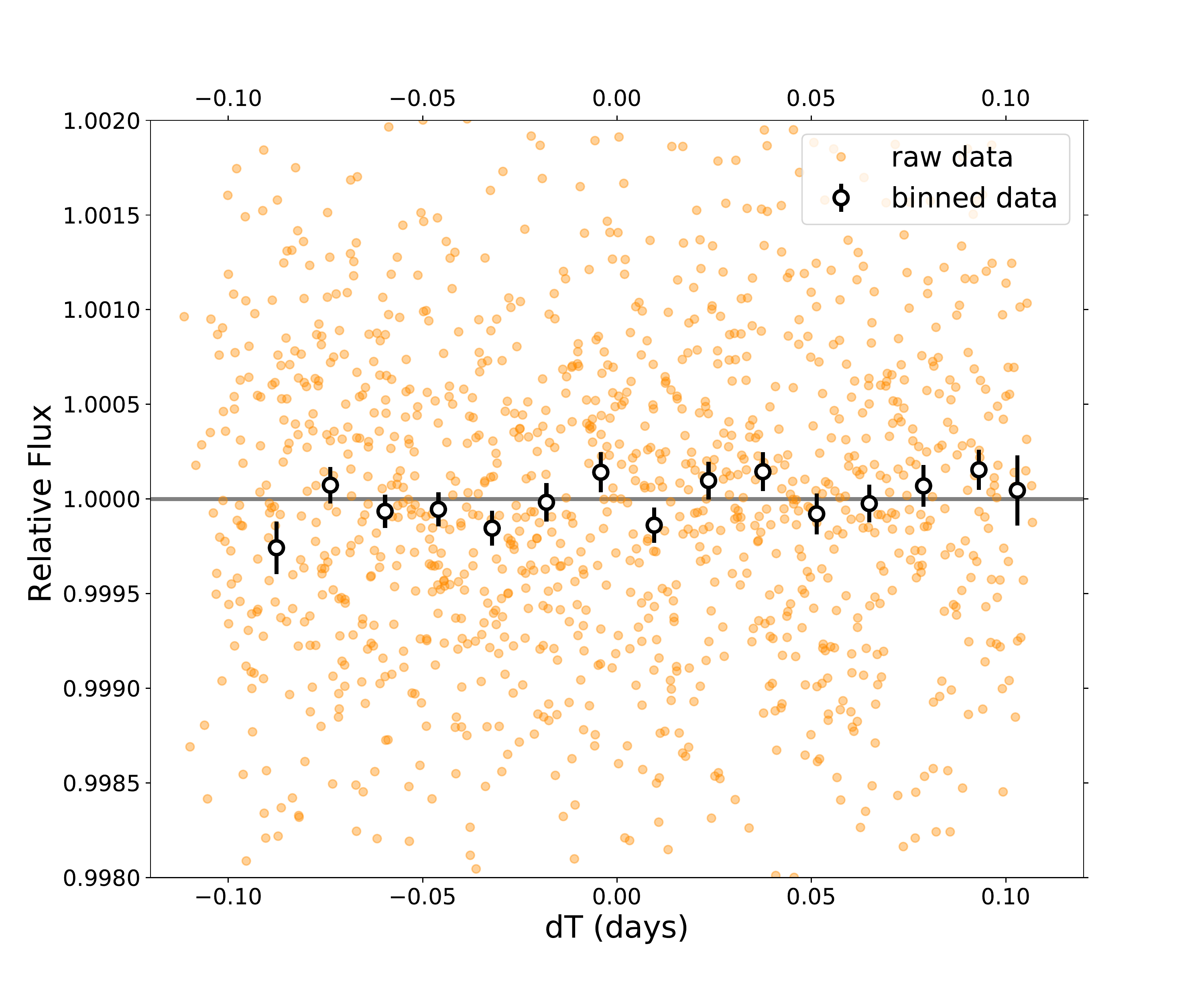}
    \caption{\textit{Left:} Period-folded photometric measurements obtained by Spitzer IRAC/channel 2 (centered in 4.5 $\mu \mathrm{m}$) near the 28 occultations of planets TRAPPIST-1b, corrected for the measured TTVs. Coloured dots show the unbinned measurements; open circles depict 20min-binned measurements for visual clarity and solid gray line is simply an horizontal line centered in 1.} \textit{Right:} Similarly but for 9 occultations of TRAPPIST-1c.
    \label{fig:occul_b_c}
\end{figure*}

In this work, we analyzed 29 predicted occultations of planet b and 8 predicted occultations of planet c, all observed in channel 2 (centered in $4.5\mu \mathrm{m}$), hereafter indexed as c2. Our aim was to derive the dayside brightness temperature $T_{p,c2}$ of the two inner planets from their occultation depths.
Unfortunately, we did not detect the occultation signal of either planet b or planet c (see Figure \ref{fig:occul_b_c}), but we were able to estimate a $3-\sigma$ upper limit on their dayside brightness temperatures. No occultation observations were taken in channel 1.

%Give constraint on thermal emission of the planets
%See want you can constraints for 1b and 1c about the temperature, talk with Franck Selsis about the possible conclusions from those constraints
%faire un tableau avec les rÃÂ©sultats de l'analyse globale pour 1b et 1c 

To derive the brightness temperature $T_{p,c2}$ from the occultation depths we used the method described in \cite{Charbonneau2005} and \cite{Deming2005}. Our starting point was to define the occultation depth as the ratio of the flux of the planet and the total flux outside of transit. This translates into equation (\ref{occultation_depth}), where $\Omega_{p}$ is the solid angle subtended by the planet, $\Omega_{\star}$ is the solid angle subtended by the star, $B_{p}$ is the surface brightness of the planet and $B_{\star}$ is the surface brightness of the star,
\begin{equation}
    \delta_{occ} = \frac{\Omega_{p}B_{p}}{\Omega_{p}B_{p}+\Omega_{\star}B_{\star}}.
    \label{occultation_depth}
\end{equation}

Then, assuming that the planet is a blackbody, its surface brightness $B_{p}$ can be expressed with Planck’s blackbody law,
\begin{equation}
    B_{p}(\nu) = \Big(\frac{2h\nu^{3}}{c^{2}}\Big)\Big(\frac{1}{e^{h\nu/k_{b}T_{p}}-1}\Big),
    \label{planck_Bp}
\end{equation}
where $\nu$ is frequency, $h$ is the Planck constant, $c$ is the speed of light in vacuum, $k_{b}$ is the Boltzmann constant, and $T_{p}$ is the brightness temperature. Equation~\ref{planck_Bp} can be re-arranged as follows:
\begin{equation}
    T_{p}(\nu) = \frac{h\nu}{k_{b}\ln\Big(\frac{2h\nu^{3}}{c^{2}B_{p}}+1\Big)}
    \label{planet_bright_T}
\end{equation}

In addition, when we develop equation (\ref{occultation_depth}) we get the ratio $ \frac{\Omega_{p}}{\Omega_{\star}}$ which we approximate as the ratio of the planet area with the star area, see equation (\ref{approx_Rp2/R*2}), such that equation (\ref{planck_Bp}) becomes equation (\ref{rewrite_Bp}) :

\begin{equation}
    \frac{R_{\star}^{2}}{R_{p}^{2}} \simeq \frac{\Omega_{\star}}{\Omega_{p}}
    \label{approx_Rp2/R*2}
\end{equation}
giving
\begin{equation}
    B_{p}(\nu) = \frac{\delta_{occ}B_{\star}(\nu)}{1-\delta_{occ}}\Big(\frac{R_{\star}}{R_{p}}\Big)^{2}.
    \label{rewrite_Bp}
\end{equation}

Finally, substituting equation (\ref{rewrite_Bp}) in equation (\ref{planet_bright_T}) we obtain the brightness temperature $T_{p}$ as a function of the occultation depth $\delta_{occ}$, the surface brightness of the star $B_{\star}$, and the frequency of the observations, see equation (\ref{planet_bright_T2}) : 

\begin{equation}
    T_{p}(\nu) = \frac{h\nu}{k_{b}\mathrm{ln}\Big(\frac{2h\nu^{3}}{c^{2}\frac{\delta_{occ}B_{\star}(\lambda)}{1-\delta_{occ}}\Big(\frac{R_{\star}}{R_{p}}\Big)^{2}}+1\Big)}
    \label{planet_bright_T2}
\end{equation}

To obtain the value of the star surface brightness $B_{\star}(\nu)$, we followed two different approaches. First, we considered the host star as a blackbody and derive $B_{\star}(\nu)$ from the value of the stellar temperature obtained in Table \ref{updated_param}. Secondly, as TRAPPIST-1 is not an ideal blackbody we computed $B_{\star}$ directly from the flux measurements in Spitzer raw images. 
To do so, we  measured the flux of  the  star in the Spitzer Basic Calibrated Data (BCDs) corrected from instrumental signatures and calibrated  in physical  units (MJy). We then followed the procedure described in the \href{https://irsa.ipac.caltech.edu/data/SPITZER/docs/irac/iracinstrumenthandbook/}{IRAC Instrument Handbook} to obtain the absolute flux density of the star in Spitzer channel 2, such that we multiplied the measured counts by $2.3504 *10^{-11}$ $\mathrm{sr}.\mathrm{arcsec}^{-2}$ x $1.22^{2}$ $\mathrm{arcsec}^{2}.\mathrm{pixel}^{-1}$, and then  divided by $\Big(\frac{R_{\star}}{d}\Big)^{2}$ to get the flux density in $\mathrm{W}.\mathrm{m}^{-2}.\mathrm{Hz}^{-1}.\mathrm{sr}^{-1}$, $R_{\star}$ being the radius of TRAPPIST-1 and $d$ the distance of the system from GAIA/DR2. Results from both approaches are presented in Table \ref{occult_Tbright}.

%To do so we extracted the flux of the star from the Basic Calibrated Date (BCDs) that is to say exposure-level data obtained after passing through the pipelines. Instrumental signatures are mostly removed from the BCDs and they are absolutely calibrated into physical units, making our life easier to derive the flux density. To get the flux per pixel we just have to multiply the measure flux by $2.3504 *10^{-11}$ $sr.arcsec^{-2}$ x $1.22^{2}$ $arcsec^{2}.pixel^{-1}$. And then dived by $\Big(\frac{R_{\star}}{d}\Big)^{2}$ to get the flux density in $W.m^{-2}.Hz^{-1}.sr^{-1}$, $R_{\star}$ being the radius of TRAPPIST-1 and d the distance of the system. Results from both approaches are shown in Table.\ref{occult_Tbright}.

\begin{table*}[h!]
\renewcommand{\arraystretch}{1.5}
\centering                          % used for centering table
\begin{tabular}{c c c c}        % centered columns (4 columns)
\hline\hline                 % inserts double horizontal lines
Planet  & \# occultations &  \thead{$3-\sigma$ upper limit \\ brightness temperature \\ from BB assumption [$K$]} & \thead{$3-\sigma$ upper limit \\ brightness temperature \\ from measured flux  [$K$]}   \\    % table heading 
\hline                        % inserts single horizontal line
  b  & 28 & 743 & 768 \\      % inserting body of the table
  c  & 9  & 812 & 842 \\  

\hline                                   %inserts single line
\end{tabular}
\caption{3$\sigma$ upper limit brightness temperatures computed from the occultation depth outputs by the MCMC analysis carried out in Section .\ref{dataanalysis} (values given in Table \ref{occultation_depth}) using equation(\ref{planet_bright_T}). The brightness temperature is a function of the surface brightness of the star that was either computed using a blackbody model (BB) or derived from the fluxes measured in the Spitzer telescope raw images of TRAPPIST-1.}             % title of Table
\label{occult_Tbright}
\end{table*}

\begin{figure*}[ht!]
    \centering
    \includegraphics[width = .49\textwidth]{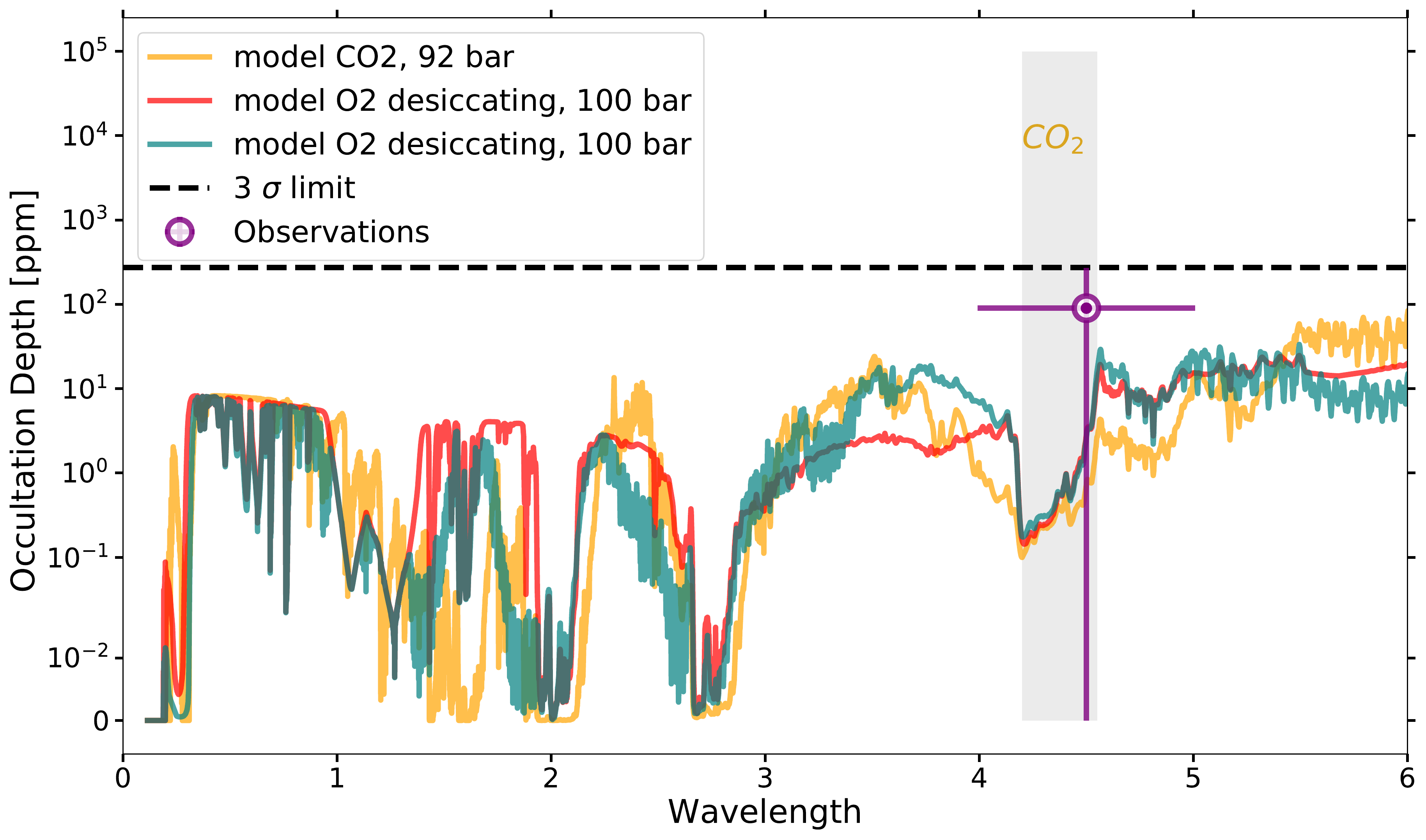}
    \includegraphics[width = .49\textwidth]{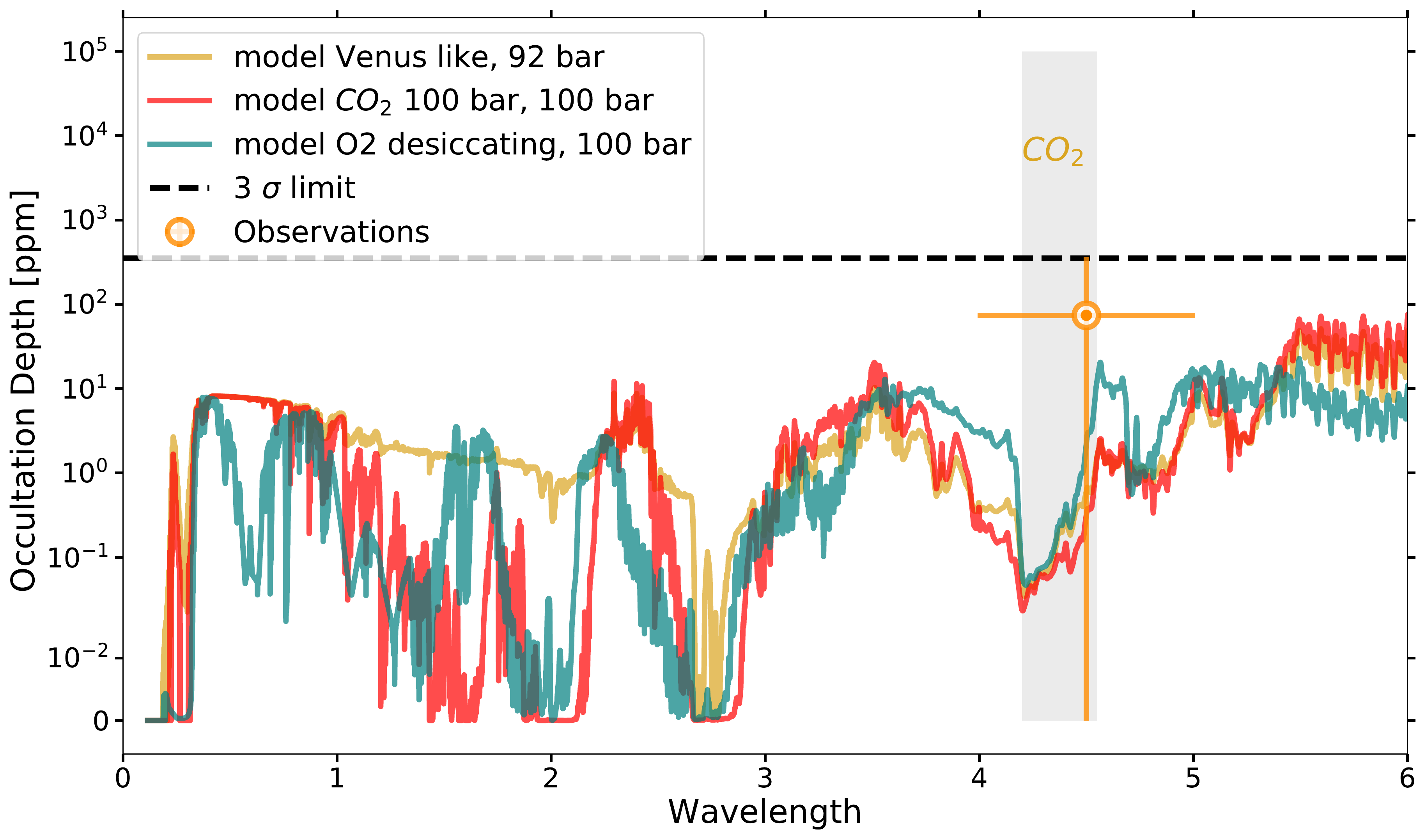}
    \caption{\textit{Left: } Secondary eclipse spectrum models of TRAPPIST-1b for different assumptions on its  atmospheric composition, as simulated by \cite{Lincowski2018}, over-plotted with our empirical value, derived from the global analysis of 28 occultations observed with Spitzer IRAC/channel 2 (centered in 4.5 $\mu \mathrm{m}$), the error bar shown corresponds to the 3-$\sigma$ confidence interval of the measurement. The y-axis has a logarithmic scale. The shade gray band stands for the zone where a spectral absorption from $\mathrm{CO}_{2}$ molecule - if present in the atmosphere - is expected. \textit{Right: }Similar but for TRAPPIST-1c.}
    \label{fig:emission_b_linconski}
\end{figure*}

Even though we did not significantly detect any occultation signal, we can compared the occultation depths outputs by the MCMC analysis and its 3-$\sigma$ uncertainty with planetary thermal emission models. On Figure~\ref{fig:emission_b_linconski}, we compare the secondary eclipse spectrum models of TRAPPIST-1b and c for different simulated atmospheric models from \cite{Lincowski2018} with the values derived from our analysis.  With Figure \ref{fig:emission_b_linconski} our intention is not to fit a model to the 3-sigma occultation depth measurement but rather to be informative on the level of signal that needs to be reached to draw conclusions from thermal occultations, and how our Spitzer occultation measurements compare to that. We observe that, for all atmospheric scenarios explored in \cite{Lincowski2018} for TRAPPIST-1b and c, the expected occultation depths are significantly smaller than the 3-$\sigma$ precision that can be reached with existing Spitzer IRAC channel 2 measurements. 

From Table~\ref{occult_Tbright} the 3-$\sigma$ upper limit brightness temperatures derived from observations are $\sim$~750~K and $\sim$~830~K for TRAPPIST-1b and c, respectively. By comparison, the equilibrium temperatures of TRAPPIST-1b and c are $\sim$~400 and $\sim$340K, respectively, assuming a null albedo. If we make the additional assumptions that the planets are (i) in synchronous rotation -- which is one of their most likely spin state \citep{Turbet:2018aa,Makarov:2018} -- and (ii) that they are devoid of atmosphere, then we calculate that equilibrium temperatures on the dayside of TRAPPIST-1b and c are $\sim$~510 and $\sim$430K, respectively.
In a thick atmosphere, the energy coming from the absorption of stellar radiation by the planet on a $\pi$R$_\text{p}^2$ area is efficiently redistributed over the surface of the planet of 4$\pi$R$_\text{p}^2$ area, leading to a dilution factor of 4. On an airless, synchronously rotating planet, the dilution factor is 1.5 due to the absence of heat redistribution combined with geometric factors. \citet{Koll:2019} provides an analytic framework to estimate this factor.

As a result, our $3-\sigma$ confidence measurements of occultation depths for both planets are not sufficient to rule out the presence or absence of an atmosphere, and cannot be used to infer the spin states of the planets. They can be used in principle to set an upper limit on the tidal heat flux of the planet, but tidal calculations have shown it should be on the order of 0.1-40~W~m$^{-2}$ \citep{Turbet:2018aa,Barr:2018,Papaloizou:2018,Dobos:2019}, depending on the tidal dissipation factor assumed and eccentricity assumed and/or calculated. These tidal heat fluxes are more than two orders of magnitude lower than the irradiation received on TRAPPIST-1b and should thus not contribute in any way to the thermal infrared flux emitted by the planet.

From the results presented in Table \ref{occult_Tbright}, we can also speculate what kinds of atmosphere would theoretically induce brightness temperatures higher than our measured upper limit in order to eliminate those scenarios for planet b and c. To maximize thermal emission between 4 and 5~$\mu \mathrm{m}$ (i.e., in the Spitzer IRAC channel 2), we can build a virtual planet with a thick atmosphere that absorbs strongly at all wavelengths (specifically at wavelengths superior to 5 $\mu \mathrm{m}$), except in the 4-5~$\mu \mathrm{m}$ spectral range. To be in agreement with our upper estimate measurements of occultation depths (see Table~\ref{occult_Tbright}), we calculate (assuming a dilution factor of 4, because here the planet needs to have a thick atmosphere producing a strong greenhouse effect and which is likely to redistribute heat efficiently) that $\sim$~76$\%$ and $\sim$~114$\%$ of the total flux absorbed (assuming a null albedo) by TRAPPIST-1b and c, respectively, need to be thermally emitted in the 4-5~$\mu \mathrm{m}$ spectral range. It is very likely impossible for planet b (and virtually impossible for planet c) to build an atmosphere which would emit nearly 100$\%$ of its thermal flux in the 4-5~$\mu \mathrm{m}$ spectral range. This justifies a posteriori why we did not detect any occultation signal of planet b nor planet c.

The most likely yet plausible atmosphere to maximize thermal flux in the 4-5~$\mu \mathrm{m}$ spectral range is a thick $\mathrm{H}_{2}\mathrm{O}$ dominated atmosphere, due to a gap between two infrared absorption bands of $\mathrm{H}_{2}\mathrm{O}$ near 4~$\mu \mathrm{m}$ \citep{MillerRicci2009,Hamano2015,Katyal:2019}. This is one of the most likely scenarios for the atmospheres of the innermost planets of the TRAPPIST-1 system if the planets formed water-rich (rich enough that they survived atmospheric erosion) as supported by some planet formation scenarios \citep{Coleman:2019aa,Schoonenberg2019,Izidoro2019,Bitsch2019,Raymond2018} and density measurements \citep{Grimm2018}. This stems from the fact that TRAPPIST-1b and c have incident fluxes beyond the runaway greenhouse limit for which water has been shown to be unstable in condensed form and should rather form a thick H2O-dominated atmosphere \citep{Turbet:2019aa}. Using the thermal emission spectra of \citealt{Hamano2015} (e.g., their Figures~1a and 3), we calculate that 15-30$\%$ of the thermal flux is emitted in the 4-5~$\mu \mathrm{m}$ spectral range, depending on the assumption made on the total water content of the planet. For TRAPPIST-1b, this corresponds to a brightness temperature of 470-530~K. These brightness temperatures are similar in magnitude to those calculated for a synchronously rotating, airless planet (equilibrium temperature on the dayside of TRAPPIST-1b of $\sim$~510~K). This demonstrates how decisive JWST occultation observations of the two TRAPPIST-1 inner planets must be to be able to constrain different realistic scenarios about the nature of these planets.

Additional gases are also likely to quantitatively change these numbers \citep{Marcq:2017,Katyal:2019}. Specifically, there is a very strong $\mathrm{CO}_{2}$ absorption band around 4.3~$\mu \mathrm{m}$ which implies that even a small amount of $\mathrm{CO}_{2}$ in the atmospheres of both planets (if any) could mitigate their 4-5~$\mu \mathrm{m}$ brightness, which would limit the ability of the Spitzer/IRAC channel 2 to detect any signal. Again, the large spectral coverage of the various instruments of JWST (NIRSpec, MIRI) combined with their expected high sensitivity will be of great use to constrain these types of atmospheres.

\color{black}{}
\subsection{Flares}\label{results_flares}

 The study of flares is essential to obtain insights into  planetary evolution and the potential presence of life on extrasolar planets. On one hand, intense flare activity can induce strong atmospheric erosion and make the surface of a planet uninhabitable \citep{Lammer2007}, but on the other hand flares could be a key element to the emergence of life \citep{Air2016,Ranjan2017}. Indeed, a minimum flaring activity seems beneficial to the formation of the ribonucleotides that will allow ribonucleic acid (RNA) synthesis and initiate prebiotic chemistry afterwards, as presented by \cite{Rimmer2018}.
This latter work outlines that further analyses of the frequencies of energetic flares around stars later than M4 are necessary to assess the habitability of temperate planets around the lowest-mass stars. \cite{Rimmer2018} also recommend concentrating on very energetic flares because of higher risks of uncertainties and contradictory findings with low energy flares. In this context, we looked for high energy flares in our extensive Spitzer data set and isolated the 5 largest-amplitude flares, 3 of them being caught during the continuous period of observations in 2016. This includes flares previously discussed by \cite{Davenport_2017} (flare \#1, \#2, \#3) and two new flares (flare \# 4 and \# 5).
We analyzed the  corresponding light curves with the  same MCMC code used to analyze the transits and occultations  (see  Section \ref{dataanalysis}), as  it also includes a  flare model represented by a instantaneous flux increase followed by an exponential flux decrease. This flare model is embodied by equation \ref{flaremodel}:
\begin{equation}
    F_{\mathrm{flare}, t} = \mathrm{Amplitude}_{\mathrm{flare}}\times e^{\Big(\frac{-dt}{\tau_{\mathrm{flare}}}\Big)}
    \label{flaremodel}
\end{equation}
where $dt = t - t_{0}$ ($t_0$ being the time of the instantaneous flux increase),  $\tau_{\mathrm{flare}}$ is the flux decrease timescale, and $\mathrm{Amplitude}_{\mathrm{flare}}$ is the flux increase amplitude.
\\

\begin{figure*}[ht!]
    \centering
    \includegraphics[width=0.95\textwidth]{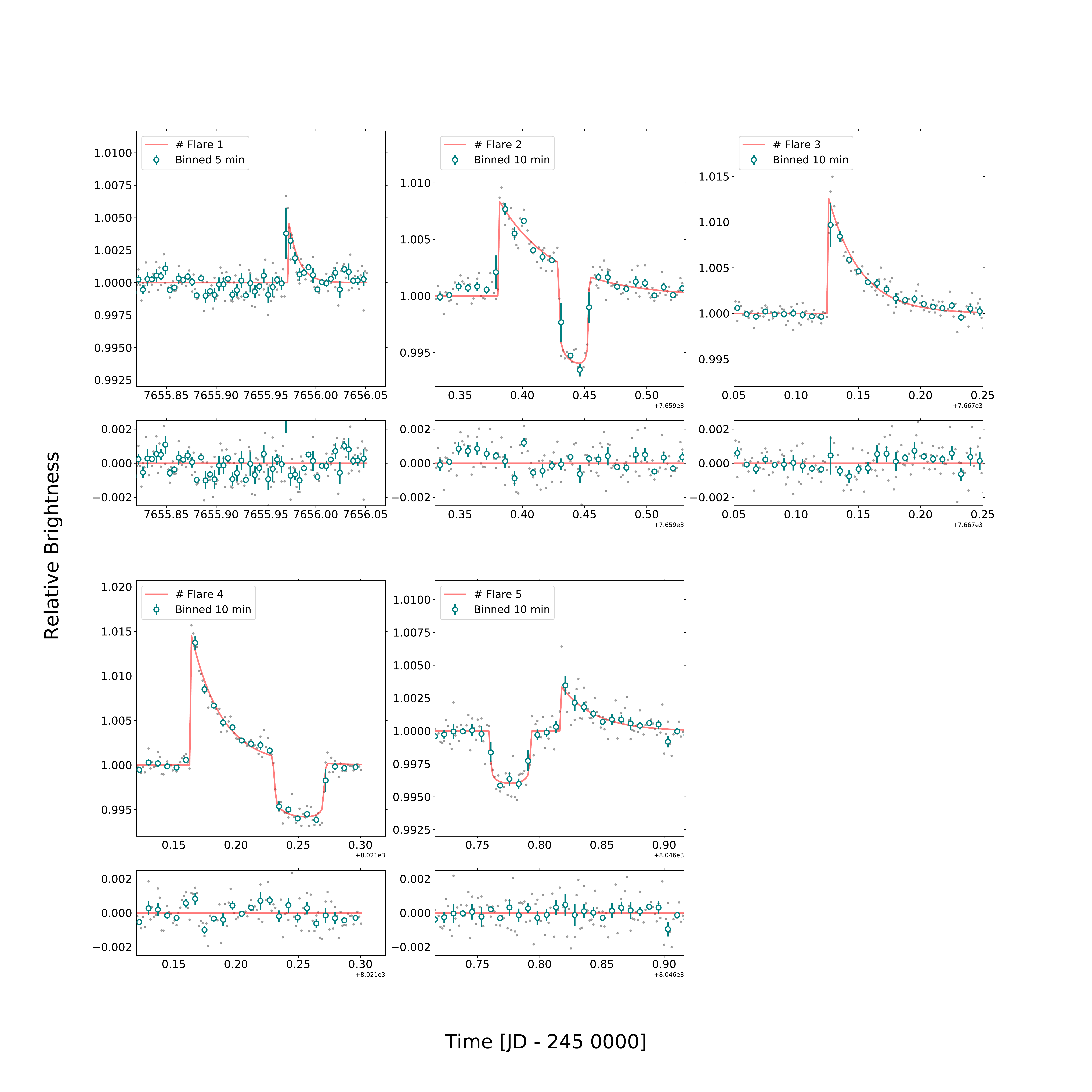}
    \caption{Light curves of the five highest amplitude flares found in our Spitzer time-series photometry. Some flares happened soon after/before a transit. The light curves are ranked in chronological order and a number is associated with each flare. }
    \label{fig:flares_plots}
\end{figure*}

The parameters resulting from our fits are presented in Table \ref{flares_stats} and the corresponding light curves are displayed in Figure \ref{fig:flares_plots}. We estimated the quality of the fit through the Gelman \& Rubin test, and for all light curves the Gelman \& Rubin coefficient was below $1.1$.\\

Table \ref{flares_stats} gives the flare parameters obtained from those fits. From those values we computed estimates of the bolometric luminosity of each flare. To do so we followed the procedure described by \cite{Shibayama2013}, that is to say we estimated the total energy of each flare from its amplitude and duration combined to the stellar luminosity (see below), by assuming that the spectrum of a flare can be described by a black body function with an effective temperature of $ T_{flare}$ $\simeq$ $9 000 K$. The justification for this assumption came from the observation of \cite{Kowalski2010}, and was reinforced in other works like \cite{Kretzschmar2011}.

Assuming that the star is a blackbody radiator, the bolometric flare luminosity can be defined as equation:
\begin{equation}
    L_{\mathrm{flare,bol}} = \sigma_{SB} T_{\mathrm{flare}}^{4} A_{\mathrm{flare}},
    \label{boloc_lumi}
\end{equation}
where $\sigma_{SB}$ is the Stefan-Boltzmann constant, $T_{\mathrm{flare}}$ is the black-body temperature of the flare, and $A_{\mathrm{flare}}$ is the area of the flare. 
%Michael: please clarify the meaning of 'area' here. 

Then, to estimate $A_{\mathrm{flare}}$, we used the observed luminosity of the star ($L_{\star}$), the luminosity of the flare ($L_{\mathrm{flare}}$) defined by equation (\ref{lum_flare}), where the integration is made for the 3.72 - 5.22 $\mu \mathrm{m}$ band pass (corresponding to IRAC/channel 2 spectral range in which all flares were observed) :

\begin{equation}
    L_{\mathrm{flare,c2}}(t) = A_{\mathrm{flare}}(t) \int R_{\lambda} B_{\lambda}(T_{\mathrm{flare}}) d\lambda
    \label{lum_flare}
\end{equation}
and the flare amplitude of the light curve $\frac{\Delta F(t)}{F(t)}$ defined as equation (\ref{ampli_flare}) :
\begin{equation}
    \frac{\Delta F(t)}{F(t)} = \frac{L_{\mathrm{flare,c2}}(t)}{L_{\star}}.
    \label{ampli_flare}
\end{equation}

In equation (\ref{lum_flare}), $R_{\lambda}$ stands for the spectral response function of Spitzer/IRAC instrument and $B_{\lambda}$ is the Planck function. From equations (\ref{lum_flare}) and (\ref{ampli_flare}), we can derive $A_{\mathrm{flare}}$, see equation (\ref{A_flare}) :

\begin{equation}
    A_{\mathrm{flare}} =  \frac{\Delta F(t)}{F(t)} \pi R_{\star}^{2}  \frac{\int R_{\lambda} B_{\lambda}(T_{\mathrm{eff}}) d\lambda}{\int R_{\lambda} B_{\lambda}(T_{\mathrm{flare}}) d\lambda}
    \label{A_flare}
\end{equation}

Finally, the total bolometric energy of the flare ($E_{\mathrm{flare}}$) is defined as the integral of $L_{\mathrm{flare,bol}}$ over the flare duration, equation (\ref{E_flare}) :

\begin{equation}
    E_{\mathrm{flare}} = \int L_{\mathrm{flare,bol}}(t) dt
    \label{E_flare}
\end{equation}

As underlined by \cite{Shibayama2013}, since the star is not a blackbody radiator, such bolometric energy estimates may have errors of a few tens of percent, too small to affect our inferences described below. The results derived from those calculations are shown in Table \ref{flares_stats}.

\begin{table*}[h!]
\centering                          % used for centering table
\begin{tabular}{|c|c c|c c|c c|c c|}        % centered columns (4 columns)                 % inserts double horizontal lines
\hline
\multicolumn{1}{|c}{\thead{\textbf{Flare \#} }} 
& \multicolumn{2}{|c|}{\thead{\textbf{Timing $\pm$ 1-$\sigma$} \\ \textbf{[JD - 2450000]}}} & \multicolumn{2}{c|}{\thead{\textbf{Amplitude $\pm$ 1-$\sigma$ } \\ \textbf{[\%]}}} 
& \multicolumn{2}{c|}{\thead{\textbf{Duration $\pm$ 1-$\sigma$ } \\ \textbf{(days)}}} & \multicolumn{2}{c|}{\thead{\textbf{Flare energy $\pm$ error } \\ \textbf{(erg)}}} \\ \hline
\hline                        % inserts single horizontal line

1&7655.97363 & 0.0038 & 0.635 & 0.39 & 0.00759 & 0.0045& 8.41 E+31 & 5.08e+31  \\
2&7659.38103 & 0.00051 & 0.846 & 0.065 & 0.04508 & 0.0072& 6.64 E+32& 4.02e+32 \\
3&7667.12545 & 0.00052 & 1.276 & 0.092 & 0.0264 & 0.0034& 5.79 E+32 & 3.50e+32\\
4&8021.16339 & 0.00052 & 0.148 & 0.093 & 0.0248 & 0.0029 & 6.41 E+31 & 3.87e+31 \\
5&8046.8164 & 0.0011 & 0.346 & 0.048 & 0.030 & 0.016& 1.81 E+32 & 1.09e+32 \\

\hline                                   %inserts single line
\end{tabular}
\caption{ Output from the individual MCMC analyses of the light curves with the 5 highest energy flares. Timing, amplitude, and duration are measured through a MCMC analysis of the corresponding light curves, while the flare energy is computed by applying equation (\ref{E_flare}) and its error is estimated to be $\pm$ 60\% of the flare energy following the recommendations of \cite{Shibayama2013} }              % title of Table
%MICHAEL: it would be interesting to know the 'area" of the flare relative to the total surface area of the star.
\label{flares_stats}
\end{table*}

\begin{figure}[ht!]
    \includegraphics[width = 0.95\columnwidth]{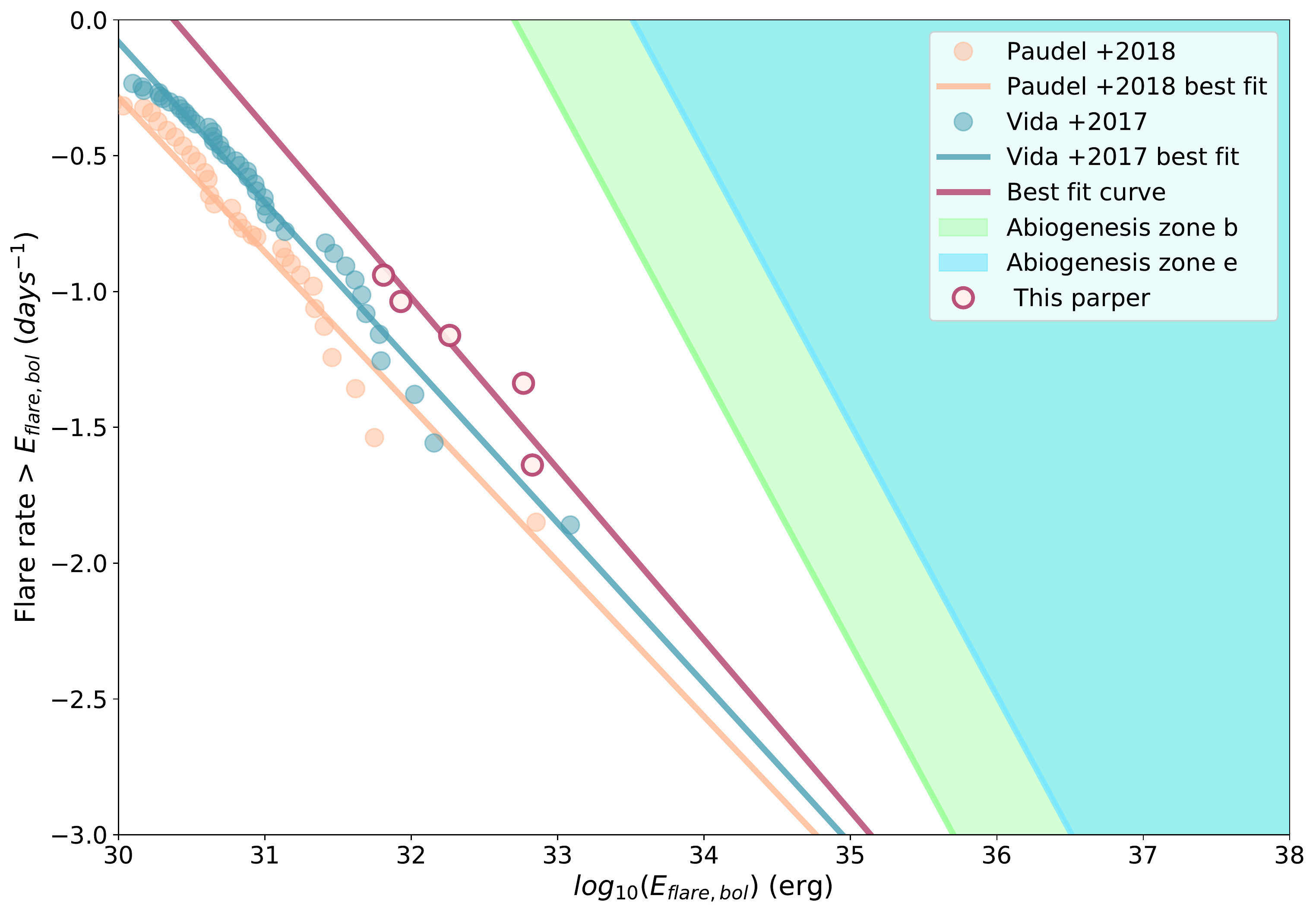}
    \caption{Flare frequency distributions (FFD) in log-log scale, x-axis is the flare energy and  y-axis is the cumulative rate of flares per day, that is to say how many flares of a corresponding energy E -or higher- happen per day. Solid lines represent the linear regressions defined by equation (\ref{power_law_flare}). The red solid line stands for the result from this work while the orange line stands for results from \cite{Paudel2018b} and the green one from \cite{Vida_2017}. The green zone denotes the abiogenesis zone for planet TRAPPIST-1b, a zone where the inequality (\ref{freq_prebio_Rimmer}) is satisfied, the green bold line on the edge of the zone represent the minimum flare rate and energy required to trigger prebiotic chemistry on this planet \citep{Rimmer2018}, which corresponds mathematically to equation (\ref{freq_prebio_Rimmer}). The blue zone is similar to the green zone but for planet TRAPPIST-1e.}
    \label{FFD_1e}
\end{figure}

The values that we obtained are consistent with flare energies amplitudes given by  \cite{Paudel2018b} and \cite{Vida_2017} (energy range from ($0.65$ - $710$) $\times 10^{30}$ erg). 
In Figure \ref{FFD_1e}, we compare the flare frequency distribution of TRAPPIST-1 from our measurements to the frequencies reported in those works. This figure is a log-log plot of cumulative frequency of flare energies; for instance, the cumulative frequency of flares with an energy E is the number of flares with energies superior or equal to E per day. 

We noticed that the flare energy distribution of TRAPPIST-1 follows a power law as equation (\ref{power_law_flare}):
\begin{equation}
    log(\nu) = \beta log(E) + \alpha,
    \label{power_law_flare}
\end{equation}
where $\nu$ is the frequency, and $\beta = -0.6303 \pm 0.1358$, fitted from our measurements. This value is consistent with the $\beta = -0.61 \pm 0.02$ derived by \cite{Paudel2018b}. 

As TRAPPIST-1 is an ultracool dwarf star and its habitable zone planets are particularly close to their host, the question of whether those planets could harbour life in such radiation environments naturally arises \citep{OMalleyJames2017,Glazier2019}. Within this context, we discuss the meaning of our results in terms of habitability; notably on how flares can promote the emergence of life. To do so, we based our discussion on the work of \cite{Rimmer2018}, where the authors explain how the synthesis of pyrimidine ribonucleotides - part of the building blocks of RNA - from hydrogen cyanide and bisulfite in liquid water is likely driven by photochemical processes in the presence of ultraviolet (UV) light. From experiments, the authors defined the “abiogenesis zones” around stars of different stellar types depending on whether their UV fluxes provide sufficient energy to build a sufficiently large prebiotic inventory \citep{Rimmer2018}. Using the flare estimates of \citet{Rimmer2018}, modified to account for the semi-major axes of the TRAPPIST-1 planets \citep[][their Eq. (10)]{Gunther2019}, we derived the flare frequencies, $\nu$, for which UV flux received by each TRAPPIST-1 planets would be sufficient for the planet to lie in the abiogenesis zone. Those frequencies are defined by the equation: 

\begin{equation}
    \nu > 25.5 \mathrm{day}^{-1} \Big(\frac{10^{34}\mathrm{erg}}{E_{\mathrm{U}}}\Big)\Big(\frac{a}{1 \mathrm{AU}}\Big)^{2}
    \label{freq_prebio_Rimmer}
\end{equation}

%\noindxent
where $\nu$ is a function of the flare’s U-band energy $E_{\mathrm{U}}$, the stellar radius, $R_{\star}$, and the stellar temperature, $T_{\star}$.
To solve inequation (\ref{freq_prebio_Rimmer}), we need the U-band energy $E_{\mathrm{U}}$ and the semi-major axis $a$. We took the semi-major axis from the refined parameters value (see Table \ref{updated_param} in Section \ref{system_parameters}) and we obtained the U-band energy from the bolometric energy through the integration of the flux density in the U-band spectral response function, like it was done by \cite{Gunther2019} (see equation \ref{E_flare_U-band}). We assumed that the flux density of the flare could be expressed as a 9000 K blackbody. We estimate that 6.6788\% of the flare’s bolometric energy belongs to the U-band, $E_{\mathrm{U}} \simeq 6.7\%E_{\mathrm{flare}}$.
\begin{equation}
    E_{\mathrm{flare}, \mathrm{U-band}} = \int_{t} A_{\mathrm{flare}} \int_{Uband} R_{\lambda} B_{\lambda}(T_{\mathrm{flare}}) d\lambda dt
    \label{E_flare_U-band}
\end{equation}

We over-plot the abiogenesis zone in terms of flare frequency and energy on Figure \ref{FFD_1e} for planet TRAPPIST-1e; in other words, a zone where the inequality \ref{freq_prebio_Rimmer} is satisfied. We chose planets b and e because e is the planet that is the most likely to harbor liquid water on its surface \citep{Wolf2017,Wolf:2018apj,Turbet:2018aa,Fauchez:2019gmd} and b is the closest to the host star. We note that a planet could lie in the abiogenesis zone while not being in the classical habitable zone \citep{Rimmer2018}, yet, by choosing to study planet e we maximize the similarities with Earth.

In Figure \ref{FFD_1e}, if TRAPPIST-1's power-law flare rates would have crossed the power law of inequality \ref{freq_prebio_Rimmer} - represented by the blue zone on the Figure - it would have potentially meant that TRAPPIST-1e was located within the abiogenesis zone of its host star. However, we see that TRAPPIST-1's power-law flare rates do not cross the blue zone. This means that TRAPPIST-1e currently does not receive enough UV flux to build up the prebiotic inventory photochemically. We note that the same interpretation can be made for planet b (the abiogenesis zone of b being the green patch on Figure \ref{FFD_1e}).

Nevertheless, TRAPPIST-1 is an old M8V type star ($7.6 \pm 2.2$ Myr old according to \citealt{Burgasser2017}) and empirical observations showed a decrease of the activity of ultracool dwarfs with age. Indeed, \cite{Paudel2019} compared the flare frequency distribution of 2M0837+2050 - a young $\simeq$ 700 Myr old M8 type star - with TRAPPIST-1 - an old 7.6 Gyr old M8 type star - and found that the highest flare energy on 2M0837+2050 are $\simeq$3 times larger than the ones on TRAPPIST-1, and that a flare of energy E=$10^{34}$erg has 10 times more chances to happen on 2M0837+2050 than on TRAPPIST-1. Considering that those two stars have a similar spectral type (M8 type) but different ages this argument could be used to hypothesize that TRAPPIST-1 used to show more energetic and more frequent flares in its youth. Both \citet{Ranjan2017} and \citet{Rimmer2018} discuss this scenario and argue that flares may be the only means to generate the building blocks of life via the pathways of \citet{Xu2018} and \citet{Patel2015}. Furthermore, contrary to the classical habitable zone, it is not required that a planet remains in the abiogenesis zone of its star to maintain the presence of life. This would imply that planet e might have received enough UV flux in its history to drive the emergence of life's building blocks.

Unfortunately, those interpretations are drawn from empirical studies and are limited by the number of flaring M8 type stars studied so far. Specifically, it remains to be seen whether there is a ``golden mean'' for flare rates, at which there are enough SEP's to form feedstock molecules needed for prebiotic chemistry, and enough NUV light to drive that prebiotic chemistry, but not so much XUV light and SEP's that the atmosphere is stripped \citep{Garraffo2017,Dong2018}, continually transformed \citep{Vida_2017,Tilley2019}, or the planet desiccated \citep{Luger2015a}. This 'golden mean' for flare rates only applies for host planets outside the abiogenesis zone as delineated by quiescent stellar NUV flux. For those planets outside the abiogenesis zone, stellar activity would be the only means to generate sufficient NUV for this prebiotic chemistry. The Earth has resided well within the abiogenesis zone throughout its history. 

It should be emphasized that the abiogenesis zones delineated by \citet{Rimmer2018, Gunther2019} and in this work are scenario-dependent. It may be that life's building blocks can arise another way, either within hydrothermal vents \citep{RimmerShorttle2019}, in surface scenarios without ultraviolet light \citep{RimmerShorttle2019}, or that they may be delivered exogenously \citep{RimmerShorttle2019}. In addition, within the scenario explored by \citet{Rimmer2018}, the threshold UV flux provides a necessary but not sufficient condition for the origins of life on a rocky planet. Hydrogen cyanide, bisulfite and phosphate must be present at high concentrations within liquid water, along with other chemical constituents \citep{Patel2015,Xu2018}. Given these added conditions, it is likely that each major category of life's building blocks: amino acids, phospholipids, nucleotides, would be present in high concentrations, along with a mechanism for joining them together to form macromolecules: proteins, phospholipid membranes, RNA and DNA \citep{Liu2019}. The problem of how life arises from this system, or any complex molecular system, remains unsolved.

\subsection{Search for a additional transiting objects}

\begin{figure}[h!]
    \centering
    \includegraphics[width=0.99\columnwidth]{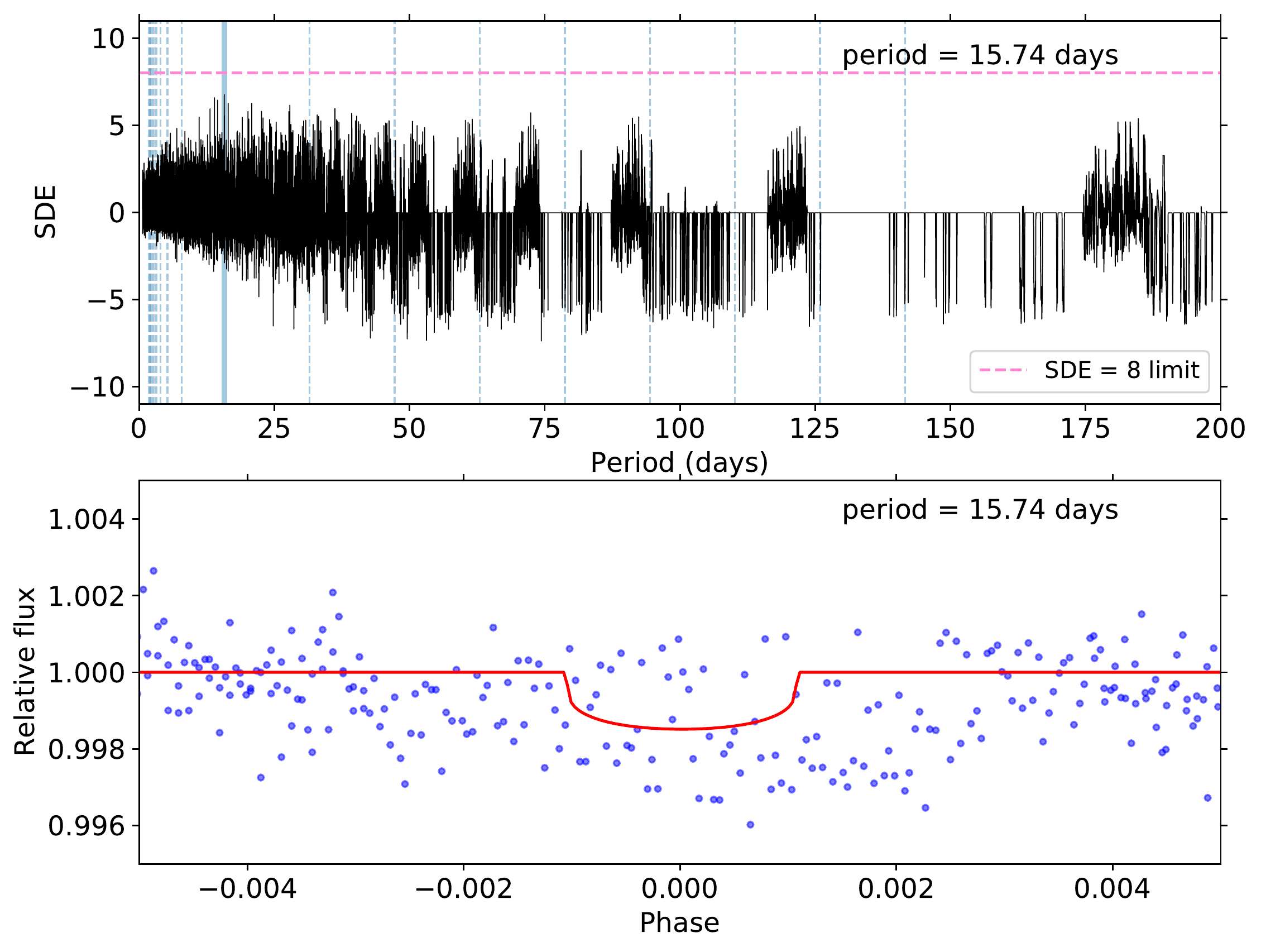}
    \caption{{\it Top panel:} Periodogram computed with the help of the Transit Least Squared (TLS) python package \citep{Hippke2019} applied to the time series made by the residuals of the entire Spitzer photometric (cleaned from all known transits). The x-axis is the period while the y-axis is the Signal Detection Efficiency (SDE) associated with each period. A steel-blue line indicates the harmonic for which the SED reached is the largest, here this output period is 15.74 days.
    {\it Bottom panel:} Phase-folded transit signal for the most probable period output by the periodogram (blue dots) + transit model computed from the parameters output by the TLS algorithm (red solid line).
    }
    \label{periodogram}
\end{figure}

One of the primary goals of the {\it Red Worlds} program was to search for additional transiting planets. In this context, we ran a Transit Least Square analysis (TLS) with period spanning from 0.2 to 200 days on the residuals of the full photometric dataset corrected from all known transits. The TLS algorithm, presented by \cite{Hippke2019}, aims to detect transit-like features from time-series photometry while taking the stellar limb-darkening, the planetary ingress and egress into account. The TLS algorithm is particularly relevant here as it is optimized for small planets and was found more reliable than the Box least Square (BLS) algorithm in finding any kind of planets by \cite{Hippke2019}. We combined this with a visual inspection of all the light curves. Results are shown on Figure \ref{periodogram}, this figure was obtained using the \href{https://github.com/hippke/tls}{Transit Least Squared (TLS)} python package by \cite{Hippke2019}.

The periodogram peaks at 15.7397 days period, yet this result must be interpreted with care as the maximum value of the SDE for this period is only 6.767 whereas \cite{Pope2016} recommend to consider planetary candidate only for Signal Detection Efficiency (SDE) > 8. The SDE being define as $SDE= max(power/std_{power})$.
Besides, the depth of the corresponding phase folded transit signal is relatively small (1487 ppm) and of the same order of magnitude as the dispersion of out-of-transit measurements (standard deviation $\simeq$ 1987 ppm) (Figure \ref{periodogram} bottom panel).
In a nutshell, those results favors a non physical explanation (most probably systematics) for this periodic signal spotted at 15.7397 days by the TLS algorithm. We did not consider any other periods in the periodogram as their SDE were always inferior to 8 \cite{Pope2016}.

\begin{figure*}[h!]
    \centering
    \includegraphics[width=0.8\textwidth]{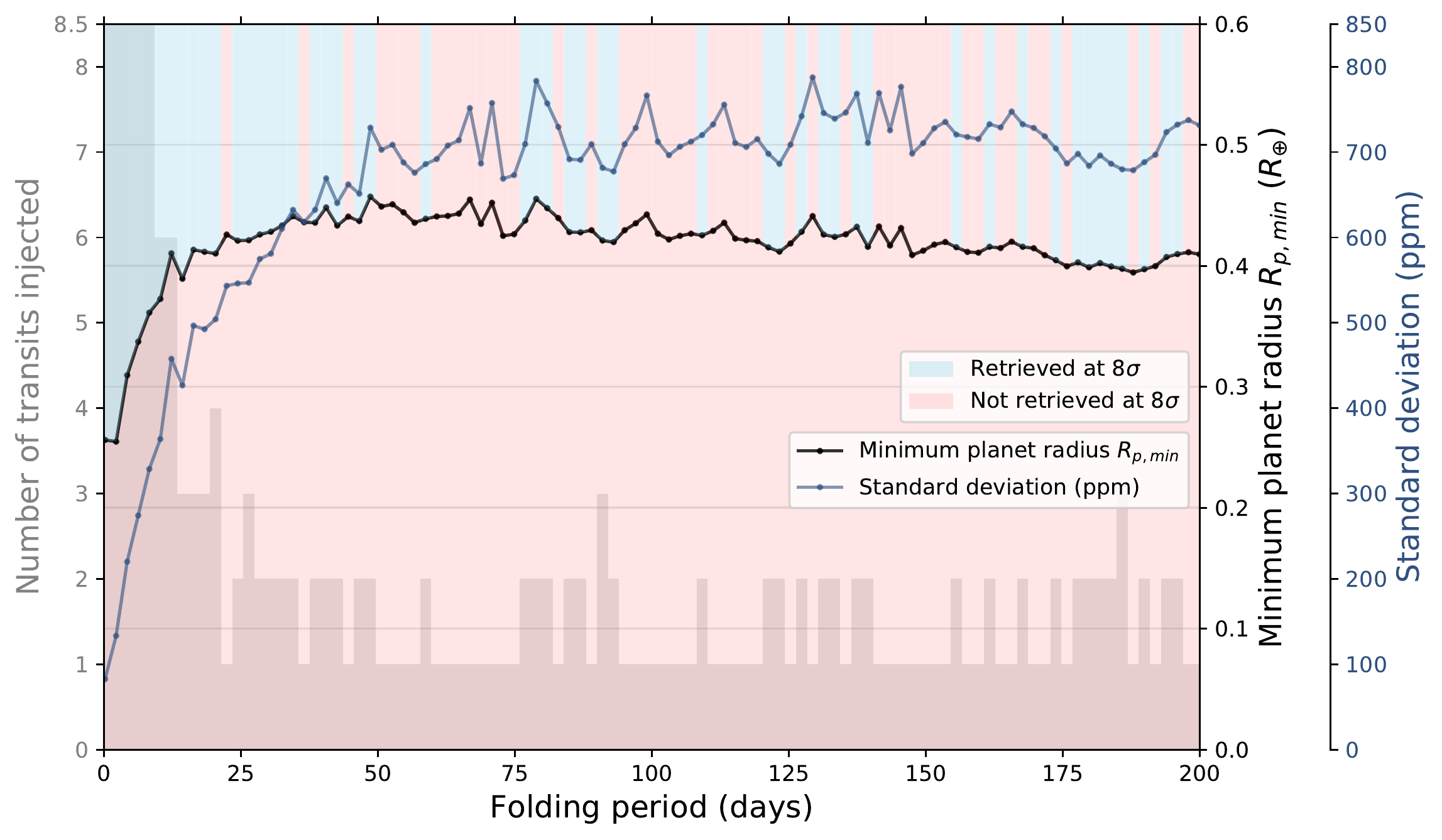}
    \caption{Evolution of the photometric precision and its corresponding planet radius for S/N = 8 as a function of the folding period. The number of injected transits for each period is given by the plot bar in gray. The blue patches show the periods for which the "injected planet" is recovered where as the red patches show the periods for which it is not. }
    \label{fig:precision_min_Rp}
\end{figure*}

Howerver, one thing we can do is to compute the photometric precision that can be reach as a function of period, then inject planets with signal-to-noise ratio (S/N) of 8 at each period and see if we recover them with TLS.  Such results can help us define which kind of hypothetical ninth planet can be discarded from Spitzer's photometry. Figure \ref{fig:precision_min_Rp} shows the precision that we get from the photometry for a set of periods going from 0.2 to 200 days. To construct this figure we have folded the data on each period, for a set of transit timings such that the full period is covered, binned them and computed the standard deviation of this binned light curve, this standard deviation is what we refer to as the photometric precision reached on a given period. The minimum planet radius (hereafter $R_{p,min}$) was then derived from a depth equivalent to a S/N of 8. Where the S/N is expressed as $S/N = \frac{dF}{\sigma}*\sqrt{N}$, with $\sigma$ the precision at a given period and N the number of points in transit, as defined by \cite{Pont2006}.

Assuming an hypothetical planets with circular orbit and null impact parameter, we observe that our precision on the dataset is good enough to detect any Mars-sized planet with period inferior to $\simeq$ 45 days (with an S/N of 8), and good enough to detect any Earth sized planet that would have a period between 0.2 and 200 days (with an S/N of 8). We observe that above P $\simeq$ 50 days the precision seems to stagnate, this is due to the fact that gaps exist in the dataset such that at some point only one event is used to construct the period-folded light curve for most of the periods. Yet, as the duration of the transit increases with the period and the precision stays more a less constant for P $\geq$ 50, the minimum planetary radius that we can detect with S/N=8  tends to decrease for P $\geq$ 50.
To second those results we performed some transits injection/retrieval, the retrieval phase being essentially the capacity to find back the injected transits with an SDE > 8 when performing a TLS analysis on the residual + injected transits. The retrieval is obviously greatly dependent on the transit timing used as reference, as a scenario where no transits fall in the observations is likely. Therefore, we imposed this reference timing to be within Spitzer's time series such that for all periods there is at least one transit in the data. The parameters of the injected transit are chosen as follow: - its period P is the main variable, - its depth $df_{p}$ is such that $S/N = 8$, with the S/N as defined above - its width $T_{14,p}$ is calculated analytically from $df_{p}$ and P assuming a circular orbit and a null impact parameter. 

As a result, TLS do recover all the injected planets with an SDE > 8 as long as at least 2 transits fall in the data, see figure \ref{fig:precision_min_Rp} . Hence, if present we should have detect any Mars-sized planet with period inferior to $\simeq$ 50 days and all Earth sized planet with P < 200 days providing at least 2 of its transits happened during the observations.

\begin{figure*}[h!]
    \centering
    \includegraphics[width=0.85\textwidth]{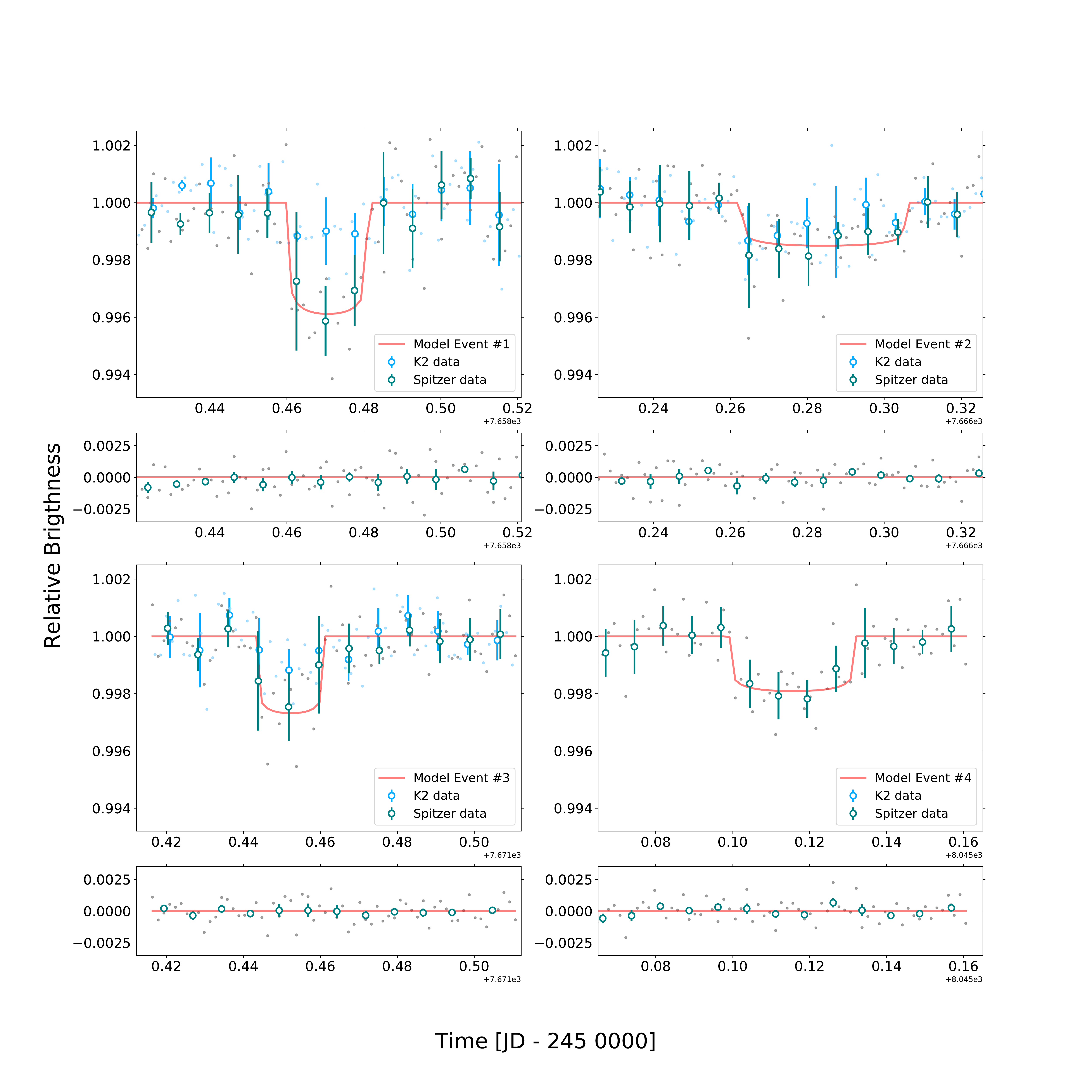}
    \caption{Visualization of the fits of the light curve of the four transit-like structures found in Spitzer photometry. Gray dots show the unbinned measurements for Spitzer; green open circles depict 10 min-binned measurements for visual clarity. The best-fit transit models are shown as red solid lines. Events are ranked and assigned a number corresponding to their chronological  order: Top-left is event \#1, top-right is event \#2, bottom-left is event \#3 and bottom-right is event \#4. When available K2 simultaneous measurements are over-plotted in blue dots with 10 min-binned measurements as blue open circles.}
    \label{fig:orphans}
\end{figure*}

To complement this analysis, we conducted a careful visual inspection of the light curves to catch any single occurrence event that could have been missed by the TLS. We found four orphan transit-like structures that did not correspond to any known planetary transit and that we could not model with any function of external parameters (e.g., x- and y-position of the star on the IRAC chip, fwhm variation and ramp effect). Therefore, we treated those events as possible transits of unknown transiting objects and tried to fit them with our MCMC code (see Section \ref{dataanalysis}). We choose to freely vary the period and the impact parameter while assuming priors on the transit depth, the eclipse duration and the transit timing with large error bars that we estimated visually. For the stellar parameters, we used the same priors as for our individual transit analysis (see Section \ref{individual_analysis}). The results from those analyses can be found in Table \ref{orphan_results} and the visualization of the fits is shown on Figure \ref{fig:orphans}.

\begin{table*}[ht!]
\centering                          % used for centering table
\renewcommand{\arraystretch}{1.2}
\begin{tabular}{|c|c c|c c|c c|c c|c c|}       
\hline
\multicolumn{1}{|c|}{\textbf{Orphan \#}} & 
\multicolumn{2}{c|}{\thead{\textbf{Timing $\pm$ 1-$\sigma$} \\ \textbf{[JD - 2450000]}}} & \multicolumn{2}{c|}{\thead{\textbf{Depth $\pm$ 1-$\sigma$} \\ \textbf{[\%]}}} & 
\multicolumn{2}{c|}{\thead{\textbf{Duration $\pm$ 1-$\sigma$} \\ \textbf{(days)}}} & 
\multicolumn{2}{c|}{\thead{\textbf{Impact parameter} \\ \textbf{$\pm$ 1-$\sigma$}}} & 
\multicolumn{2}{c|}{\thead{\textbf{Period $\pm$ 1-$\sigma$} \\ \textbf{(days) }}}\\ \hline 
\hline                        % inserts single horizontal line
\#1 & 7658.47094 & 0.00110 & 0.463 & 0.091 & 0.0287 & 0.0039 & 0.920  & 0.058 & 17 & 10 \\
\#2 & 7666.28113 & 0.00058 & 0.151 & 0.068 & 0.048 & 0.012 & 0.83  & 0.36 & 59 & 42 \\
\#3 & 7671.45227 & 0.00053 & 0.249 & 0.146 & 0.0185 & 0.0017 & 0.65 & 0.42 & 1 & 2\\
\#4 & 8045.11500 & 0.00230 & 0.198 & 0.091 & 0.0331 & 0.006 & 0.903  & 0.062 & 34 & 28 \\

\hline                                   %inserts single line
\end{tabular}
\caption{Outputs from the individual MCMC analysis of four transit-like structures found in Spitzer's photometric observations of TRAPPIST-1. Convergence of our analyses was assessed with the Gelman \& Rubin test \citep{Gelman1992} (lower that 1.1 for all jump parameters).}             
\label{orphan_results}
\end{table*}

From Table \ref{orphan_results}, we note that if those events were associated with one or more transiting objects this object would be highly grazing as the impact parameters output from our fit are all larger than 0.6. Yet, we observe that the differences in duration and amplitude between events \# 1, 2, 3 and \# 4 tend to discard a common origin scenario. Event \# 3 was caught in a particularly noisy AOR so even if the light curve structure can not be removed with any baseline detrending, we are doubtful this is a physical transit. As a general comment, event \# 1, 2 and 3 were caught during the second campaigns which, as mentioned before, had some known drifting issues due to the use of inaccurate pointing coordinates, weakening our confidence in the detrending performed. Besides, none of the timings of those orphan transits are included in the transits timings associated to the TLS most favored period of 15.74 days. Furthermore, at the time of event \# 1, 2 and 3 we found out that some K2 observations were carried out simultaneously but the data do not confirm any of the structures we identifies, which strongly weaken any astrophysical origin scenario. Finally, event \# 4 is rather shallow $\simeq$ 0.2 \% and of similar order than the out-of-transit dispersion ($\simeq$ 0.12\%), the event being significant at the $1.6\sigma$ level only.

In a nut shell, the 4 orphans structures identified seem to be emerging from a nonperfect correction of systematics effects rather than an eighth transiting object. 

\section{Conclusion} \label{conclusion}

The Spitzer Exploration Program {\it{Red Worlds}} is among the largest programs undertaken with the Spitzer Space Telescope and has gathered more than 1000 hours of observation of the touchstone TRAPPIST-1 system. 
First, we emphasize that this program has largely met its expectations, notably through the discovery of 4 new planets orbiting the TRAPPIST-1 star, all well-suited for detailed atmospheric characterization with next-generation telescopes \citep{Gillon2017}, as well as for the assessment of the variability of the host star \citep{Delrez2018}, and the determination of planet masses and orbital parameters through the transit timing variations method \citep{Grimm2018}.
In this work, we presented the analysis of all the transits of the TRAPPIST-1 planets observed with Spitzer/IRAC from February 2016 to March 2018 within the framework of the \emph{Red Worlds} program. Our approach to analyze this exquisite data-set and our most relevant results are summarized as followed:\\
- We refined both the stellar and transit parameters through global analyses of the entire dataset, which enabled us to revise the planets' physical parameters (see table \ref{updated_param}).\\
- We also performed a global analysis for each planet, and individual analyses of each transit, to search for signs of stellar contamination either of spectral or photometric nature. We did not find clear evidence of stellar contamination in the transit light curves, in agreement with \cite{Morris2018c}. \\
- Comparing our individual and global analyses of the transits, we estimate for TRAPPIST-1 transit depth measurements mean noise floors of $\sim$35 and 25 ppm in channels 1 and 2 of Spitzer/IRAC, respectively. We estimate that most of this noise floor is of instrumental origins and due to the large inter-pixel inhomogeneity of IRAC InSb arrays, and that the much better inter-pixel homogeneity of JWST instruments should result in noise floors  as low as 10ppm, low enough to enable the atmospheric characterization of the planets by transit transmission spectroscopy. 
- On the spectral side, similarly to \cite{Ducrot2018} and \cite{Burdanov2019}, we concluded that the transmission spectra of the TRAPPIST-1 planets are not flat and that \cite{Zhang2018}'s fit derived from the stellar contamination model from \cite{Rackham2018} agrees relatively well with the 7 planets' combined observations, apart from the Spitzer observational point at $3.6\mu \mathrm{m}$. Nevertheless, because of existing discrepancies in the near-IR between measurements from different instruments it is still too early to confirm that those spectral features originate from stellar contamination. \\
- Then, we searched for an occultation signal but did not detect any, even with 28 occultations of b and 9 occultations of c. Yet, we were able to set an upper constraint on the dayside brightness temperature of b and  c. Those upper limits on the temperatures were so high that it appears impossible for planet b and c to build virtual atmospheres that would be in agreement with our upper estimate measurements of  occultation  depths. This justifies a posteriori why we did not detect any occultation signal of either planet b or planet c. \\
- We then compared realistic atmospheric models with our derived transmission/emission spectra and emphasized that even with a large number of transits observed with Spitzer/IRAC, interpretations are still limited by the current precision on the measurements. Only the combined transit transmission spectrum of several planets (b+c+d+e+f+g) lowered our error bars enough to open discussions on the presence/absence of certain molecular absorbers. This combined spectrum suggests that it is unlikely that most of the TRAPPIST-1 planets possess $\mathrm{CH}_{4}$-dominated atmospheres, yet this interpretation was made from only two observational points (Spitzer IRAC channels 1 and 2) and therefore requires further investigation. This observation highlights the even more crucial role of the future James Webb Space Telescope to shed light on the presence or absence of atmospheres around the TRAPPIST-1 planets, and their subsequent characterization. \\
- We also computed the flare frequency distribution of the planets and observed that none of them currently lies in the abiogenenis zone of their host, meaning that the actual flaring activity is currently too weak to initiate prebiotic chemistry via the mechanism detailed in \citet{Xu2018,Rimmer2018}. However, TRAPPIST-1 is believed to be old and flaring activity is believed to decrease with star's age meaning that the planets could have received the appropriate amount of UV energy in their past. \\
- Last but not least, we identified four orphan transit structures in the dataset that we could not link to any known planet. However, three of those structures are not confirmed by simultaneous K2 data and the last one is not significant at more that the 2$\sigma$ level. These structure could be the result of unknown external systematical effect.

The most recent observations of the TRAPPIST-1 system with Spitzer were performed in October 2019, and will unfortunately be the last. But on a brighter note, the James Webb Space Telescope is on its way to take over and yield even more insight into this extraordinary system.

\begin{acknowledgements}

    We thank Benjamin Rackham for useful discussions and the anonymous referee for very constructive remarks that improved the paper in a significant way.
	This work is based in part on observations made with the Spitzer Space Telescope, which is operated by the Jet Propulsion Laboratory, California Institute of Technology under a contract with NASA. The research leading to these results has received funding from the European Research Council under the FP/2007-2013 ERC Grant Agreement n$\deg$ 336480, the European Union's Horizon 2020 research and innovation programme (grant agreement n$^\circ$ 679030/WHIPLASH), and from the ARC grant for Concerted Research Actions, financed by the Wallonia-Brussels Federation. This project has received funding from the European Union’s Horizon 2020 research and innovation program under the Marie Sklodowska-Curie Grant Agreement n$^\circ$ 832738/ESCAPE. M.T. and L.D. acknowledge funding from the Gruber Foundation. MG, EJ, and VVG are F.R.S.-FNRS senior Research Associates. PBR acknowledges support from the Simons Foundation (SCOL Grant \#599634). MNG acknowledges support from MIT's Kavli Institute as a Juan Carlos Torres Fellow.
	This work has been carried out within the framework of the National Centre of Competence in Research PlanetS supported by the Swiss National Science Foundation. The authors acknowledge the financial support of the SNSF. 
\end{acknowledgements}


\bibliographystyle{aa} % style aa.bst
\bibliography{biblio}
%----------------------------------------------------------------------------------------
%	 APPENDICES
%----------------------------------------------------------------------------------------

% Include the appendices of the thesis as separate files from the Appendices folder
% Uncomment the lines as you write the Appendices

% Appendix A
% Appendix A
\onecolumn
 % Main appendix title
\begin{appendix} %First appendix
% Change X to a consecutive letter; for referencing this appendix elsewhere, use \ref{AppendixX}
\section{Description of the data}\label{AppendixA} 

\renewcommand{\arraystretch}{1.2}
\setlength{\tabcolsep}{6pt} % Default value: 6pt
\begin{longtable}{|c|c|c|c|c|c|c|c|}
\caption{\label{baseline_indiv} Baseline for the individual analysis of each transit}  \\
\hline
Date & Number of Points & Epoch & Baseline & $\beta_{w}$ & $\beta_{r}$ & CF&Channel\\
\hline 
\endfirsthead
\caption{continued.}\\
\hline\hline
Date & Number of Points & Epoch&Baseline & $\beta_{w}$ & $\beta_{r}$ & CF&Channel\\
\hline \hline
\endhead
\hline
\endfoot
\hline
\textbf{TRAPPIST-1b} \\
\hline
2016-02-21&108&78&$p(t^3)$+$p(fwhm_x^1)$+$p(xy^1)$&1.09&1.0&1.09&c2\\
2016-03-04&132&86&$p(xy^1)$&0.84&1.12&1.06&c2\\
2016-03-15&164&93&$p(t^3)$+$p(xy^1)$&1.01&1.07&1.08&c2\\
2016-09-20&114&218&$p(fwhm_y^1)$+$p(xy^1)$&1.05&1.08&1.13&c2\\
2016-09-21&665&219&$p(xy^1)$&1.04&1.21&1.25&c2\\
2016-09-26&132&222&$p(fwhm_y^1)$+$p(xy^1)$+$p(r^1)$&1.03&1.31&1.35&c2\\
2016-09-29&135&224&$p(t^1)$+$p(xy^1)$&0.97&1.04&1&c2\\
2016-09-30&56&225&$p(t^1)$+$p(xy^1)$&0.75&1.21&0.9&c2\\
2016-10-05&141&228&$p(xy^1)$&0.87&1.04&0.91&c2\\
2016-10-07&126&229&$p(xy^1)$&0.99&1.13&1.12&c2\\
2016-10-08&127&230&$p(t^2)$+$p(xy^1)$&0.96&1.09&1.05&c2\\
2017-02-18&67&318&$p(xy^1)$&0.99&1.0&0.99&c2\\
2017-02-21&70&320&$p(xy^1)$&0.91&1.0&0.91&c2\\
2017-02-23&67&321&$p(t^1)$+$p(fwhm_y^1)$+$p(xy^1)$+$p(r^1)$&0.97&1.0&0.97&c2\\
2017-02-24&67&322&$p(t^1)$+$p(fwhm_y^1)$+$p(xy^1)$&0.86&1.15&0.93&c1\\
2017-02-27&105&324&$p(xy^1)$+$p(r^1)$&0.92&1.0&0.92&c2\\
2017-03-01&74&325&$p(fwhm_x^1)$+$p(xy^1)$&0.81&1.09&0.9&c2\\
2017-03-02&67&326&$p(fwhm_y^1)$+$p(xy^1)$&0.72&1.28&0.92&c1\\
2017-03-04&67&327&$p(fwhm_y^1)$+$p(xy^1)$+$p(r^1)$&0.93&1.0&0.93&c1\\
2017-03-05&74&328&$p(xy^1)$&1.03&1.0&1.03&c2\\
2017-03-07&67&329&$p(fwhm_y^1)$+$p(xy^1)$&0.87&1.18&1.03&c1\\
2017-03-08&67&330&$p(t^1)$+$p(fwhm_y^1)$+$p(xy^1)$&0.87&1.14&0.97&c1\\
2017-03-11&68&332&$p(xy^1)$&1.13&1.0&1.13&c2\\
2017-03-13&67&333&$p(t^2)$+$p(fwhm_x^1)$+$p(fwhm_y^1)$+$p(xy^1)$&0.84&1.15&0.96&c1\\
2017-03-14&67&334&$p(fwhm_y^1)$+$p(xy^1)$&0.79&1.39&1.11&c1\\
2017-03-16&67&335&$p(t^1)$+$p(fwhm_y^1)$+$p(xy^1)$&0.87&1.51&1.31&c1\\
2017-03-20&67&338&$p(t^1)$+$p(fwhm_y^1)$+$p(xy^1)$&0.66&1.07&0.71&c1\\
2017-03-22&67&339&$p(t^1)$+$p(fwhm_y^1)$+$p(xy^1)$&0.76&1.0&0.76&c1\\
2017-03-25&67&341&$p(t^2)$+$p(fwhm_x^2)$+$p(fwhm_y^1)$+$p(xy^1)$&0.96&1.0&0.96&c1\\
2017-03-26&67&342&$p(t^2)$+$p(fwhm_y^1)$+$p(xy^2)$&0.61&1.0&0.61&c1\\
2017-09-13&67&455&$p(t^1)$+$p(fwhm_y^1)$+$p(xy^1)$+$p(r^1)$&0.84&1.21&1.01&c1\\
2017-09-14&118&456&$p(t^1)$+$p(xy^1)$&1.14&1.24&1.42&c1\\
2017-09-16&67&457&$p(t^1)$+$p(fwhm_y^1)$+$p(xy^1)$&0.86&1.0&0.86&c1\\
2017-09-17&95&458&$p(t^4)$+$p(fwhm_y^3)$+$p(xy^1)$+$p(r^1)$&1.1&1.0&1.1&c1\\
2017-09-19&67&459&$p(t^1)$+$p(fwhm_x^1)$+$p(xy^1)$&1.06&1.0&1.06&c1\\
2017-09-21&67&460&$p(t^1)$+$p(fwhm_y^1)$+$p(xy^1)$&1.17&1.15&1.34&c1\\
2017-09-24&67&462&$p(fwhm_x^1)$+$p(fwhm_y^1)$+$p(xy^1)$+$p(r^1)$&0.9&1.17&1.05&c1\\
2017-09-28&67&465&$p(t^2)$+$p(xy^1)$&0.93&1.0&0.93&c1\\
2017-10-01&67&467&$p(t^1)$+$p(fwhm_y^1)$+$p(xy^1)$&0.84&1.33&1.12&c1\\
2017-10-03&67&468&$p(fwhm_x^1)$+$p(fwhm_y^1)$+$p(xy^1)$&0.85&1.21&1.03&c1\\
2017-10-09&67&472&$p(xy^1)$&0.88&1.59&1.4&c1\\
2017-10-10&67&473&$p(fwhm_y^1)$+$p(xy^1)$&1.01&1.42&1.44&c1\\
2017-10-13&67&475&$p(t^1)$+$p(fwhm_y^1)$+$p(xy^1)$&0.66&1.05&0.69&c1\\
2017-10-15&67&476&$p(xy^1)$+$p(r^1)$&0.87&1.0&0.87&c1\\
2017-10-16&52&477&$p(t^2)$+$p(xy^1)$+$p(r^1)$&0.9&1.0&0.9&c1\\
2017-10-18&70&478&$p(t^2)$+$p(fwhm_x^1)$+$p(fwhm_y^1)$+$p(xy^1)$&1.13&1.0&1.13&c1\\
2017-10-19&66&479&$p(t^1)$+$p(fwhm_x^1)$+$p(fwhm_y^1)$+$p(xy^1)$&0.84&1.0&0.84&c1\\
2018-03-09&129&572&$p(fwhm_y^2)$+$p(xy^1)$&0.97&1.17&1.14&c2\\
2018-03-10&96&573&$p(fwhm_x^1)$+$p(xy^1)$&0.99&1.0&0.99&c2\\
2018-03-19&80&579&$p(xy^1)$&0.94&1.22&1.16&c2\\
2018-03-25&68&583&$p(t^2)$+$p(xy^1)$&1.1&1.0&1.1&c2\\
2019-10-01&182&950&$p(t^1)$+$p(fwhm_x^1)$+$p(fwhm_y^1)$+$p(xy^1)$&1.06&1.0&1.06&c2\\
2019-10-10&150&956&$p(t^2)$+$p(fwhm_x^1)$+$p(fwhm_y^1)$+$p(xy^1)$+$p(r^1)$&0.97&1.0&0.97&c2\\
2019-10-13&180&958&$p(fwhm_x^1)$+$p(fwhm_y^1)$+$p(xy^1)$&1.05&1.0&1.05&c2\\
\hline
\textbf{TRAPPIST-1c}   \\
\hline
2016-03-04&66.0&70&$p(xy^1)$&1.01&1.0&1.01&c2\\
2016-09-19&118.0&152&$p(xy^1)$+$p(r^1)$&0.88&1.38&1.21&c2\\
2016-09-21&82.0&153&$p(xy^1)$+$p(r^1)$&0.83&1.0&0.83&c2\\
2016-09-24&108.0&154&$p(xy^1)$&1.12&1.58&1.77&c2\\
2016-09-26&111.0&155&$p(xy^1)$&1.0&1.21&1.22&c2\\
2016-10-01&134.0&157&$p(fwhm_x^2)$+$p(fwhm_y^1)$+$p(xy^1)$+$p(r^1)$&0.86&1.44&1.24&c2\\
2016-10-06&156.0&159&$p(xy^1)$&0.98&1.06&1.04&c2\\
2016-10-08&153.0&160&$p(xy^1)$&0.82&1.25&1.02&c2\\
2017-02-18&67.0&215&$p(xy^1)$&1.03&1.0&1.03&c2\\
2017-02-21&77.0&216&$p(t^2)$+$p(xy^1)$&0.84&1.0&0.84&c2\\
2017-02-23&67.0&217&$p(xy^1)$&1.05&1.26&1.32&c2\\
2017-02-26&67.0&218&$p(fwhm_x^1)$+$p(xy^1)$+$p(r^1)$&1.0&1.0&1.0&c2\\
2017-02-28&67.0&219&$p(fwhm_x^1)$+$p(xy^1)$&0.93&1.0&0.93&c2\\
2017-03-03&67.0&220&$p(t^1)$+$p(fwhm_x^1)$+$p(xy^1)$&0.95&1.0&0.95&c2\\
2017-03-05&52.0&221&$p(xy^1)$&0.95&1.02&0.97&c2\\
2017-03-07&59.0&222&$p(xy^1)$&0.9&1.0&0.9&c2\\
2017-03-10&67.0&223&$p(fwhm_x^1)$+$p(xy^1)$+$p(r^1)$&0.9&1.11&0.99&c2\\
2017-03-12&95.0&224&$p(fwhm_x^1)$+$p(xy^1)$+$p(r^1)$&0.99&1.02&1.01&c2\\
2017-03-15&95.0&225&$p(xy^1)$+$p(r^1)$&0.92&1.0&0.92&c2\\
2017-03-20&67.0&227&$p(t^1)$+$p(fwhm_x^1)$+$p(xy^1)$&0.94&1.0&0.94&c2\\
2017-03-22&67.0&228&$p(fwhm_y^1)$+$p(xy^1)$&1.01&1.11&1.12&c2\\
2017-03-24&110.0&229&$p(t^2)$+$p(xy^1)$&0.99&1.0&0.99&c2\\
2017-03-27&67.0&230&$p(fwhm_x^1)$+$p(fwhm_y^1)$+$p(xy^1)$&0.95&1.47&1.4&c2\\
2017-09-15&94.0&301&$p(fwhm_x^1)$+$p(xy^1)$+$p(r^1)$&1.04&1.54&1.6&c2\\
2017-09-17&113.0&302&$p(t^2)$+$p(xy^1)$&0.95&1.41&1.34&c1\\
2017-09-24&105.0&305&$p(xy^1)$&1.04&1.0&1.04&c2\\
2017-09-27&67.0&306&$p(fwhm_y^1)$+$p(xy^1)$&0.91&1.21&1.11&c1\\
2017-10-07&108.0&310&$p(t^2)$+$p(xy^1)$&1.07&1.0&1.07&c2\\
2017-10-11&100.0&312&$p(fwhm_y^1)$+$p(xy^1)$+$p(r^1)$&0.82&1.56&1.38&c1\\
2017-10-14&74.0&313&$p(t^3)$+$p(xy^1)$+$p(r^1)$&0.91&1.46&1.33&c2\\
2017-10-16&59.0&314&$p(fwhm_x^1)$+$p(fwhm_y^1)$+$p(r^1)$&1.09&1.02&1.11&c1\\
2017-10-19&60&315&$p(t^2)$+$p(fwhm_y^2)$+$p(xy^1)$+$p(r^1)$&0.74&1.07&0.8&c1\\
2018-03-13&65.0&375&$p(t^1)$+$p(fwhm_y^1)$+$p(xy^1)$+$p(r^1)$&0.9&1.03&0.93&c1\\
2018-03-25&83.0&380&$p(fwhm_y^1)$+$p(xy^1)$&1.07&1.07&1.15&c2\\
2018-03-28&98.0&381&$p(t^1)$+$p(fwhm_y^1)$+$p(xy^1)$&1.11&1.27&1.4&c1\\
2019-10-01&182&609&$p(fwhm_x^1)$+$p(fwhm_y^1)$+$p(xy^1)$&1.06&1.0&1.06&c2\\
2019-10-08&190&612&$p(t^3)$+$p(fwhm_x^1)$+$p(r^1)$&0.95&1.44&1.37&c2\\
2019-10-13&180&614&$p(fwhm_x^1)$+$p(fwhm_y^1)$+$p(xy^1)$&1.05&1.0&1.05&c2\\
2019-10-20&188&617&$p(t^2)$+$p(fwhm_x^1)$+$p(fwhm_y^1)$+$p(xy^1)$+$p(r^1)$&0.97&1.0&0.97&c2\\
\hline
\textbf{TRAPPIST-1d}  \\
\hline
2016-09-22&134&-4&$p(fwhm_x^1)$+$p(fwhm_y^1)$+$p(xy^1)$&1.05&1.34&1.4&c2\\
2016-09-26&114&-3&$p(fwhm_y^1)$+$p(xy^1)$&1.1&1.1&1.21&c2\\
2016-09-30&154&-2&$p(xy^1)$&1.08&1.59&1.71&c2\\
2016-10-04&145&-1&$p(fwhm_x^2)$+$p(xy^1)$&0.76&1.36&1.04&c2\\
2016-10-08&133&0&$p(xy^1)$&0.86&1.56&1.35&c2\\
2017-02-19&122&33&$p(xy^1)$+$p(r^1)$&0.94&1.06&0.97&c2\\
2017-02-23&122&34&$p(xy^1)$+$p(r^1)$&0.94&1.0&0.94&c2\\
2017-02-27&134&35&$p(fwhm_x^1)$+$p(fwhm_y^1)$+$p(xy^1)$+$p(r^1)$&0.96&1.04&1&c2\\
2017-03-03&122&36&$p(xy^1)$&0.89&1.0&0 ,89&c2\\
2017-03-07&121&37&$p(xy^1)$&1.01&1.0&1.01&c2\\
2017-03-11&120&38&$p(xy^1)$+$p(r^1)$&1.09&1.0&1.09&c2\\
2017-03-15&142&39&$p(t^2)$+$p(fwhm_y^1)$+$p(xy^1)$&1.01&1.0&1.01&c2\\
2017-03-19&122&40&$p(fwhm_x^1)$+$p(xy^1)$&0.99&1.0&0.99&c2\\
2017-09-17&120&85&$p(t^3)$+$p(xy^2)$+$p(r^1)$&1.03&1.4&1.44&c1\\
2017-09-25&122&87&$p(t^1)$+$p(fwhm_y^1)$+$p(xy^1)$&0.76&1.76&1.33&c1\\
2017-10-08&122&90&$p(t^3)$+$p(fwhm_y^2)$+$p(xy^2)$&0.75&1.32&0.98&c1\\
2017-10-16&122&92&$p(t^2)$+$p(xy^2)$&0.99&1.69&1.67&c1\\
2017-10-20&129&93&$p(xy^1)$+$p(r^1)$&0.98&1.0&0.98&c2\\
2018-03-06&183&127&$p(fwhm_x^1)$+$p(xy^1)$+$p(r^2)$&1.04&1.21&1.25&c2\\
2018-03-31&106&133&$p(xy^1)$&1.01&1.33&1.33&c2\\
\hline
\textbf{TRAPPIST-1e}  \\
\hline
2016-09-22&97&-1&$p(xy^1)$+$p(fwhm_x^1)$+$p(fwhm_y^1)$+$p(r^1)$&1.00&1.12&1.12&c2\\
2016-09-28&154&0&$p(t^1)$+$p(xy^1)$&0.99&1.0&0.99&c2\\
2017-02-22&106&24&$p(fwhm_x^1)$+$p(xy^1)$&1.03&1.24&1.28&c2\\
2017-02-28&99&25&$p(xy^1)$+$p(r^1)$&0.85&1.02&0.87&c2\\
2017-03-06&141&26&$p(fwhm_x^1)$+$p(xy^1)$&1.06&1.09&1.15&c2\\
2017-03-12&124&27&$p(t^3)$&1.05&1.0&1.05&c2\\
2017-03-18&134&28&$p(t^1)$+$p(xy^1)$+$p(r^1)$&0.88&1.01&0.89&c2\\
2017-03-24&127&29&$p(xy^1)$&0.99&1.0&0.99&c2\\
2017-09-17&88&58&$p(t^2)$+$p(xy^1)$&1.01&1.02&1.03&c1\\
2017-09-23&122&59&$p(fwhm_x^1)$+$p(fwhm_y^1)$+$p(xy^1)$+$p(r^1)$&0.9&1.0&0.9&c1\\
2017-10-12&113&62&$p(xy^1)$+$p(r^2)$&0.93&1.07&1.01&c1\\
2018-03-07&122&86&$p(fwhm_y^1)$+$p(xy^1)$+$p(r^2)$&0.99&1.53&1.5&c1\\
2018-03-13&71&87&$p(fwhm_y^1)$+$p(xy^1)$&0.89&1.36&1.22&c1\\
2018-03-19&161&88&$p(fwhm_x^1)$+$p(xy^1)$&1.01&1.1&1.12&c2\\
2018-03-25&113&89&$p(xy^1)$&0.97&1.03&1.0&c2\\
2019-10-01&122&180&$p(t^1)$+$p(fwhm_x^1)$+$p(fwhm_y^1)$+$p(xy^1)$&1.04&1.42&1.48&c2\\
2019-10-13&122&182&$p(t^1)$+$p(fwhm_x^2)$+$p(fwhm_y^2)$&1.06&1.09&1.16&c2\\
\hline
\textbf{TRAPPIST-1f}   \\
\hline
2016-09-30&170&-1&$p(xy^1)$&0.93&1.66&1.54&c2\\
2016-10-09&200&0&$p(fwhm_x^1)$+$p(fwhm_y^1)$+$p(xy^1)$&0.87&1.66&1.45&c2\\
2017-02-24&150&15&$p(t^1)$+$p(fwhm_x^1)$+$p(xy^1)$&1.04&1.74&1.81&c2\\
2017-03-06&124&16&$p(fwhm_x^1)$+$p(xy^1)$+$p(r^1)$&0.96&1.06&1.02&c2\\
2017-03-15&173&17&$p(t^1)$+$p(fwhm_y^1)$+$p(xy^1)$+$p(r^1)$&1.04&1.01&1.05&c2\\
2017-03-24&138&18&$p(fwhm_x^1)$+$p(xy^1)$+$p(r^1)$&0.96&1.48&1.42&c2\\
2017-09-24&106&38&$p(fwhm_x^1)$+$p(xy^1)$+$p(r^1)$&0.88&1.03&0.91&c2\\
2018-03-09&160&56&$p(fwhm_x^1)$+$p(xy^1)$&1.02&1.03&1.06&c2\\
2018-03-18&150&57&$p(fwhm_x^1)$+$p(xy^1)$+$p(r^2)$&0.85&1.23&1.04&c1\\
2018-03-27&148&58&$p(t^2)$+$p(fwhm_y^1)$+$p(xy^1)$&1.15&1.31&1.5&c1\\
2019-10-01&182&118&$p(fwhm_x^1)$+$p(fwhm_y^1)$+$p(xy^1)$&1.06&1.0&1.06&c2\\
2019-10-10&150&119&$p(t^2)$+$p(fwhm_x^1)$+$p(fwhm_y^1)$+$p(xy^1)$+$p(r^1)$&0.97&1.0&0.97&c2\\
2019-10-28&150&121&$p(t^1)$+$p(fwhm_x^1)$+$p(fwhm_y^1)$+$p(r^1)$&0.88&1.01&0.89&c2\\
\hline
\textbf{TRAPPIST-1g}  \\
\hline
2016-10-03&147&30&$p(fwhm_x^1)$+$p(fwhm_y^1)$+$p(xy^1)$&0.76&1.64&1.24&c2\\
2017-03-01&86&12&$p(fwhm_x^1)$+$p(fwhm_y^1)$+$p(xy^1)$&0.93&1.0&0.93&c2\\
2017-03-13&150&13&$p(t^2)$+$p(fwhm_y^1)$+$p(xy^1)$&0.98&1.0&0.98&c2\\
2017-03-25&150&14&$p(fwhm_x^1)$+$p(xy^2)$&1.08&1.0&1.08&c2\\
2017-09-14&158&28&$p(fwhm_y^2)$+$p(xy^2)$+$p(r^1)$&0.94&1.34&1.26&c1\\
2018-03-06&156&42&$p(t^2)$+$p(xy^1)$&1.09&1.01&1.1&c2\\
2018-03-31&147&44&$p(fwhm_x^1)$+$p(xy^1)$+$p(r^1)$&0.97&1.0&0.97&c2\\
2019-10-08&190&89&$p(t^3)$+$p(fwhm_x^1)$+$p(r^1)$&0.95&1.44&1.37&c2\\
2019-10-20&188&90&$p(t^2)$+$p(fwhm_x^1)$+$p(fwhm_y^1)$+$p(xy^1)$+$p(r^1)$&0.97&1.0&0.97&c2\\
\hline
\textbf{TRAPPIST-1h} \\
\hline
2016-10-01&174&0&$p(fwhm_x^1)$+$p(fwhm_y^1)$+$p(xy^2)$&0.93&1.7&1.59&c2\\
2017-03-18&139&9&$p(fwhm_x^1)$+$p(xy^1)$&1.02&1.15&1.18&c2\\
2017-09-22&136&19&$p(xy^1)$&0.84&1.38&1.15&c1\\
2017-10-11&156&20&$p(t^1)$+$p(fwhm_x^1)$+$p(xy^1)$&0.98&1.02&1.01&c2\\
2018-03-10&132&28&$p(xy^1)$+$p(r^1)$&0.88&1.35&1.18&c2\\
2018-03-29&150&29&$p(fwhm_x^1)$+$p(fwhm_y^1)$+$p(xy^1)$+$p(r^1)$&0.95&1.31&1.24\&c2\\
2019-10-13&180&59&$p(fwhm_x^1)$+$p(fwhm_y^1)$+$p(xy^1)$&1.05&1.0&1.05&c2\\
\hline

\end{longtable}

\renewcommand{\arraystretch}{1.2}
\setlength{\tabcolsep}{14pt} % Default value: 6pt
\begin{longtable}{|c|cc|cc|c|}
\caption{\label{Table_indiv} Transit Timings and Depths Obtained from Individual Analyses of each transit. Blended and partial transits are presented in a separate table \ref{Table_indiv_blended}} \\
\hline
\multicolumn{1}{|c|}{\textbf{Epoch}} & \multicolumn{2}{c|}{\thead{\textbf{Transit timing + 1$\sigma$ error } \\\textbf{ [$\mathrm{BJD}_{\mathrm{TDB}}-2450000$]}}} & \multicolumn{2}{c|}{\thead{\textbf{Transit depth} \\\textbf{+ 1$\sigma$ error (\%)}}}&
\multicolumn{1}{|c|}{\textbf{Channel}}\\
\hline \hline
\endfirsthead
\caption{continued.}\\
\hline\hline
\multicolumn{1}{|c|}{\textbf{Epoch}} & \multicolumn{2}{c|}{\thead{\textbf{Transit timing + 1$\sigma$ error} \\\textbf{ [$\mathrm{BJD}_{\mathrm{TDB}}-2450000$]}}} & \multicolumn{2}{c|}{\thead{\textbf{Transit depth} \\\textbf{+ 1$\sigma$ error (\%)}}} &
\multicolumn{1}{|c|}{\textbf{Channel}}\\
\hline \hline
\endhead
\hline
\endfoot
\textbf{TRAPPIST-1b} &&&&\\
\hline
78&7440.36499&0.00019&0.75&0.031&c2\\
86&7452.45225&0.00017&0.759&0.028&c2\\
93&7463.02846&0.00019&0.684&0.025&c2\\
218&7651.88731&0.0002&0.759&0.035&c2\\
219&7653.39799&0.00027&0.696&0.034&c2\\
222&7657.93126&0.00023&0.728&0.034&c2\\
224&7660.95216&0.00017&0.689&0.031&c2\\
225&7662.46363&0.00027&0.717&0.038&c2\\
228&7666.99561&0.00013&0.702&0.022&c2\\
229&7668.50665&0.00018&0.725&0.027&c2\\
230&7670.01776&0.00018&0.727&0.027&c2\\
318&7802.9756&0.00015&0.753&0.025&c2\\
320&7805.99698&0.00014&0.705&0.024&c2\\
321&7807.50727&0.00017&0.714&0.029&c2\\
322&7809.01832&0.0002&0.749&0.023&c1\\
324&7812.04041&0.00019&0.695&0.028&c2\\
325&7813.55125&0.00014&0.713&0.025&c2\\
326&7815.06275&0.00019&0.726&0.025&c1\\
327&7816.57335&0.00012&0.666&0.021&c1\\
328&7818.08384&0.00015&0.715&0.026&c2\\
329&7819.59477&0.00018&0.705&0.029&c1\\
330&7821.10556&0.00015&0.72&0.026&c1\\
332&7824.12734&0.00016&0.734&0.028&c2\\
333&7825.63815&0.00014&0.731&0.027&c1\\
334&7828.66083&0.00017&0.74&0.031&c1\\
335&7828.66036&0.00023&0.728&0.03&c1\\
338&7833.19286&0.00017&0.658&0.024&c1\\
339&7834.70398&0.00014&0.694&0.018&c1\\
341&7837.72528&0.00014&0.741&0.028&c1\\
342&7839.23684&0.00022&0.792&0.033&c1\\
455&8009.96629&0.00024&0.701&0.024&c1\\
456&8011.47742&0.00024&0.695&0.031&c1\\
457&8012.98805&0.00012&0.707&0.02&c1\\
458&8014.49882&0.00017&0.68&0.028&c1\\
459&8016.0104&0.00014&0.782&0.023&c1\\
460&8017.52126&0.00023&0.712&0.029&c1\\
462&8020.54237&0.00013&0.739&0.021&c1\\
465&8025.0754&0.0002&0.711&0.027&c1\\
467&8028.09738&0.0002&0.686&0.028&c1\\
468&8029.60816&0.00016&0.703&0.029&c1\\
472&8035.65155&0.00023&0.757&0.038&c1\\
473&8037.16251&0.00023&0.71&0.03&c1\\
475&8040.18429&0.00015&0.727&0.021&c1\\
476&8041.69509&0.00014&0.709&0.021&c1\\
477&8043.20589&0.00012&0.708&0.028&c1\\
478&8044.71651&0.00015&0.754&0.027&c1\\
479&8046.22749&0.00016&0.736&0.025&c1\\
572&8186.74006&0.00018&0.782&0.028&c2\\
573&8188.25135&0.00016&0.738&0.027&c2\\
579&8197.31644&0.00021&0.699&0.03&c2\\
583&8203.35999&0.00017&0.701&0.033&c2\\
950&8757.85509&0.0002&0.784&0.028&c2\\
956&8766.92066&0.00018&0.656&0.027&c2\\
958&8769.94191&0.00016&0.786&0.026&c2\\
\hline
\textbf{TRAPPIST-1c}&&&& \\
\hline
70&7452.33467&0.00015&0.714&0.028&c2\\
152&7650.92394&0.00024&0.698&0.029&c2\\
153&7653.34548&0.00017&0.69&0.021&c2\\
154&7655.768&0.00038&0.676&0.046&c2\\
155&7658.18964&0.00022&0.685&0.03&c2\\
157&7663.03331&0.00038&0.719&0.041&c2\\
159&7667.87729&0.00017&0.69&0.023&c2\\
160&7670.29871&0.00019&0.733&0.022&c2\\
215&7803.49754&0.00017&0.675&0.025&c2\\
216&7805.91881&0.00015&0.642&0.026&c2\\
217&7808.3412&0.00026&0.689&0.03&c2\\
218&7810.76281&0.00019&0.668&0.027&c2\\
219&7813.18456&0.00024&0.669&0.024&c2\\
220&7815.60585&0.00017&0.72&0.025&c2\\
221&7818.02833&0.00018&0.75&0.028&c2\\
222&7820.45018&0.00018&0.688&0.023&c2\\
223&7822.87186&0.00021&0.757&0.028&c2\\
224&7825.29382&0.0002&0.694&0.023&c2\\
225&7827.71521&0.00015&0.73&0.021&c2\\
227&7832.55892&0.00014&0.73&0.026&c2\\
228&7834.98115&0.00023&0.689&0.028&c2\\
229&7837.40276&0.00017&0.713&0.024&c2\\
230&7839.8241&0.00025&0.686&0.043&c2\\
301&8011.7715&0.00036&0.681&0.044&c2\\
302&8014.19267&0.0002&0.735&0.027&c1\\
305&8021.45847&0.00017&0.75&0.025&c2\\
306&8023.87959&0.00021&0.715&0.028&c1\\
310&8033.56753&0.00017&0.738&0.026&c2\\
312&8038.41064&0.00028&0.712&0.035&c1\\
313&8040.83258&0.00032&0.779&0.052&c2\\
314&8043.25402&0.00017&0.739&0.024&c1\\
315&8045.67653&0.00017&0.762&0.027&c1\\
375&8190.98264&0.00022&0.675&0.023&c1\\
380&8203.09199&0.0002&0.698&0.028&c2\\
381&8205.51293&0.00021&0.748&0.027&c1\\
609&8757.6834&0.00019&0.696&0.024&c2\\
612&8764.94945&0.00024&0.719&0.032&c2\\
614&8769.79254&0.00018&0.619&0.03&c2\\
617&8777.0583&0.00021&0.699&0.024&c2\\
\hline
\textbf{TRAPPIST-1d} &&&&\\
\hline
-4&7653.94267&0.00036&0.437&0.031&c2\\
-3&7657.99196&0.00054&0.324&0.025&c2\\
-2&7662.04263&0.00063&0.397&0.037&c2\\
-1&7666.09187&0.00048&0.35&0.03&c2\\
0&7670.14194&0.00039&0.359&0.029&c2\\
33&7803.79084&0.00046&0.367&0.019&c2\\
34&7807.8403&0.0003&0.385&0.02&c2\\
35&7811.89102&0.00039&0.388&0.021&c2\\
36&7815.94061&0.00029&0.349&0.018&c2\\
37&7819.99047&0.00054&0.313&0.02&c2\\
38&7824.04153&0.00079&0.395&0.023&c2\\
39&7828.0908&0.00034&0.375&0.028&c2\\
40&7832.14042&0.00028&0.334&0.023&c2\\
85&8014.37932&0.00095&0.329&0.031&c1\\
87&8022.48021&0.00038&0.364&0.028&c1\\
90&8022.47826&0.00033&0.354&0.025&c1\\
92&8042.72676&0.00047&0.362&0.032&c1\\
93&8046.77637&0.00028&0.376&0.023&c2\\
127&8184.45805&0.00043&0.386&0.027&c2\\
133&8208.75644&0.0005&0.333&0.029&c2\\
\hline
\textbf{TRAPPIST-1e} &&&&\\
\hline
-1&7654.27853&0.00042&0.573&0.043&c2\\
0&7660.3803&0.00026&0.507&0.018&c2\\
24&7806.75764&0.00047&0.46&0.03&c2\\
25&7812.85751&0.00032&0.447&0.018&c2\\
26&7818.95509&0.0003&0.478&0.022&c2\\
27&7825.05304&0.00035&0.439&0.025&c2\\
28&7831.15206&0.00025&0.521&0.019&c2\\
29&7837.2497&0.00027&0.503&0.019&c2\\
58&8014.13087&0.0002&0.509&0.021&c2\\
59&8020.23323&0.00023&0.485&0.019&c1\\
62&8038.5351&0.00032&0.518&0.021&c1\\
86&8184.94895&0.00036&0.415&0.028&c1\\
87&8191.04813&0.00051&0.475&0.033&c1\\
88&8197.14651&0.00034&0.52&0.022&c1\\
89&8203.24763&0.00024&0.501&0.021&c2\\
180&8758.28125&0.00053&0.498&0.034&c2\\
182&8770.47845&0.00036&0.486&0.026&c2\\
\hline
\textbf{TRAPPIST-1f}&&&& \\
\hline
-1&7662.18743&0.42&0.605&0.03&c2\\
0&7671.39266&0.00045&0.622&0.046&c2\\
15&7809.47541&0.0004&0.656&0.037&c2\\
16&7818.68262&0.00028&0.633&0.021&c2\\
17&7827.88676&0.00024&0.604&0.02&c2\\
18&7837.10322&0.00049&0.577&0.03&c2\\
38&8021.25068&0.00021&0.623&0.019&c2\\
56&8186.91882&0.00026&0.623&0.022&c2\\
57&8196.12561&0.00024&0.631&0.019&c2\\
58&8205.32761&0.00027&0.668&0.029&c1\\
118&8757.76211&0.00056&0.662&0.026&c1\\
119&8766.96813&0.00024&0.626&0.025&c2\\
121&8785.38901&0.00022&0.671&0.018&c2\\
\hline
\textbf{TRAPPIST-1g}&&&& \\
\hline
0&7665.35136&0.00048&0.602&0.036&c2\\
12&7813.6068&0.00026&0.776&0.024&c2\\
13&7825.96112&0.0002&0.8&0.02&c2\\
14&7838.30652&0.00026&0.706&0.024&c2\\
28&8011.24018&0.0003&0.705&0.029&c2\\
42&8184.21905&0.00023&0.735&0.023&c1\\
44&8208.93037&0.0002&0.716&0.019&c2\\
89&8764.82751&0.00032&0.713&0.031&c2\\
90&8777.17395&0.00026&0.75&0.021&c2\\
\hline
\textbf{TRAPPIST-1h}&&&& \\
\hline
0&7662.55449&0.0012&0.309&0.044&c2\\
9&7831.46614&0.0006&0.342&0.02&c2\\
19&8019.16844&0.0006&0.31&0.02&c2\\
20&8037.93276&0.00051&0.377&0.019&c1\\
28&8188.05067&0.00052&0.361&0.025&c2\\
29&8206.81914&0.00071&0.334&0.023&c2\\
59&8769.83809&0.00054&0.334&0.024&c2\\
\hline
\end{longtable}

\renewcommand{\arraystretch}{1.2}
\setlength{\tabcolsep}{14pt} % Default value: 6pt
\begin{longtable}{|c|cc|cc|c|}
\caption{\label{Table_indiv_blended} Transit timings and depths Obtained from Individual Analyses of each blended or partial transit} \\
\hline
\multicolumn{1}{|c|}{\textbf{Epoch}} & \multicolumn{2}{c|}{\thead{\textbf{Transit timing + 1$\sigma$ error} \\\textbf{ [$\mathrm{BJD}_{\mathrm{TDB}}-2450000$]}}} & \multicolumn{2}{c|}{\thead{\textbf{Transit depth} \\\textbf{+ 1$\sigma$ error (\%)}}} &
\multicolumn{1}{|c|}{\textbf{Channel}}\\
\hline \hline
\endfirsthead
\caption{continued.}\\
\hline\hline
\multicolumn{1}{|c|}{\textbf{Epoch}} & \multicolumn{2}{c|}{\thead{\textbf{Transit timing + 1$\sigma$ error} \\\textbf{ [$\mathrm{BJD}_{\mathrm{TDB}}-2450000$]}}} & \multicolumn{2}{c|}{\thead{\textbf{Transit depth} \\\textbf{+ 1$\sigma$ error (\%)}}} &
\multicolumn{1}{|c|}{\textbf{Channel}}\\
\hline \hline
\endhead
\hline
\endfoot
\textbf{TRAPPIST-1b} &&&&\\
\hline
226&7663.97530&0.00120&0.642&8.300&c2\\
227&7665.48546&0.00030&0.761&0.036&c2\\
231&7671.52791&0.00068&0.696&0.046&c2\\
336&7830.17083&0.00020&0.729&0.035&c2\\
340&7836.21439&0.00018&0.703&0.026&c2\\
461&8019.03167&0.00027&0.662&0.067&c1\\
464&8023.56458&0.00015&0.847&0.028&c1\\
469&8031.11892&0.00012&0.796&1.100&c1\\
474&8038.67292&0.00017&0.752&0.033&c1\\
566&8177.67496&0.00027&0.707&0.027&c2\\
\hline
\textbf{TRAPPIST-1c} &&&&\\
\hline
71&7454.75685&0.00058&0.680&0.030&c2\\
156&7660.611680&0.00051&0.698&0.036&c2\\
158&7665.45539&0.00032&0.662&0.037&c2\\
226&7830.13725&0.00024&0.733&0.034&c2\\
304&8019.03635&0.00027&0.744&0.063&c1\\
309&8031.14517&0.00015&0.755&0.024&c1\\
311&8035.98910&0.00017&0.688&0.023&c1\\
370&8178.87407&0.00015&0.729&0.020&c1\\
\hline
\textbf{TRAPPIST-1d} &&&&\\
\hline
41&7836.19171&0.00041&0.344&0.023&c2\\
91&8038.67921&0.00033&0.330&0.030&c1\\
130&8196.60651&0.00065&0.413&0.030&c2\\
\hline
\textbf{TRAPPIST-1e} &&&&\\
\hline
85&8178.84731&0.00019&0.536&0.017&c1\\
\hline
\textbf{TRAPPIST-1f} &&&&\\
\hline
-2&7652.98592&0.00035&0.743&0.050&c2\\
0&7671.39268&0.00041&0.621&0.043&c2\\
55&8177.71567&0.00026&0.647&0.026&c2\\
\hline
\textbf{TRAPPIST-1g} &&&&\\
\hline
-1&7652.99505&0.00037&0.734132&0.051&c2\\
29&8023.59087&0.00023&0.778&0.021&c1\\
30&8035.94551&0.00025&0.729&0.020&c1\\
43&8196.57292&0.00031&0.750&0.026&c2\\
\end{longtable}

\renewcommand{\arraystretch}{1.2}
\setlength{\tabcolsep}{14pt} % Default value: 6pt
\begin{longtable}{|c|cc|cc|}
\caption{\label{global_3-6} Transit Timings and Depths Obtained from Global Analyses of each transit with ddf variations allowed for $3.6\mu \mathrm{m}$ channel} \\
\hline
\multicolumn{1}{|c|}{\textbf{Epoch}} & \multicolumn{2}{c|}{\thead{\textbf{Transit timing + 1$\sigma$ error} \\\textbf{ [$\mathrm{BJD}_{\mathrm{TDB}}-2450000$]}}} & \multicolumn{2}{c|}{\thead{\textbf{Transit depth} \\\textbf{+ 1$\sigma$ error (\%)}}} \\
\hline \hline
\endfirsthead
\caption{continued.}\\
\hline\hline
\multicolumn{1}{|c|}{\textbf{Epoch}} & \multicolumn{2}{c|}{\thead{\textbf{Transit timing + 1$\sigma$ error} \\\textbf{ [$\mathrm{BJD}_{\mathrm{TDB}}-2450000$]}}} & \multicolumn{2}{c|}{\thead{\textbf{Transit depth} \\\textbf{+ 1$\sigma$ error (\%)}}} \\
\hline \hline
\endhead
\hline
\endfoot
\textbf{TRAPPIST-1b} &&&& \\
\hline
322&7809.01833&0.00022&0.730&0.025\\
326&7815.06277&0.00020&0.729&0.026\\
327&7816.57334&0.00012&0.658&0.021\\
329&7819.59475&0.00015&0.704&0.025\\
330&7821.10556&0.00016&0.719&0.026\\
333&7825.63814&0.00012&0.729&0.027\\
334&7827.14996&0.00014&0.723&0.027\\
335&7828.66039&0.00019&0.743&0.022\\
338&7833.19283&0.00021&0.657&0.026\\
339&7834.70397&0.00016&0.699&0.019\\
341&7837.72530&0.00018&0.735&0.032\\
342&7839.23688&0.00020&0.784&0.027\\
455&8009.96628&0.00023&0.724&0.023\\
456&8011.47739&0.00021&0.698&0.029\\
457&8012.98803&0.00013&0.706&0.023\\
458&8014.49878&0.00017&0.692&0.03\\
459&8016.01031&0.00014&0.761&0.024\\
460&8017.52126&0.00020&0.711&0.027\\
462&8020.54236&0.00014&0.739&0.022\\
465&8025.07537&0.00020&0.705&0.027\\
467&8028.09740&0.00023&0.679&0.027\\
468&8029.60819&0.00016&0.702&0.027\\
472&8035.65151&0.00025&0.759&0.039\\
473&8037.16249&0.00028&0.709&0.027\\
475&8040.18409&0.00018&0.740&0.027\\
476&8041.6951&00.00013&0.715&0.022\\
477&8043.20589&0.00016&0.762&0.027\\
478&8044.71646&0.00024&0.754&0.04\\
479&8046.22749&0.00013&0.735&0.021\\
\hline
\textbf{TRAPPIST-1c} &&&&\\
\hline
302&8014.19266&0.00021&0.734&0.024\\
306&8023.87966&0.00020&0.701&0.029\\
312&8038.41062&0.00024&0.707&0.034\\
314&8043.25404&0.00021&0.737&0.025\\
315&8045.67667&0.00037&0.762&0.049\\
375&8190.98265&0.00021&0.674&0.021\\
381&8205.51292&0.00023&0.731&0.027\\
\hline
\textbf{TRAPPIST-1d}&&&& \\
\hline
85&8014.37913&0.00040&0.333&0.0190\\
87&8022.48019&0.00031&0.362&0.0200\\
90&8034.62830&0.00031&0.323&0.0230\\
92&8042.72672&0.00033&0.353&0.0180\\
\hline
\textbf{TRAPPIST-1e} &&&&\\
\hline
58&8014.13087&0.00024&0.513&0.022\\
59&8020.23322&0.00024&0.463&0.016\\
62&8038.53515&0.00032&0.513&0.020\\
86&8184.94890&0.00032&0.439&0.025\\
87&8191.04817&0.00052&0.507&0.026\\
\hline
\textbf{TRAPPIST-1f} &&&&\\
\hline
57&8196.12562&0.00025&0.636&0.019\\
58&8205.32761&0.00029&0.665&0.030\\
\hline
\textbf{TRAPPIST-1g}&&&& \\
28&8011.24018&0.00034&0.710&0.033\\
\hline
\textbf{TRAPPIST-1h}&&&& \\
\hline
19&8019.16846&0.00064&0.312&0.021\\
\hline
\end{longtable}

\renewcommand{\arraystretch}{1.2}
\setlength{\tabcolsep}{14pt} % Default value: 6pt
\begin{longtable}{|c|cc|cc|}
\caption{\label{global_4-5} Transit Timings and Depths Obtained from Global Analyses of each transit with ddf variations allowed for $4.5\mu \mathrm{m}$ channel} \\
\hline
\multicolumn{1}{|c|}{\textbf{Epoch}} & \multicolumn{2}{c|}{\thead{\textbf{Transit timing + 1$\sigma$ error} \\\textbf{ [$\mathrm{BJD}_{\mathrm{TDB}}-2450000$]}}} & \multicolumn{2}{c|}{\thead{\textbf{Transit depth} \\\textbf{+ 1$\sigma$ error (\%)}}} \\
\hline \hline
\endfirsthead
\caption{continued.}\\
\hline\hline
\multicolumn{1}{|c|}{\textbf{Epoch}} & \multicolumn{2}{c|}{\thead{\textbf{Transit timing + 1$\sigma$ error} \\\textbf{ [$\mathrm{BJD}_{\mathrm{TDB}}-2450000$]}}} & \multicolumn{2}{c|}{\thead{\textbf{Transit depth} \\\textbf{+ 1$\sigma$ error (\%)}}} \\
\hline \hline
\endhead
\hline
\endfoot
\textbf{TRAPPIST-1b} &&&&\\
\hline
78&7440.36517&0.00036&0.746&0.048\\
86&7452.45225&0.00017&0.751&0.027\\
93&7463.02843&0.00024&0.689&0.034\\
218&7651.88734&0.00022&0.755&0.036\\
219&7653.39799&0.00028&0.692&0.032\\
222&7657.93138&0.00022&0.736&0.034\\
224&7660.95209&0.00024&0.694&0.032\\
225&7662.46362&0.00036&0.726&0.046\\
228&7666.99560&0.00014&0.703&0.021\\
229&7668.50662&0.00018&0.726&0.027\\
230&7670.01772&0.00019&0.732&0.027\\
318&7802.97561&0.00015&0.749&0.025\\
320&7805.99698&0.00014&0.707&0.024\\
321&7807.50727&0.00020&0.708&0.031\\
324&7812.04032&0.00016&0.702&0.022\\
325&7813.55123&0.00013&0.710&0.025\\
328&7818.08384&0.00016&0.718&0.027\\
332&7824.12735&0.00018&0.734&0.032\\
572&8186.74003&0.00018&0.783&0.027\\
573&8188.25135&0.00015&0.741&0.027\\
579&8197.31644&0.00023&0.693&0.027\\
583&8203.36001&0.00018&0.697&0.033\\
950&8757.85493&0.00024&0.788&0.038\\
956&8766.92069&0.00020&0.778&0.039\\
958&8769.94187&0.00022&0.688&0.030\\
\hline
\textbf{TRAPPIST-1c} &&&&\\
\hline
70&7452.33466&0.00014&0.711&0.027\\
152&7650.92393&0.00027&0.699&0.033\\
153&7653.34547&0.00022&0.690&0.030\\
154&7655.76801&0.00051&0.673&0.047\\
155&7658.18964&0.00023&0.679&0.029\\
157&7663.03333&0.00040&0.709&0.039\\
159&7667.87731&0.00018&0.687&0.021\\
160&7670.29871&0.00019&0.727&0.023\\
215&7803.49753&0.00018&0.664&0.025\\
216&7805.91881&0.00017&0.636&0.030\\
217&7808.34117&0.00028&0.681&0.030\\
218&7810.76273&0.00019&0.673&0.031\\
219&7813.18463&0.00037&0.689&0.033\\
220&7815.60587&0.00019&0.721&0.029\\
221&7818.02836&0.00029&0.745&0.039\\
222&7820.45018&0.00018&0.681&0.022\\
223&7822.87187&0.00026&0.753&0.030\\
224&7825.29385&0.00035&0.707&0.039\\
225&7827.71522&0.00016&0.726&0.021\\
227&7832.55893&0.00019&0.733&0.036\\
228&7834.98112&0.00024&0.687&0.028\\
229&7837.40275&0.00019&0.704&0.025\\
230&7839.82416&0.00029&0.683&0.047\\
301&8011.77148&0.00032&0.668&0.040\\
305&8021.45848&0.00017&0.750&0.025\\
310&8033.56754&0.00018&0.732&0.029\\
313&8040.83258&0.00035&0.653&0.046\\
380&8203.09196&0.00021&0.696&0.029\\
609&8757.68343&0.00021&0.690&0.028\\
612&8764.94945&0.00023&0.615&0.042\\
614&8769.79242&0.00028&0.641&0.041\\
617&8777.05833&0.00022&0.707&0.028\\
\hline
\textbf{TRAPPIST-1d} &&&&\\
\hline
-4&7653.94271&0.00032&0.432&0.022\\
-3&7657.99205&0.00049&0.327&0.020\\
-2&7662.04269&0.00028&0.412&0.021\\
-1&7666.09182&0.00057&0.377&0.034\\
0&7670.14197&0.0003&0.357&0.021\\
33&7803.79079&0.00047&0.367&0.018\\
34&7807.84031&0.00033&0.385&0.020\\
35&7811.89086&0.00037&0.391&0.020\\
36&7815.94057&0.00030&0.352&0.020\\
37&7819.99084&0.00065&0.313&0.020\\
38&7824.04150&0.00038&0.383&0.022\\
39&7828.09075&0.00037&0.377&0.027\\
40&7832.14033&0.00032&0.330&0.022\\
93&8046.77628&0.00026&0.368&0.023\\
127&8184.45806&0.00034&0.388&0.021\\
133&8208.75641&0.00033&0.329&0.021\\
\hline
\textbf{TRAPPIST-1e} &&&&\\
\hline
-1&7654.27828&0.00049&0.567&0.044\\
0&7660.3803&0.00027&0.504&0.018\\
24&7806.75787&0.00045&0.449&0.030\\
25&7812.85749&0.00033&0.445&0.019\\
26&7818.95509&0.00031&0.476&0.022\\
27&7825.05294&0.00053&0.452&0.032\\
28&7831.15205&0.00028&0.521&0.021\\
29&7837.24969&0.00027&0.500&0.019\\
88&8197.14652&0.00036&0.521&0.022\\
89&8203.24762&0.00025&0.498&0.020\\
180&8758.28132&0.00048&0.496&0.038\\
182&8770.47851&0.00032&0.486&0.025\\
\hline
\textbf{TRAPPIST-1f}&&&& \\
\hline
-1&7662.18741&0.00044&0.606&0.030\\
0&7671.39267&0.00045&0.62&0.044\\
15&7809.47544&0.00039&0.662&0.034\\
16&7818.68262&0.00031&0.632&0.023\\
17&7827.88679&0.00027&0.598&0.021\\
18&7837.10323&0.00046&0.567&0.027\\
38&8021.25084&0.00030&0.627&0.027\\
56&8186.91882&0.00025&0.618&0.020\\
118&8757.76211&0.00026&0.645&0.022\\
119&8766.96815&0.00029&0.576&0.070\\
121&8785.38907&0.00021&0.673&0.018\\
\hline
\textbf{TRAPPIST-1g} &&&&\\
\hline
0&7665.35136&0.0005&0.696&0.024\\
12&7813.60685&0.00025&0.605&0.037\\
13&7825.96111&0.00021&0.772&0.024\\
14&7838.30656&0.00027&0.793&0.018\\
42&8184.21900&0.00037&0.784&0.032\\
44&8208.93034&0.00018&0.727&0.019\\
89&8764.82746&0.00035&0.701&0.038\\
90&8777.17382&0.00033&0.687&0.045\\
\hline
\textbf{TRAPPIST-1h} &&&&\\
\hline
0&7662.55444&0.0019&0.307&0.045\\
9&7831.46615&0.0006&0.343&0.021\\
20&8037.93277&0.00052&0.376&0.019\\
28&8188.05070&0.00059&0.360&0.024\\
29&8206.81913&0.00075&0.333&0.022\\
59&8769.83900&0.00077&0.313&0.059\\
\end{longtable}

\renewcommand{\arraystretch}{1.2}
\setlength{\tabcolsep}{14pt} % Default value: 6pt
\begin{longtable}{|cc|cc|c|}
\caption{\label{transit_timing_variations} Transit Timings and transit timing variation calculated as the difference of the transit timing from the value given by the linear regression calculated with the reference timing, the period of the planet and the epoch of the transit. } \\
\hline
 \multicolumn{2}{|c|}{\thead{\textbf{Transit timing + 1$\sigma$ error} \\\textbf{ [$\mathrm{BJD}_{\mathrm{TDB}}-2450000$]}}} & \multicolumn{2}{c|}{\thead{\textbf{TTV + 1$\sigma$ error} \\\textbf{ (min)}}} &
\multicolumn{1}{|c|}{\textbf{Channel}}\\
\hline \hline
\endfirsthead
\caption{continued.}\\
\hline\hline
 \multicolumn{2}{|c|}{\thead{\textbf{Transit timing + 1$\sigma$ error} \\\textbf{ [$\mathrm{BJD}_{\mathrm{TDB}}-2450000$]}}} & \multicolumn{2}{c|}{\thead{\textbf{TTV + 1$\sigma$ error} \\\textbf{ (min)}}} &
\multicolumn{1}{|c|}{\textbf{Channel}}\\
\hline \hline
\endhead
\hline
\endfoot
\textbf{TRAPPIST-1b} &&&\\
\hline 
7440.36516&0.00037&0.37&0.53&c2\\
7452.45228&0.00018&0.53&0.26&c2\\
7463.02844&0.00023&0.58&0.33&c2\\
7651.88733&0.00022&-0.37&0.32&c2\\
7653.39799&0.00034&-0.68&0.49&c2\\
7657.93137&0.00021&0.4&0.3&c2\\
7660.95214&0.00022&-1.02&0.32&c2\\
7662.46368&0.00041&-0.07&0.59&c2\\
7666.99562&0.00013&-1.06&0.19&c2\\
7668.50666&0.00019&-0.82&0.27&c2\\
7670.01776&0.00019&-0.5&0.27&c2\\
7802.97561&0.00016&0.55&0.23&c2\\
7805.99699&0.00015&0.02&0.22&c2\\
7807.50726&0.00017&-0.86&0.24&c2\\
7809.01834&0.0002&-0.56&0.29&c1\\
7812.04034&0.00016&-0.21&0.23&c2\\
7813.55122&0.00015&-0.2&0.22&c2\\
7815.06274&0.0002&0.73&0.29&c1\\
7816.57338&0.00019&0.38&0.27&c1\\
7818.08384&0.00016&-0.22&0.23&c2\\
7819.59475&0.00021&-0.17&0.3&c1\\
7821.10555&0.00018&-0.27&0.26&c1\\
7824.12732&0.00017&-0.25&0.24&c2\\
7825.63815&0.00014&-0.32&0.2&c1\\
7827.14995&0.00012&1.01&0.17&c1\\
7828.66035&0.00023&0.33&0.33&c1\\
7833.19292&0.00024&0.24&0.35&c1\\
7834.70398&0.00015&0.5&0.22&c1\\
7837.72528&0.00017&-0.15&0.24&c1\\
7839.23687&0.00035&0.89&0.5&c1\\
8009.96629&0.00021&1.44&0.3&c1\\
8011.47739&0.00021&1.77&0.3&c1\\
8012.98805&0.00012&1.46&0.17&c1\\
8014.49881&0.00019&1.29&0.27&c1\\
8016.01032&0.00016&2.2&0.23&c1\\
8017.52125&0.0002&2.27&0.29&c1\\
8020.54237&0.00015&1.36&0.22&c1\\
8025.07536&0.0002&1.89&0.29&c1\\
8028.09734&0.00029&2.21&0.42&c1\\
8029.60817&0.00016&2.14&0.23&c1\\
8035.65153&0.00029&1.93&0.42&c1\\
8037.16248&0.00024&2.04&0.35&c1\\
8040.18409&0.00017&1.84&0.24&c1\\
8041.6951&0.00013&2.03&0.19&c1\\
8043.20589&0.00017&1.9&0.24&c1\\
8044.71648&0.00016&1.5&0.23&c1\\
8046.22747&0.00013&1.66&0.19&c1\\
8186.74006&0.00031&3.22&0.45&c2\\
8188.25135&0.00017&3.82&0.24&c2\\
8197.31645&0.00021&3.58&0.3&c2\\
8203.35999&0.00016&3.63&0.23&c2\\
8757.85481&0.00032&8.23&0.46&c2\\
8766.92065&0.00019&9.08&0.27&c2\\
8769.94192&0.00017&8.38&0.24&c2\\
\hline
\textbf{TRAPPIST-1c} &&\\
\hline 
7452.33468&0.00014&-3.24&0.2&c2\\
7650.92393&0.00022&-0.09&0.32&c2\\
7653.3455&0.0002&-0.41&0.29&c2\\
7655.76806&0.00023&0.69&0.33&c2\\
7658.18964&0.0002&0.38&0.29&c2\\
7663.0333&0.00031&0.49&0.45&c2\\
7667.8773&0.00017&1.08&0.24&c2\\
7670.29872&0.00017&0.55&0.24&c2\\
7803.49753&0.00017&0.79&0.24&c2\\
7805.91883&0.00018&0.08&0.26&c2\\
7808.34124&0.00022&0.97&0.32&c2\\
7810.76274&0.00018&0.54&0.26&c2\\
7813.18458&0.00026&0.61&0.37&c2\\
7815.60585&0.00018&-0.14&0.26&c2\\
7818.02835&0.00018&0.88&0.26&c2\\
7820.4502&0.00018&0.96&0.26&c2\\
7822.87188&0.00021&0.79&0.3&c2\\
7825.2938&0.0002&0.98&0.29&c2\\
7827.71523&0.00018&0.45&0.26&c2\\
7832.55892&0.00015&0.6&0.22&c2\\
7834.98113&0.0002&1.2&0.29&c2\\
7837.40274&0.00018&0.94&0.26&c2\\
7839.8241&0.00018&0.31&0.26&c2\\
8011.77142&0.00022&0.29&0.32&c2\\
8014.19266&0.00016&-0.51&0.23&c1\\
8021.45845&0.00017&0.08&0.24&c2\\
8023.87965&0.00018&-0.77&0.26&c1\\
8033.56754&0.00017&0.26&0.24&c2\\
8038.41063&0.0002&-0.46&0.29&c1\\
8040.83248&0.00024&-0.38&0.35&c2\\
8043.25404&0.00016&-0.71&0.23&c1\\
8045.67663&0.00023&0.44&0.33&c1\\
8190.98262&0.00024&-1.9&0.35&c1\\
8203.09199&0.00019&-1.32&0.27&c2\\
8205.51296&0.00019&-2.5&0.27&c1\\
8757.68344&0.00019&-0.24&0.27&c2\\
8764.94941&0.0002&0.61&0.29&c2\\
8769.79241&0.00018&-0.24&0.26&c2\\
8777.05833&0.00019&0.54&0.27&c2\\ 
\hline
\textbf{TRAPPIST-1d} &&&\\
\hline 
7653.9427&0.00042&-6.45&0.6&c2\\
7657.99196&0.00069&-7.2&0.99&c2\\
7662.04264&0.00076&-5.9&1.09&c2\\
7666.09183&0.00054&-6.76&0.78&c2\\
7670.14193&0.00037&-6.3&0.53&c2\\
7803.79081&0.00045&2.53&0.65&c2\\
7807.84029&0.00032&2.1&0.46&c2\\
7811.891&0.00049&3.44&0.71&c2\\
7815.94059&0.0003&3.16&0.43&c2\\
7819.99043&0.00071&3.25&1.02&c2\\
7824.0417&0.0011&5.39&1.58&c2\\
7828.09082&0.00036&4.44&0.52&c2\\
7832.1404&0.00034&4.15&0.49&c2\\
8014.37954&0.00099&2.75&1.43&c1\\
8022.48021&0.00044&4.35&0.63&c1\\
8034.62828&0.00041&2.52&0.59&c1\\
8042.72684&0.00049&1.08&0.71&c1\\
8046.77634&0.00032&0.67&0.46&c2\\
8184.45808&0.00056&-14.87&0.81&c2\\
8208.75643&0.00065&-15.35&0.94&c2\\
\hline 
\textbf{TRAPPIST-1e} &&&\\
\hline 
7654.27839&0.00053&2.4&0.76&c2\\
7660.3803&0.00037&18.2&0.53&c2\\
7806.75784&0.00044&0.9&0.63&c2\\
7812.85752&0.00044&1.06&0.63&c2\\
7818.95511&0.00032&-1.78&0.46&c2\\
7825.05293&0.00048&-4.29&0.69&c2\\
7831.15209&0.0003&-4.88&0.43&c2\\
7837.24972&0.00028&-7.66&0.4&c2\\
8014.13085&0.00021&-16.66&0.3&c2\\
8020.23322&0.00024&-12.62&0.35&c2\\
8038.53517&0.00036&-7.93&0.52&c1\\
8184.94893&0.00032&26.92&0.46&c1\\
8191.04818&0.00058&26.47&0.84&c1\\
8197.14657&0.0004&24.78&0.58&c1\\
8203.24765&0.00026&26.96&0.37&c1\\
8758.28133&0.00047&-11.51&0.68&c2\\
8770.47849&0.00032&-14.35&0.46&c2\\
\hline 
\textbf{TRAPPIST-1f} &&&\\
\hline 
7662.18742&0.00043&29.21&0.62&c2\\
7671.39269&0.00044&27.3&0.63&c2\\
7809.47546&0.00046&4.06&0.66&c2\\
7818.68263&0.00027&4.89&0.39&c2\\
7827.88681&0.00029&1.41&0.42&c2\\
7837.10334&0.00053&15.72&0.76&c2\\
8021.25083&0.00025&38.2&0.36&c2\\
8186.91883&0.00025&-34.8&0.36&c2\\
8196.12562&0.00025&-34.51&0.36&c2\\
8205.32762&0.00028&-41.13&0.4&c2\\
8757.76211&0.00027&14.82&0.39&c1\\
8766.96815&0.0003&14.02&0.43&c1\\
8785.38907&0.00035&25.15&0.5&c2\\
\hline
\textbf{TRAPPIST-1g} &&&\\
\hline
7665.35141&0.00086&-16.46&1.24&c2\\
7813.60688&0.00026&1.97&0.37&c2\\
7825.96111&0.00022&2.94&0.32&c2\\
7838.30658&0.00028&-8.7&0.4&c2\\
8011.24017&0.00031&-32.02&0.45&c2\\
8184.219&0.00054&9.82&0.78&c2\\
8208.93033&0.00018&15.89&0.26&c1\\
8764.82748&0.00031&-2.63&0.45&c2\\
8777.17377&0.00033&-13.09&0.48&c2\\
\hline
\textbf{TRAPPIST-1h} &&&\\
\hline
7662.55448&0.0016&-28.32&2.3&c2\\
7831.46617&0.0006&-19.37&0.86&c2\\
8019.16844&0.00058&23.15&0.84&c2\\
8037.93275&0.0006&18.88&0.86&c2\\
8188.05076&0.00057&-10.19&0.82&c1\\
8206.81914&0.00073&-8.59&1.05&c2\\
8769.83907&0.00083&-6.15&1.2&c2\\
\end{longtable}

\renewcommand{\arraystretch}{1.2}
\setlength{\tabcolsep}{8pt} % Default value: 6pt
\begin{longtable}{|c|cc|cc|c|c|}
\caption{\label{table_dev_res} Median values ($median_{in}$, $median_{out}$) and median absolute deviations ($\sigma_{in}$ and $\sigma_{out}$) of the residuals in and out of transit, using the residuals from the global analyses planet-by-planet. The last column gives the significance of the difference between $median_{in}$ and $median_{out}$, computed as $\frac{\abs{median_{in}-median_{out}}}{\sqrt{\sigma_{out}^{2}+\sigma_{in}^{2}}}$} \\
\hline
\multicolumn{1}{|c|}{\thead{\textbf{Epoch}}}&
\multicolumn{1}{c}{\thead{\textbf{$median_{in}$} \\ \textbf{[ppm]}}}& 
\multicolumn{1}{c|}{\thead{\textbf{ $\sigma_{in}$} \\ \textbf{[ppm]}}} &
\multicolumn{1}{c}{\thead{\textbf{$median_{out}$} \\ \textbf{[ppm]}}} & \multicolumn{1}{c|}{\thead{\textbf{ $\sigma_{out}$} \\ \textbf{[ppm]}}} &
\multicolumn{1}{c|}{\textbf{Significance}} &
\multicolumn{1}{|c|}{\textbf{Channel}}\\
\hline \hline
\endfirsthead
\caption{continued.}\\
\hline\hline
\multicolumn{1}{|c|}{\thead{\textbf{Epoch}}}&
\multicolumn{1}{c}{\thead{\textbf{$median_{in}$} \\ \textbf{[ppm]}}}& 
\multicolumn{1}{c|}{\thead{\textbf{ $\sigma_{in}$} \\ \textbf{[ppm]}}} &
\multicolumn{1}{c}{\thead{\textbf{$median_{out}$} \\ \textbf{[ppm]}}} & \multicolumn{1}{c|}{\thead{\textbf{ $\sigma_{out}$} \\ \textbf{[ppm]}}} &
\multicolumn{1}{c|}{\textbf{Significance}}&
\multicolumn{1}{|c|}{\textbf{Channel}} \\
\hline \hline
\endhead
\hline
\endfoot
\textbf{TRAPPIST-1b} &&&&&\\
\hline
78&348&649&44&630&0.19&c2\\
86&59&396&17&620&0.64&c2\\
93&-168&477&40&590&0.69&c2\\
218&-207&389&6&555&0.16&c2\\
219&247&446&8&644&0.45&c2\\
222&-363&939&178&591&0.21&c2\\
224&-245&384&-83&599&0.01&c2\\
225&20&1076&-8&535&0.4&c2\\
228&199&367&-4&565&0.28&c2\\
229&308&646&-49&513&0.18&c2\\
230&11&596&41&574&0.11&c2\\
318&-99&478&-22&560&0.01&c2\\
320&185&696&-121&428&0.02&c2\\
321&-122&549&12&577&0.45&c2\\
322&-31&347&-154&545&0.45&c1\\
324&-14&526&-62&458&0.55&c2\\
325&158&406&12&562&0.03&c2\\
326&-379&504&75&502&0.65&c1\\
327&337&393&-180&632&0.12&c1\\
328&-256&515&73&516&0.07&c2\\
329&-59&656&80&542&0.29&c1\\
330&214&504&-130&574&0.21&c1\\
332&90&550&-158&814&0.04&c2\\
333&-6&416&117&424&0.2&c1\\
334&-15&535&-5&484&0.5&c1\\
335&103&478&-196&563&0.26&c1\\
338&-178&386&14&555&0.38&c1\\
339&-72&269&17&428&0.23&c1\\
341&45&426&113&436&0.28&c1\\
342&118&558&126&356&0.34&c1\\
455&49&480&36&410&0.06&c1\\
456&278&401&-22&528&0.27&c1\\
457&126&420&-132&383&0.31&c1\\
458&-266&472&146&590&0.31&c1\\
459&-94&380&-110&518&0.49&c1\\
460&-290&427&125&480&0.23&c1\\
462&26&401&-42&410&0.02&c1\\
465&-54&650&-114&538&0.3&c1\\
467&95&444&-89&458&0.43&c1\\
468&34&372&164&479&0.04&c1\\
472&87&299&111&475&0.1&c1\\
473&-30&404&92&456&0.37&c1\\
475&42&442&-243&365&0.17&c1\\
476&3&424&-169&515&0.07&c1\\
477&-226&464&5&388&0.21&c1\\
478&152&354&2&553&0.45&c1\\
479&126&568&-78&458&0.25&c1\\
572&17&304&69&476&0.09&c2\\
573&180&486&-21&550&0.27&c2\\
579&-111&641&164&516&0.33&c2\\
583&237&649&-160&576&0.46&c2\\
950&-82&613&203&573&0.34&c2\\
956&-20&518&34&641&0.33&c2\\
958&92&528&-165&582&0.06&c2\\
\hline
\textbf{TRAPPIST-1c} &&&&&\\
\hline
70&111&344&-26&718&0.02&c2\\
152&-243&418&62&586&0.27&c2\\
153&-223&420&112&475&0.07&c2\\
154&220&561&-76&645&0.45&c2\\
155&107&563&22&644&0.04&c2\\
157&701&782&-124&761&0.41&c2\\
159&376&659&-21&501&0.07&c2\\
160&14&395&75&531&0.17&c2\\
215&-51&660&35&563&0.42&c2\\
216&11&539&99&471&0.53&c2\\
217&-185&428&-59&831&0.35&c2\\
218&-212&393&-7&652&0.1&c2\\
219&-268&738&-14&622&0.76&c2\\
220&-15&401&-137&697&0.48&c2\\
221&-9&552&-16&388&0.09&c2\\
222&-70&564&24&471&0.1&c2\\
223&-98&545&92&666&0.12&c2\\
224&232&556&-3&661&0.13&c2\\
225&20&602&-78&468&0.27&c2\\
227&-16&452&118&690&0.26&c2\\
228&106&594&64&598&0.15&c2\\
229&96&511&-82&468&0.01&c2\\
230&-30&644&8&586&0.13&c2\\
301&-198&721&-38&579&0.22&c2\\
302&-22&494&-37&486&0.27&c1\\
305&124&605&-14&726&0.13&c2\\
306&128&395&-60&561&0.16&c1\\
310&-1&790&6&598&0.05&c2\\
312&20&547&-30&425&0.26&c1\\
313&367&676&-2&521&0.04&c2\\
314&260&356&3&454&0.17&c1\\
315&44&516&24&283&0.15&c1\\
375&86&274&-128&438&0.01&c1\\
380&-142&570&-3&652&0.43&c2\\
381&-30&700&30&534&0.16&c1\\
609&88&708&-65&706&0.15&c2\\
612&-117&889&22&439&0.48&c2\\
614&376&675&-106&730&0.14&c2\\
617&286&816&-168&546&0.46&c2\\
\hline
\textbf{TRAPPIST-1d} &&&&&\\
\hline
-4&-73&594&-24&699&0.07&c2\\
-3&-14&501&128&565&0.29&c2\\
-2&76&824&60&584&0.17&c2\\
-1&-57&698&56&709&0.01&c2\\
0&-96&450&-13&519&0.05&c2\\
33&85&647&14&468&0.19&c2\\
34&124&332&61&606&0.02&c2\\
35&-105&800&91&532&0.11&c2\\
36&143&490&49&582&0.12&c2\\
37&93&610&33&576&0.09&c2\\
38&196&582&-111&653&0.09&c2\\
39&-44&510&68&617&0.2&c2\\
40&99&576&-60&563&0.12&c2\\
85&-144&675&-84&516&0.07&c1\\
87&62&326&-95&439&0.35&c1\\
90&89&280&-3&445&0.14&c1\\
92&-24&508&-17&483&0.2&c1\\
93&124&632&-34&560&0.19&c2\\
127&94&728&26&630&0.07&c2\\
133&344&661&81&586&0.3&c2\\
\hline
\textbf{TRAPPIST-1e} &&\\
\hline
-1&-327&544&121&730&0.42&c2\\
0&-73&657&67&584&0.29&c2\\
24&70&686&94&580&0.06&c2\\
25&61&505&101&535&0.19&c2\\
26&54&506&-26&587&0.21&c2\\
27&19&626&128&682&0.16&c2\\
28&128&653&8&560&0.03&c2\\
29&-13&678&176&466&0.05&c2\\
58&258&456&-22&479&0.1&c2\\
59&57&420&-122&464&0.12&c1\\
62&62&449&20&513&0.14&c1\\
86&72&519&-66&501&0.23&c1\\
87&125&399&-15&516&0.18&c1\\
88&-6&745&158&508&0.12&c1\\
89&-147&499&-54&562&0.01&c2\\
180&33&815&45&612&0.16&c2\\
182&-207&630&-80&501&0.49&c2\\
\hline
\textbf{TRAPPIST-1f} &&&&&\\
\hline
-1&37&558&-23&559&0.1&c2\\
0&-220&824&-15&738&0.23&c2\\
15&-220&552&12&689&0.08&c2\\
16&-75&552&-18&558&0.19&c2\\
17&92&536&30&570&0.26&c2\\
18&-68&390&38&533&0.07&c2\\
38&-36&482&-102&438&0.08&c2\\
56&205&555&-66&450&0.16&c2\\
57&-78&590&-8&407&0.1&c2\\
58&82&545&-86&502&0.38&c1\\
118&89&386&287&746&0.24&c1\\
119&-3&580&92&555&0.12&c2\\
121&54&534&-2&528&0.07&c2\\
\hline
\textbf{TRAPPIST-1g} &&&&&\\
\hline
0&-123&429&75&737&0.04&c2\\
12&46&391&-24&597&0.23&c2\\
13&7&590&146&469&0.1&c2\\
14&-129&588&-7&622&0.18&c2\\
28&-74&554&-42&480&0.14&c2\\
42&-200&569&105&705&0.34&c1\\
44&-130&606&6&575&0.16&c2\\
89&-59&548&117&643&0.21&c2\\
90&21&358&30&583&0.01&c2\\
\hline
\textbf{TRAPPIST-1h} &&&&&\\
\hline
0&-2&828&178&802&0.01&c2\\
9&12&553&11&584&0.16&c2\\
19&6&549&-1&541&0.0&c2\\
20&28&419&99&577&0.1&c1\\
28&1&469&50&517&0.07&c2\\
29&-161&484&-2&530&0.22&c2\\
59&101&433&358&497&0.39&c2\\
\hline

\end{longtable}

\end{appendix}

\end{document}